\documentclass[twocolumn,preprintnumbers,superscriptaddress,nofootinbib,prd,floatfix]{revtex4}

\usepackage{datetime}
\usepackage{color}

\usepackage{graphicx,amsmath,amssymb}
\usepackage{slashed}
\usepackage{bbm}
\usepackage{footmisc}
\usepackage{booktabs}
\usepackage{ulem}

\usepackage{hyperref}
\hypersetup{colorlinks=true, citecolor=blue, urlcolor=blue, linkcolor=blue}

\newcommand{\be}{\begin{eqnarray*}}
\newcommand{\ee}{\end{eqnarray*}}

\newcommand{\bee}{\begin{eqnarray}}
\newcommand{\eee}{\end{eqnarray}}
\newcommand{\beeq}{\begin{equation}}
\newcommand{\eeeq}{\end{equation}}

\newcommand{\beq}{\begin{eqnarray}} 
\newcommand{\eeq}{\end{eqnarray}} 
\newcommand{\non}{\nonumber} 
\newcommand{\MSb}{\overline{\text{MS}}}
\newcommand{\DRb}{\overline{\text{DR}}}

\newcommand{\ti}{\tilde}

\begin{document}

\title{Showcasing HH production: Benchmarks for the (HL-)LHC}

\begin{abstract}
Current projections suggest that the LHC will have only limited
sensitivity to di-Higgs production in the Standard Model (SM),
possibly even after the completion of its high luminosity
phase. Multi-Higgs final states play a fundamental role in many
extensions of the SM as they are intrinsically sensitive to modifications
of the Higgs sector. Therefore, any new observation in multi-Higgs
final states could be linked to a range of beyond the SM (BSM)
phenomena that are not sufficiently addressed by the SM. Extensions of
the Higgs sector typically lead to new phenomenological signatures in
multi-Higgs final states that are vastly different from the SM
expectation. In  this work, we provide a range of signature-driven
benchmark points for resonant and non-resonant BSM di-Higgs production
that motivate non-SM kinematic correlations and multi-fermion discovery
channels. Relying on theoretically well-motivated assumptions, special
attention is devoted to the particular case where the presence of new physics will dominantly manifest itself in multi-Higgs final states.
\end{abstract}

\author{Philipp Basler}\email{philipp.basler@kit.edu}
\affiliation{Institute for Theoretical Physics, Karlsruhe Institute of Technology, 76128 Karlsruhe, Germany}

\author{Sally Dawson}\email{dawson@bnl.gov}
\affiliation{Department of Physics, Brookhaven National Laboratory, Upton, N.Y., 11973, U.S.A.}

\author{Christoph Englert}\email{christoph.englert@glasgow.ac.uk}
\affiliation{SUPA, School of Physics and Astronomy, University of Glasgow, Glasgow G12 8QQ, U.K.}

\author{Margarete M\"uhlleitner}\email{milada.muehlleitner@kit.edu}
\affiliation{Institute for Theoretical Physics, Karlsruhe Institute of Technology, 76128 Karlsruhe, Germany}

\maketitle

\section{Introduction}
\label{sec:intro}
The Higgs precision spectroscopy program that ensued after the discovery of the Higgs boson in 2012~\cite{Aad:2012tfa,Chatrchyan:2012xdj} has assumed a central role in particle physics over the past years. One reason why the Higgs couplings and properties 
measurements move increasingly into the focus of searches for physics
beyond the Standard Model (BSM) is the lack of conclusive hints for
new interactions in the plethora of BSM searches performed
by the ATLAS and CMS experiments. The Higgs boson as the direct implication of electroweak symmetry breaking is typically
considered as a harbinger of new physics due to its special role in the unitarization of scattering amplitudes at high energy~\cite{Cornwall:1973tb,Cornwall:1974km,Lee:1977yc,Lee:1977eg} and its relation to naturalness of
the electroweak scale~\cite{Veltman:1980mj}, to only name a couple of examples.
 
Although modifications of Higgs physics at the TeV scale of this size
are still well within the limits set by recent 13 TeV LHC
measurements, now that the SM can be considered as complete no
additional ultra-violet (UV) energy scale can be predicted from the SM alone. This becomes even more pressing when coupling measurements stay consistent with the SM expectation in the future. 

Additional requirements are more conveniently imposed in
model-specific approaches, which try to mend apparent shortcomings of
the SM such as the lack of a viable dark matter candidate or an
insufficiently first-order electroweak phase transition to address
criteria of baryogenesis~\cite{Sakharov:1967dj}. While
model-independent approaches based on effective field theory
(EFT)~\cite{Grzadkowski:2010es} can inform UV completions that address
these questions through matching calculations, the appearance of novel
phenomenological signatures such as resonances or thresholds within
the LHC's kinematic coverage typically falls outside the region of
reliability of these techniques. 

\begin{figure}[!b]
\includegraphics[width=0.4\textwidth]{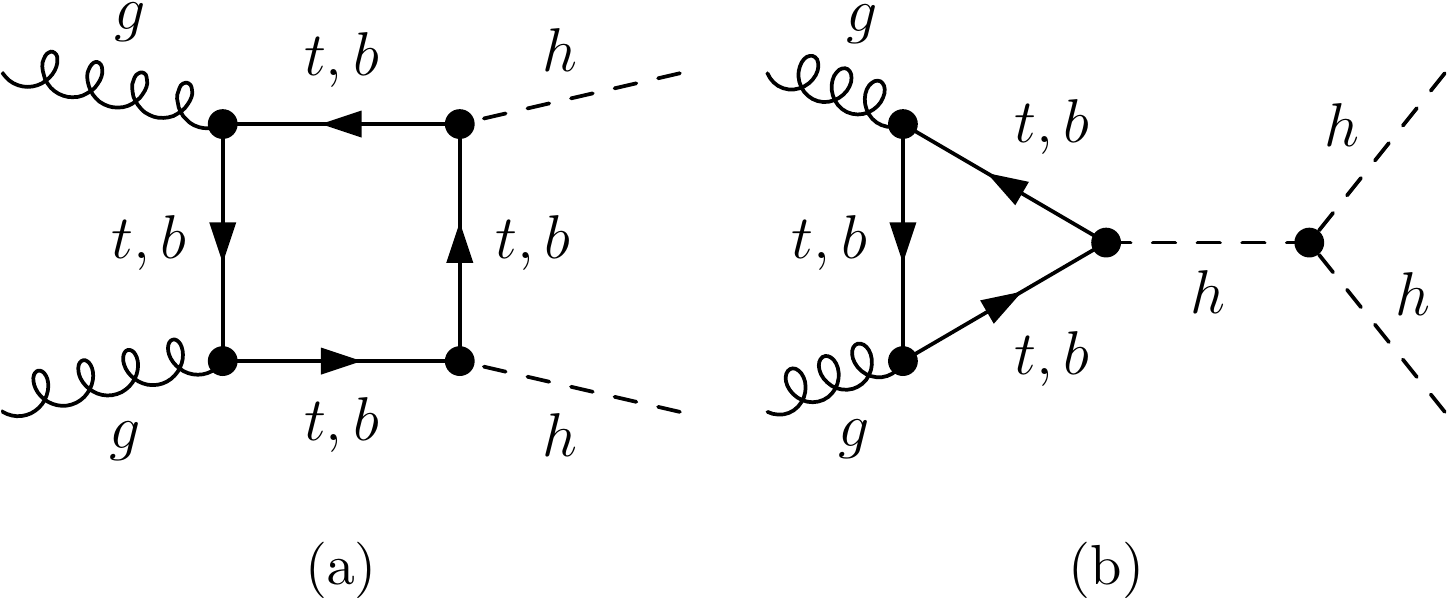}
\caption{Feynman diagrams contributing to $gg\to hh$ production. Although the bottom quark contribution is 
included, it is entirely negligible.}
\label{fig:hhfey}
\end{figure}

A process that highlights the shortcomings of EFT methods in the
presence of thresholds already in the SM context is di-Higgs
production $pp\to hh$ (see {\it e.g.}~\cite{deFlorian:2016spz,Dawson:2018dcd} for recent summaries). 
  Therefore, EFT methods for $gg h^n$ that can be summarized as~\cite{Ellis:1975ap,Shifman:1979eb,Vainshtein:1980ea,Voloshin:1985tc,Kniehl:1995tn} 
\begin{equation}
\begin{split}
{\cal{L}}=&{\alpha_s\over 12\pi} G^{a\,\mu\nu}G^{a}_{\mu\nu} \log \left(1+ {h\over v} \right)\\
&={\alpha_s\over 12\pi} G^{a\,\mu\nu}G^{a}_{\mu\nu} \left( {h\over v} - {h^2\over v^2} \right) + {\cal{O}}(h^3)
\end{split}
\end{equation}
reflect the destructive interference between the top triangle and box
diagrams of Fig.~\ref{fig:hhfey}. While it can be assumed that the
Higgs-top coupling can be accessed at the LHC in the range of $\sim
10\%$ and the intermediate top mass scale is under theoretical control
at the next-to-leading order (NLO)
level\footnote{A recent summary of higher-order corrections to
  Higgs pair production can be found in \cite{Baglio:2018lrj}.}~\cite{Borowka:2016ehy,Borowka:2016ypz,Heinrich:2017kxx,Baglio:2018lrj,Davies:2018qvx,Bonciani:2018omm,Chen:2014ask},
the situation for the
trilinear coupling is less clear as it probes a direction in the
dimension-6 linear EFT space $\sim c_6(H^\dagger H)^3/\Lambda^2$ (we
denote the SM Higgs doublet with $H$ in the following)~\cite{Barger:2003rs}.
As $c_6$ is essentially a free parameter unless a matching calculation is
performed, its size is only limited by technical considerations related
to perturbation theory, on which we need to rely to make
phenomenological predictions. This has raised the question of how
large trilinear coupling modifications can
be~\cite{Baglio:2012np,Gupta:2013zza,DiVita:2017eyz,DiLuzio:2017tfn} 
to inform di-Higgs investigations. Current CMS projections show that a sensitivity of $\lesssim 0.5~\sigma$ at $3~{\text{ab}}^{-1}$~\cite{CMS:2017cwx} seems a realistic target at the LHC. Recent theoretical studies are slightly more
        optimistic \cite{Adhikary:2017jtu,Goncalves:2018yva,Homiller:2018dgu}, but the Higgs trilinear coupling will only understood in great detail even at large luminosity, in particular in the light of possible top quark Yukawa coupling modifications~\cite{Azatov:2015oxa}. As a consequence, di-Higgs production is a main motivation for considering a major energy upgrade of the LHC or a future hadron collider~\cite{Yao:2013ika,Barr:2014sga,Papaefstathiou:2015iba,Zhao:2016tai,Contino:2016spe,Banerjee:2018yxy}.

In concrete UV scenarios that address fundamental BSM questions,
typically a number of exotic states appear in the spectrum that will
not only impact the ``standard'' single Higgs phenomenology at the LHC but can lead to even more dramatic changes in the production of multiple Higgs bosons. For instance, the triangle and box diagrams probe different aspects of top physics in extended top sectors~\cite{Grober:2010yv,Dolan:2012ac,Chen:2014xwa,Grober:2016wmf,Vryonidou:2018eyv}. 

Interleaving modifications of single Higgs physics with theoretically well-motivated UV considerations can therefore turn di-Higgs production into a strong probe of new physics: new kinematic features can appear that motivate new final states and search strategies, which might currently be unconsidered; di-Higgs production can be enhanced or suppressed. Such phenomenological modifications become particularly relevant when extrapolations of standard single Higgs channels do not show a significant departure from their SM expectation in these scenarios. 

We will address these questions in this work using a particular set of
models that allow us to contrast precise theoretical and phenomenological
requirements with concrete predictions of single- and multi-Higgs
production. Imposing, {\it e.g.}, a strong first-order electroweak phase
transition, dark matter constraints, electric dipole measurements and
consistency with current Higgs coupling measurements as well as an extrapolation thereof, we discuss the results of a comprehensive scan of the models' parameter space with a particular emphasis on the relevance of multi-Higgs final states.
We distill this scan into a number of representative benchmark points of
BSM theories that highlight the importance of di-Higgs measurements in
the future. In passing, we discuss how non-SM signatures are
correlated with modifications of single Higgs physics BSM effects (or
a lack of the latter). In turn, this also allows us to formulate an
upper limit of the SM-like di-Higgs production cross section in these
models when there are no conclusive hints for new physics in single
Higgs phenomenology. 

This work is structured as follows: In Sec.~\ref{sec:models}, we
outline the models that we consider for the purpose of this work. 
These are the CP-violating 2-Higgs-Doublet Model (C2HDM) and the
Next-to-Minimal Supersymmetric Extension (NMSSM). Both models are
special in the sense that they feature
extended Higgs sectors that allow for Higgs-to-Higgs decays, also into final
states with different Higgs bosons, or even cascade Higgs-to-Higgs decays.
In Sec.~\ref{sec:scan} we outline the details of our scan over the model
space 
and discuss the overall scan results.
We
present our benchmarks for the C2HDM and NMSSM together with their phenomenological properties in Sec.~\ref{sec:bench}.
We conclude in Sec.~\ref{sec:conc}.

\section{The models} 
\label{sec:models}
\subsection{The C2HDM \label{sec:c2dhm}}
\label{sec:c2hd2m}
The 2-Higgs-Doublet Model (2HDM) \cite{Lee:1973iz,Gunion:1989we,Branco:2011iw} is obtained from the SM by adding a second $SU(2)_L$ Higgs doublet. The
Higgs potential of a general 2HDM with a softly broken $\mathbb{Z}_2$
symmetry, under which $\Phi_1 \to \Phi_1$ and $\Phi_2 \to - \Phi_2$,
can be written as
\beq
V &=& m_{11}^2 |\Phi_1|^2 + m_{22}^2 |\Phi_2|^2 \non\\
&-& (m_{12}^2 \Phi_1^\dagger
\Phi_2 + h.c.) + \frac{\lambda_1}{2} (\Phi_1^\dagger \Phi_1)^2 \non\\
&+& \frac{\lambda_2}{2} (\Phi_2^\dagger \Phi_2)^2 
+ \lambda_3 (\Phi_1^\dagger \Phi_1) (\Phi_2^\dagger \Phi_2) \non\\
&+&  \lambda_4 (\Phi_1^\dagger \Phi_2) (\Phi_2^\dagger \Phi_1) +
\left[\frac{\lambda_5}{2} (\Phi_1^\dagger \Phi_2)^2 + h.c.\right] \; .
\eeq
The absence of Flavour Changing Neutral Currents (FCNC) is guaranteed
by extending the $\mathbb{Z}_2$ symmetry to the fermions. 
Depending on the $\mathbb{Z}_2$ charge assignments, there are four
phenomenologically different types of 2HDMs that are summarized in
Tab.~\ref{tab:2hdmtypes}. 
\begin{table}
\begin{center}
\begin{tabular}{r|ccc} \toprule
& $u$-type & $d$-type & leptons \\ \hline
type I (T1) & $\Phi_2$ & $\Phi_2$ & $\Phi_2$ \\
type II (T2) & $\Phi_2$ & $\Phi_1$ & $\Phi_1$ \\
lepton-specific & $\Phi_2$ & $\Phi_2$ & $\Phi_1$ \\
flipped & $\Phi_2$ & $\Phi_1$ & $\Phi_2$ \\ \bottomrule
\end{tabular}
\caption{The four Yukawa types of the $\mathbb{Z}_2$-symmetric 2HDM
  defined by the Higgs doublet that couples to each kind of fermions. \label{tab:2hdmtypes}}
\end{center}
\end{table}
Hermiticity of the potential requires all parameters to be real, except for
$\lambda_5$ and $m_{12}^2$. If they have different unrelated complex
phases we are in the framework of the complex or CP-violating
2HDM~\cite{Ginzburg:2002wt}, which depends on ten real parameters. In
the description of the C2HDM we follow the conventions
of~\cite{Fontes:2014xva}. The vacuum expectation values (VEVs)
developed by the Higgs doublets after 
electroweak symmetry breaking (EWSB) could in principle be complex in
the C2HDM. Since the phase can be removed by a basis change~\cite{Ginzburg:2002wt},
we set it to zero without loss of generality. In terms of the complex
charged fields $\phi_i^+$ and the real neutral CP-even and CP-odd
fields $\rho_i$ and $\eta_i$ ($i=1,2$), respectively, the Higgs
doublets are given by
\beq
\Phi_1 = \begin{pmatrix} \phi_1^+ \\ \frac{v_1+\rho_1 + i \eta_1}{\sqrt{2}} 
\end{pmatrix}  \quad \mbox{and} \quad
\Phi_2 = \begin{pmatrix} \phi_2^+ \\ \frac{v_2+\rho_2 + i \eta_2}{\sqrt{2}} 
\end{pmatrix} \, ,
\eeq
where $v_1$ and $v_2$ denote the VEVs of the two Higgs doublets $\Phi_1$
and $\Phi_2$, respectively, and $v_1^2+ v_2^2 = v^2$ with the SM VEV
$v \approx 246$~GeV. The ratio of the VEVs is parametrised by the
mixing angle $\beta$,
\beq
\tan\beta \equiv t_\beta= \frac{v_2}{v_1} \;.
\eeq
The minimum conditions obtained from the 
requirement that the minimum of the potential is given by
$\langle \Phi_i \rangle = (0, v_i/\sqrt{2})^T$, can be used to trade
the parameters $m_{11}^2$ and $m_{22}^2$ for $v_1$ and $v_2$. They
also yield a relation between the imaginary parts of $m_{12}^2$
and $\lambda_5$, so that one of the ten parameters is fixed. The
neutral Higgs mass eigenstates $H_i$ ($i=1,2,3$) are obtained from the
neutral components of the C2HDM basis, $\rho_{1,2}$ and $\rho_3 \equiv
(1/\sqrt{2}) (- \sin\beta 
\,\eta_1 + \cos \beta \,\eta_2)$ via the rotation\footnote{Actually, the field $\rho_3$
  is equal to the CP-odd component of the second Higgs doublet in the
  Higgs basis~\cite{Lavoura:1994fv,Botella:1994cs}.} 
\beq
\left( \begin{array}{c} H_1\\ H_2\\ H_3 \end{array} \right) = R
\left( \begin{array}{c} \rho_1 \\ \rho_2 \\ \rho_3 \end{array} \right) \, .
\label{eq:rotc2hdm}
\eeq
The matrix $R$ diagonalizes the mass matrix ${\cal M}$ of the neutral states, 
\beq
R\, {\cal M}^2\, R^T = \textrm{diag} \left(m_{H_1}^2, m_{H_2}^2,
  m_{H_3}^2 \right) \;,
\eeq
where $m_{H_1} \leq m_{H_2} \leq m_{H_3}$ denote the masses of the neutral Higgs
bosons. Introducing the abbreviations $s_{i} \equiv \sin \alpha_i$ and
$c_{i} \equiv \cos \alpha_i$ with
\beq
-\frac{\pi}{2} \le \alpha_i < \frac{\pi}{2} \,,
\label{eg:alpharanges}
\eeq
the mixing matrix $R$ can be parametrised as
\begin{equation}
R =\left( \begin{array}{ccc}
c_{1} c_{2} & s_{1} c_{2} & s_{2}\\
-(c_{1} s_{2} s_{3} + s_{1} c_{3})
& c_{1} c_{3} - s_{1} s_{2} s_{3}
& c_{2} s_{3} \\
- c_{1} s_{2} c_{3} + s_{1} s_{3} &
-(c_{1} s_{3} + s_{1} s_{2} c_{3})
& c_{2}  c_{3}
\end{array} \right) \;.
\label{eq:c2hdmmatrix}
\end{equation}
In total, the C2HDM has nine independent parameters that we choose to
be~\cite{ElKaffas:2007rq}
\begin{equation}
v \;, \quad t_\beta \;, \quad \alpha_{1,2,3}
\;, \quad m_{H_i} \;, \quad m_{H_j} \;, \quad m_{H^\pm} \;, \quad 
\mbox{Re}(m_{12}^2) \;.
\label{eq:2hdminputset}
\end{equation}
The $m_{H_i}$ and $m_{H_j}$ denote any of the three neutral Higgs
boson masses. The third mass is not independent and  is calculated from the
other parameters~\cite{ElKaffas:2007rq}. For further details, in
particular also on the couplings of the C2HDM, see~\cite{Fontes:2017zfn}.

\subsection{The NMSSM \label{sec:nmssm}}
\label{sec:mssm}
Supersymmetric (SUSY) extensions require the introduction of at least a second
Higgs doublet. In the NMSSM, the minimal field content with the doublet
superfields $\hat{H}_u$ and $\hat{H}_d$ is extended by a complex
superfield $\hat{S}$ (for reviews on the NMSSM, see for example
\cite{Ellwanger:2009dp,Maniatis:2009re}). 
The NMSSM Higgs potential is derived from the
superpotential, the soft SUSY breaking Lagrangian and the $D$-term
contributions. The scale-invariant NMSSM superpotential reads in
terms of the hatted superfields
\beq
{\cal W} &=& \lambda \widehat{S} \widehat{H}_u \widehat{H}_d +
\frac{\kappa}{3} \, \widehat{S}^3 + h_t
\widehat{Q}_3\widehat{H}_u\widehat{t}_R^c \non \\
&-& h_b \widehat{Q}_3
\widehat{H}_d\widehat{b}_R^c  - h_\tau \widehat{L}_3 \widehat{H}_d
\widehat{\tau}_R^c \; .
\label{eq:superpotential}
\eeq
For simplicity, we only included here the third generation fermion
superfields, given by the left-handed doublet quark ($\hat{Q}_3$), and
lepton ($\hat{L}_3$) superfields as well as right-handed singlet quark
($\hat{t}^c_R,\hat{b}^c_R$) and lepton ($\hat{\tau}^c_R$)
superfields. 
The soft SUSY breaking Lagrangian 
\beq
\label{eq:Lagmass}
 -{\cal L}_{\mathrm{mass}} &=&
 m_{H_u}^2 | H_u |^2 + m_{H_d}^2 | H_d|^2 + m_{S}^2| S |^2 +
 m_{{\tilde Q}_3}^2|{\tilde Q}_3^2| \nonumber \\
  && + m_{\tilde t_R}^2 |{\tilde t}_R^2|
 +  m_{\tilde b_R}^2|{\tilde b}_R^2| + m_{{\tilde L}_3}^2|{\tilde L}_3^2| +
 m_{\tilde  \tau_R}^2|{\tilde \tau}_R^2| \, , \non \\
\eeq
contains the mass terms $m_x$ for the Higgs ($x=H_u,H_d,S$) and
sfermion
($x=\tilde{Q}_3,\tilde{t}_R,\tilde{b}_R,\tilde{L}_3,\tilde{\tau}_R$)
fields, obtained from the complex scalar components of the
superfields. The Lagrangian with the trilinear soft SUSY breaking
interactions $A_{\lambda,\kappa,t,b,\tau}$ between the sfermion and Higgs fields reads
\beq
\label{eq:Trilmass}
-{\cal L}_{\mathrm{tril}}&=&  \lambda A_\lambda H_u H_d S + \frac{1}{3}
\kappa  A_\kappa S^3 + h_t A_t \tilde Q_3 H_u \tilde t_R^c \non\\
&-& h_b A_b \tilde Q_3 H_d \tilde b_R^c - h_\tau A_\tau \tilde L_3 H_d \tilde \tau_R^c
+ h.c. \;.
\eeq
The contribution to soft SUSY breaking from the gaugino mass
parameters $M_{1,2,3}$ of the bino ($\tilde{B}$), winos ($\tilde{W}$)
and gluinos ($\tilde{G}$), respectively, is given by
\beq
-{\cal L}_\mathrm{gauginos}&=& \frac{1}{2} \bigg[ M_1 \tilde{B}
\tilde{B}+M_2 \sum_{a=1}^3 \tilde{W}^a \tilde{W}_a \non \\
&+& M_3 \sum_{a=1}^8 \tilde{G}^a \tilde{G}_a  \ + \ h.c.
\bigg] \;.
\eeq
Expanding the tree-level scalar potential around the non-vanishing
VEVs of the Higgs doublet and singlet fields,
\beq
H_d &=& \left( \begin{array}{c} (v_d + h_d + i a_d)/\sqrt{2} \\
   h_d^- \end{array} \right) \,, \;  \\
H_u &=& \left( \begin{array}{c} h_u^+ \\ (v_u + h_u + i a_u)/\sqrt{2}
 \end{array} \right) \,, \;
S= \frac{v_s+h_s+ia_s}{\sqrt{2}} \;, \non
\eeq
leads to the Higgs mass matrices for the three scalar ($h_d,h_u,h_s$),
the three pseudoscalar ($a_d,a_u,a_s$) and the charged Higgs states
($h_u^\pm,h_d^\mp$). We choose the VEVs $v_u,v_d$ and $v_s$ to be real
and positive. The three CP-even mass eigenstates $H_i$ ($i=1,2,3$) are
obtained from the interaction states through rotation with the
orthogonal matrix ${\cal R}^S$ that diagonalizes the $3\times 3$ mass
matrix squared of the CP-even fields,
\beq
(H_1, H_2, H_3)^T = {\cal R}^S (h_d,h_u,h_s)^T \;.
\label{eq:scalarrotation}
\eeq
The mass eigenstates are ordered by ascending mass, $M_{H_1} \le
M_{H_2} \le M_{H_3}$. The CP-odd mass eigenstates $A_1$ and $A_2$ are
obtained from a rotation ${\cal R}^G$ that separates the Goldstone
boson, followed by a rotation ${\cal R}^P$ into the mass eigenstates,
\beq
(A_1,A_2,G)^T = {\cal R}^P {\cal R}^G (a_d,a_u,a_s)^T \; .
\label{eq:pseudorot}
\eeq
They are also ordered by ascending mass, $M_{A_1} \le M_{A_2}$. 
Altogether, the NMSSM Higgs spectrum consists of seven
physical Higgs states, three neutral CP-even, two neutral CP-odd and
two charged Higgs bosons. We use
the three minimisation conditions of the scalar potential to replace
the soft SUSY breaking masses squared for $H_u,H_d$ and $S$ in ${\cal
  L}_{\text{mass}}$ by the remaining parameters of the tree-level
potential so that the set of six parameters parametrising the
tree-level NMSSM Higgs sector is given by
\begin{equation}
\lambda\ , \ \kappa\ , \ A_{\lambda} \ , \ A_{\kappa}, \
\tan \beta =v_u/ v_d \ , \
\mu_\mathrm{eff} = \lambda v_s/\sqrt{2}\; .
\end{equation}
The sign conventions are such that $\lambda$ and $\tan\beta$ are
positive, while $\kappa,A_\lambda,A_\kappa$ and $\mu_{\text{eff}}$ can
take both signs. Note that the Higgs boson masses are not input
parameters, but dependent parameters calculated from the input
values. The inclusion of higher-order corrections in the Higgs boson
masses is crucial here to shift the mass of the SM-like Higgs boson to
the observed value of 125~GeV. 

\section{Details of the Scan}
\label{sec:scan}
\subsection{The C2HDM Scan}
The benchmark points{\color{blue}
\footnote{2HDM benchmarks for double Higgs production can
be found in Refs. \cite{Haber:2015pua,Baglio:2014nea}.}}
provided in this paper have to satisfy
theoretical and experimental constraints. In order to find valid
points, we perform a scan in the C2HDM parameter space and
additionally require the mass of one of the Higgs bosons, to be identified with the SM-like one and denoted by $h$, to be $m_h=125.09$~GeV~\cite{Aad:2015zhl}.
The scan ranges are summarized in Tab.~\ref{tab:c2hdmscan}. 
For
simplicity, we only consider the C2HDM Type 1 (T1) and Type 2 (T2),
which cover to a large extent the phenomenological effects to be expected
in the C2HDM.
\begin{table}
\begin{center}
\begin{tabular}{l|c|c|c|c|c} \toprule
& $t_\beta$ & $\alpha_{1,2,3}$ & $\mbox{Re}(m_{12}^2)$ [TeV$^2$] & $m_{H^\pm}$ [TeV] & $m_{H_i\ne h}$ [TeV] \\ \hline
min & 0.8 & $-\frac{\pi}{2}$ & 0 & 0.15/0.59 & 0.01 \\
max & 20 & $\frac{\pi}{2}$ & 0.5 & 1.5 & 1.5 \\ \bottomrule
\end{tabular}
\caption{Input parameters for the C2HDM scan, all parameters varied
  independently between the given minimum and maximum values. The two
  minimum values of the charged Higgs mass range refer to the scan in
  the C2HDM T1 and T2, respectively. For more details, see text. 
\label{tab:c2hdmscan}}
\end{center}
\end{table}
Since physical parameter points with $\mbox{Re} (m_{12}^2) <0$ are
extremely rare, though possible, we neglect them in our scan. We test
for compatibility with the flavour constraints on
$R_b$~\cite{Haber:1999zh,Deschamps:2009rh} and $B \to X_s
\gamma$~\cite{Deschamps:2009rh,Mahmoudi:2009zx,Hermann:2012fc,Misiak:2015xwa,Misiak:2017bgg}
as 2$\sigma$ exclusion bounds in the $m_{H^\pm}-\tan\beta$ plane. In
accordance with~\cite{Misiak:2017bgg} we therefore require $m_{H^\pm}$
to be above 580~GeV in the C2HDM T2, whereas in  the C2HDM T1 this
bound is much weaker and depends more strongly on $\tan\beta$. We
verify agreement with the electroweak precision data by using the
oblique parameters $S$, $T$ and $U$ - the 2HDM formulae are given
in~\cite{Branco:2011iw,Dawson:2017jja} - for which we demand $2\sigma$
compatibility with the SM fit~\cite{Baak:2014ora}, including the full
correlation among the three parameters. We require perturbative
unitarity to hold at tree level.  
The third neutral Higgs boson mass $m_{H_j \ne H_i,h}$, which is not an independent input parameter in the C2HDM, but calculated from the other input values, is demanded to lie in the interval
\beq
10 \mbox{ GeV } \le m_{H_j} < 1.5 \mbox{ TeV}.
\eeq
To avoid degenerate Higgs signals, we additionally impose $m_{H_i \ne h}$ to be 5 GeV away from 125~GeV. For the SM input parameters we use~\cite{PhysRevD.98.030001,Dennerlhcnote}
\begin{equation}
\begin{tabular}{lcllcl}
\quad $\alpha(M_Z)$ &=& 1/127.92, & $\alpha^{\MSb}_s(M_Z)$ &=&
0.118, \\
\quad $M_Z$ &=& 91.187~GeV, & $M_W$ &=& 80.358~GeV,  \\
\quad $m_t$ &=& 172.5~GeV, & $m^{\MSb}_b(m_b^{\MSb})$ &=& 4.18~GeV, \\
\quad $m_\tau$ &=& 1.777~GeV. 
\end{tabular}
\end{equation}
The remaining light quark  and lepton masses have been set to \cite{PhysRevD.98.030001,Dennerlhcnote}
\beq
\begin{array}{lcllcl}
m_e &=& 0.510998928 \mbox{ MeV} \;, & m_\mu &=& 105.6583715
\mbox{ MeV} \;, \\
m_u &=& 100 \mbox{ MeV} \;, & m_d &=& 100 \mbox{ MeV} \;, \\
m_s &=& 100 \mbox{ MeV} \;.
\end{array}
\eeq

\begin{table*}[!t]
\begin{center}
\begin{tabular}{l|ccc|cccccccccccc} \toprule
& $t_\beta$ & $\lambda$ & $\kappa$ & $M_1$ & $M_2$ & $M_3$ & $A_t$ &
$A_b$ & $A_\tau$ & $m_{\tilde{Q}_3}$ & $m_{\tilde{L}_3}$ & $A_\lambda$
& $A_\kappa$ & $\mu_{\text{eff}}$ \\ 
& \multicolumn{3}{|c|}{} & \multicolumn{11}{|c}{in TeV} \\ \hline
min & 1 & 0 & -0.7 & 0.1 & 0.2 & 1.3 & -6 & -6 & -3 & 0.6 & 0.6 & -2 &
-2 & -5 \\
max & 50 & 0.7 & 0.7 & 1 & 2 & 7 & 6 & 6 & 3 & 4 & 4 & 2 & 2 & 5 \\ \bottomrule
\end{tabular}
\caption{Input parameters for the NMSSM scan, all parameters varied independently between the given minimum and maximum values. \label{tab:nmssmscan}}
\end{center}
\end{table*}

In order to perform the scans and find valid parameter points we use
the program {\tt
  ScannerS}~\cite{Coimbra:2013qq,Ferreira:2014dya}. Besides the above
mentioned constraints it also tests for the potential to be bounded
from below and uses the tree-level discriminant
of~\cite{Ivanov:2015nea} to enforce the electroweak vacuum to be the
global minimum of the tree-level Higgs potential. Agreement with the
Higgs exclusion limits from LEP, Tevatron and LHC is checked by using
{\tt HiggsBounds5.2.0}
\cite{Bechtle:2008jh,Bechtle:2011sb,Bechtle:2013wla}  and with the
Higgs rates by using {\tt 
  HiggsSignals2.2.1}~\cite{Bechtle:2013xfa}. The required decay widths and
branching ratios are obtained from the C2HDM implementation {\tt
  C2HDM\_HDECAY}~\cite{Fontes:2017zfn} in {\tt
  HDECAY}~\cite{Djouadi:1997yw,Djouadi:2018xqq}. In the production
cross sections we included the QCD corrections taken over from the SM
and the Minimal Supersymmetric Extension (MSSM), where
available. Electroweak corrections are consistently neglected both in
production and decays. For more details on the production cross
sections, see~\cite{Fontes:2017zfn,Muhlleitner:2017dkd}. 

Working in
the C2HDM, we also make sure to be in agreement with the measurements
of the electric dipole moment (EDM), where the strongest constraint
originates from the electron EDM~\cite{Inoue:2014nva}. We take the
experimental limit given by the ACME
collaboration~\cite{Andreev:2018ayy}. Finally, we also investigate for
the C2HDM if the parameters of the final data set induce a strong first order phase
transition, a necessary condition for successful
baryogenesis~\cite{Sakharov:1967dj,Quiros:1994dr,Moore:1998swa}, by
using the {\tt C++} code {\tt BSMPT}~\cite{Basler:2018cwe}.  

\subsection{The NMSSM Scan}
In order to find benchmark points\footnote{For NMSSM benchmarks 
  for double Higgs production from Higgs-to-Higgs decays, see
  \cite{King:2014xwa}.} that are compatible with the recent
experimental constraints we proceed as described
in~\cite{Costa:2015llh,King:2014xwa,Azevedo:2018llq}, where also
further details can be found.    
We perform a scan in the NMSSM parameter range with scan ranges summarized in Tab.~\ref{tab:nmssmscan}. The remaining mass parameters of the third generation sfermions not listed in the table are chosen as
\begin{equation}   
m_{\tilde{t}_R} = m_{\tilde{Q}_3} \;, \quad m_{\tilde{\tau}_R} =
m_{\tilde{L}_3} \quad \mbox{ and } \; m_{\tilde{b}_R} = 3 \mbox{ TeV} \;.
\end{equation}
The mass parameters of the first and second generation sfermions are
set to
\begin{equation}   
m_{\tilde{u}_R,\tilde{c}_R} = 
m_{\tilde{d}_R,\tilde{s}_R} =
m_{\tilde{Q}_{1,2}}= m_{\tilde L_{1,2}} =m_{\tilde e_R,\tilde{\mu}_R}
= 3\;\mbox{TeV} \;.
\end{equation}
The soft SUSY breaking trilinear couplings of the first two generations are set equal to the corresponding values of the third generation.
In order to ensure perturbativity we applied the rough constraint 
\beq
\lambda^2 + \kappa^2 < 0.7^2 \;.
\eeq
In accordance with the SUSY Les Houches Accord (SLHA) format~\cite{Skands:2003cj,Allanach:2008qq} the
soft SUSY breaking masses and trilinear couplings are understood as
$\DRb$ parameters at the scale 
\be 
\mu_R = M_s= \sqrt{m_{\ti Q_3} m_{\ti t_R}} \;. 
\ee
The SM input parameters have been chosen as in the C2HDM scan,
with
the exception of the top quark mass which has been set to $m_t=
173.5$~GeV. The small difference of 1~GeV has no effect on the scan results.
%

The spectrum of the Higgs and SUSY particles including higher-order corrections is calculated with {\tt NMSSMTools5.2.0}~\cite{Ellwanger:2004xm,Ellwanger:2005dv,Ellwanger:2006rn,Das:2011dg,Muhlleitner:2003vg,Belanger:2005kh} which also checks for the constraints from flavour and low-energy observables. It provides the input for {\tt HiggsBounds}5.2.0~\cite{Bechtle:2008jh,Bechtle:2011sb,Bechtle:2013wla} to check for  compatibility with the exclusion bounds from the Higgs searches. The mass of one of the neutral
CP-even Higgs bosons, identified with the SM-like Higgs boson denoted by $h$, has to lie in the range 
\beq
124 \mbox{ GeV } \le m_h \le 126 \mbox{ GeV} \;,
\eeq
and the masses of all other Higgs bosons are demanded to be separated by at least
1~GeV in order to avoid two overlapping signals.
The signal strengths of this Higgs boson have to be in agreement with the
signal strength fit of~\cite{Khachatryan:2016vau} at the
$2 \times 1\sigma$ level. For the computation of the signal strengths
we need the production cross section and branching ratios for the
NMSSM Higgs bosons.  
To compute production through gluon fusion and $b\bar{b}$
annihilation, we take the SM cross sections and multiply them with the
effective couplings obtained from {\tt NMSSMTools}. The SM values are
calculated with {\tt SusHi}~\cite{Harlander:2012pb,Harlander:2016hcx}
and include in gluon fusion the NLO corrections with the full top quark mass
dependence~\cite{Spira:1995rr} as well as the next-to-next-to-leading order (NNLO)
corrections in the heavy quark effective theory~\cite{Harlander:2002wh,Anastasiou:2002yz,Harlander:2002vv,Anastasiou:2002wq,Ravindran:2003um}. The next-to-next-to-next-to-leading order (N$^3$LO) corrections are taken  into account in a threshold expansion~\cite{Anastasiou:2014lda,Anastasiou:2015yha,Anastasiou:2016cez,Mistlberger:2018etf}
for Higgs masses below 300~GeV. For
masses above 50~GeV, $b\bar{b}$ annihilation cross sections that match
between the five- and four-flavor scheme are used, obtained in the soft-collinear
effective theory~\cite{Bonvini:2015pxa,Bonvini:2016fgf}. They are in accordance with 
the results from~\cite{Forte:2015hba,Forte:2016sja}.  For masses below
50~GeV, cross sections obtained in the Santander matching~\cite{Harlander:2011aa} are used, with the five-flavor scheme cross
sections from~\cite{Harlander:2003ai} and the four-flavor scheme ones
from~\cite{Dittmaier:2003ej,Dawson:2003kb,Wiesemann:2014ioa}. 
The branching ratios are taken from {\tt NMSSMTools} and cross-checked against 
against {\tt NMSSMCALC}~\cite{Baglio:2013iia}. 

The parameter points also have to satisfy the bounds from SUSY searches at the LHC. The gluino mass and the lightest squark mass of the second generation are demanded to lie above 1.85~TeV, respectively, see~\cite{Aad:2015iea}. The stop and sbottom masses are required to be above 800~GeV, respectively,~\cite{Aaboud:2016lwz, Aad:2015iea}, the slepton masses above 400~GeV~\cite{Aad:2015iea} and the absolute value of the lightest chargino mass above 300~GeV~\cite{Aad:2015jqa}. 

Through an interface with {\tt micrOMEGAS}~\cite{Belanger:2005kh} we obtain the relic
density that we require not to exceed the value measured by the PLANCK collaboration~\cite{Ade:2013zuv}. The spin-independent nucleon-dark matter direct
detection cross section, also provided by {\tt micrOMEGAS}, is
required not to violate the upper bound from the LUX
experiment~\cite{Akerib:2016vxi}.
We furthermore test for compatibility with the direct detection limits from XENON1T
\cite{Aprile:2018dbl} and check the Dark Matter annihilation cross
section against the results provided by FermiLat~\cite{Ackermann:2015zua}. 

\subsection{Extrapolations \label{subsec:extrapol}}
We include a range of extrapolations of single Higgs measurements to our discussion to identify an approximate ``exclusion luminosity'' (see below) at which single Higgs measurements will start to become sensitive to a particular scenario and spectrum. This notion will allow us to put multi-Higgs final states in direct comparison with single Higgs measurement expectations and identify interesting regions of parameter space.

In particular, we include projections for the
$m_h\simeq125~\text{GeV}$ standard single Higgs production modes, $g g
\to h$ (gluon fusion), $q q \to h j j$ (weak boson fusion), $qq \to
Vh$, $V=W^\pm,Z$ (Higgs radiation), $pp\to t\bar t h$ (associated
production with top quarks), and consider decays $h\to ZZ$, $h\to WW$,
$h\to \gamma \gamma$, $h\to b\bar b $ and $h\to \tau^+\tau^-$. For the
decays $h\to \gamma\gamma,ZZ$ we use the CMS projections provided in
Ref.~\cite{CMS:2017cwx}; these include the production modes gluon
fusion, weak boson fusion and $t\bar th$. We interpolate between different
luminosities using a $\sqrt{\cal{L}}$ luminosity dependence at all
times. 

Projections for $h\to WW$ are obtained using Ref.~\cite{ATLAS:2014aga} and we rescale these results taking into account the cross section differences between 13 TeV and 8 TeV using the results provided by the Higgs cross section working group~\cite{Dittmaier:2011ti}. For $h\to b\bar b$ we consider extrapolations based on $Vh$ production~\cite{Aaboud:2017xsd}, $t\bar t h$ production~\cite{Sirunyan:2017elk} as well as weak boson fusion~\cite{Khachatryan:2015bnx}. $h\to \tau \tau$ is based on~\cite{Sirunyan:2017khh}, which agrees with the ECFA results of~\cite{CMS:2017cwx} upon projecting to 3~ab$^{-1}$.

The improved determination of the SM-like Higgs boson with $m_h\simeq
125~\text{GeV}$ needs to be contrasted with additional coverage of
Higgs-like searches for masses $m_h\neq 125~\text{GeV}$. We include
projections of existing resonance searches in
$\gamma\gamma$~\cite{Aaboud:2017yyg},
$\tau\tau$~\cite{Aaboud:2017sjh}, $WW$~\cite{Aaboud:2017gsl} and
$ZZ$~\cite{Aaboud:2017rel} final states. By far the most constraining
exotic searches result from $t\bar t$ resonance searches, and we
extrapolate the results of Ref.~\cite{Aaboud:2018mjh}. This analysis
is performed in the context of a $Z'$ model and can therefore be
interpreted as only a rough estimate of the search potential of $t\bar
t$ Higgs resonances. To our knowledge no comprehensive analysis of
exotic heavy Higgs masses is publicly available. This is mostly due to
the dedicated interference between the background and the
Higgs signal that also depends on the CP character of the produced
state~\cite{Aaboud:2017hnm}. This has a significant impact on the
sensitivity of $t\bar t$ final states. Including such effects is
beyond the scope of this work. 
%

\subsection{Results}
Both in the C2HDM and the NMSSM the enlarged Higgs sector leads to a
plethora of di-Higgs production processes. In particular, they feature
processes with two different Higgs bosons in the final state. Compared
to the SM, the cross sections can be largely enhanced in case of
resonant production of a heavy Higgs boson that subsequently decays
into a pair of lighter Higgs bosons, provided the Higgs self-coupling
is not too small. Moreover, the different Higgs self-couplings
themselves can enhance the cross section in view of the well-known
fact that in the SM the triangle and box diagrams interfere
destructively. Additionally, loops with bottom quarks may play a role
in scenarios with enhanced down-type Yukawa couplings for large
$\tan\beta$ in the NMSSM or the C2HDM T2. In the
NMSSM  loops with stop and sbottom quarks  also contribute to Higgs pair
production, and we furthermore have the possibility to produce 
a di-Higgs final state with pseudoscalars.
These
processes can yield even larger rates as has been discussed in detail
in~\cite{Costa:2015llh}. However, due to supersymmetry, the Higgs
self-couplings are given in terms of the gauge couplings
limiting to some extent deviations from the SM.
This is not the case for the C2HDM, so that effects different from the
NMSSM may be expected here. In the
NMSSM, on the other hand, we have the possibility of the final state
Higgs bosons to decay into non-SM final states like {\it
  e.g.}~neutralinos, inducing signatures
with phenomenologically interesting features. Altogether, both models
provide a large playground for possible BSM effects in Higgs pair
production that can be rather different.  

Furthermore, we restrict ourselves to SM final states. Most of our
results show the leading order (LO) Higgs pair production cross sections. For the benchmark points we also computed the NLO QCD corrections in the limit of heavy loop particles. They typically increase the cross section by about a factor two. We have implemented the NLO QCD corrections both for the NMSSM\footnote{They only include the corrections to the top quark loops. For NLO QCD corrections including also the squarks in the limit of vanishing external momenta, see~\cite{Agostini:2016vze}.}~\cite{Costa:2015llh} and C2HDM~\cite{Grober:2017gut} in the Fortran code {\tt HPAIR}\footnote{See M.~Spira's website, \url{http://tiger.web.psi.ch/proglist.html}.} that was originally designed to compute the SM and MSSM Higgs pair production at NLO QCD. All Higgs pair production processes have been computed at a c.m.~energy of $\sqrt{s}=14$~TeV, and we have adopted the CT14 parton densities~\cite{Dulat:2015mca} for the LO and NLO cross sections with $\alpha_s (M_Z) =0.118$ at LO and NLO.
 The renormalisation scale has been set equal to $M_{HH}/2$, where
 $M_{HH}$ generically denotes the invariant mass of the final state
 Higgs pair. Consistent with the application of the heavy top quark
 limit in the NLO QCD corrections, we neglect the bottom quark loops
 in the LO cross section.

In view of the possibility of (largely) enhanced production of a 
pair of SM-like Higgs bosons, in the selection of valid scenarios, we
also took into account limits set by LHC  
$4b$ \cite{Aaboud:2018knk,CMS:2017xxp,Sirunyan:2017isc}, $(2b)(2\tau)$
\cite{CMS:2017vdr,Sirunyan:2017djm,Aaboud:2018sfw} and $(2b)(2\gamma)$
\cite{CMS:2017ihs} final states from the  production of a heavy scalar
resonance that decays into two 125~GeV Higgs bosons.

\subsubsection{C2HDM}
We start by discussing the possible sizes of Higgs pair production
that are compatible with all present experimental
constraints. Table~\ref{tab:maxcxn} summarizes the maximum cross
section values found in the sample of valid parameter points where we
additionally applied the extrapolations of
Subsection~\ref{subsec:extrapol}. Taking these into account, we
only kept the points that are not excluded at 64~fb$^{-1}$, which
corresponds approximately to the present luminosity acquired by the
LHC experiments. We will come back to the role of the extrapolations
below. In the following, we  denote the SM-like Higgs boson with a
mass of 125~GeV $h$, the lighter 
of the non-SM like neutral Higgs bosons is called $H_\downarrow$, the
heavier one $H_\uparrow$. All cross sections are calculated at LO QCD
and hence still increase by approximately a factor two when QCD
corrections are included.  

\begin{table}[b!]
\vspace*{0.2cm}
\begin{tabular}{c|c|c} \toprule
$H_iH_j$/model & T1 & T2 \\ \hline
$hh$ & 794 & 34.2 \\
$h H_\downarrow$ & 49.17 & 11.38 \\
$h H_\uparrow$ & 17.65 & 10.84 \\
$H_\downarrow H_\downarrow$ & 3196 & 0.18 \\
$H_\downarrow H_\uparrow$ & 12.58 & 0.11 \\
$H_\uparrow H_\uparrow$ & 7.10 & 0.18 \\ \bottomrule
\end{tabular}
\caption{Maximum cross section values in fb for LO gluon fusion into
  Higgs pairs, $\sigma( gg \to H_i H_j)$, in the C2HDM T1 and T2, with
  an exclusion luminosity $\ge 64$~fb$^{-1}$ that satisfy all
  theoretical and experimental constraints described above.
 \label{tab:maxcxn}}
\end{table}

\begin{figure*}[t!]
  \centering
  \includegraphics[width=0.47\linewidth]{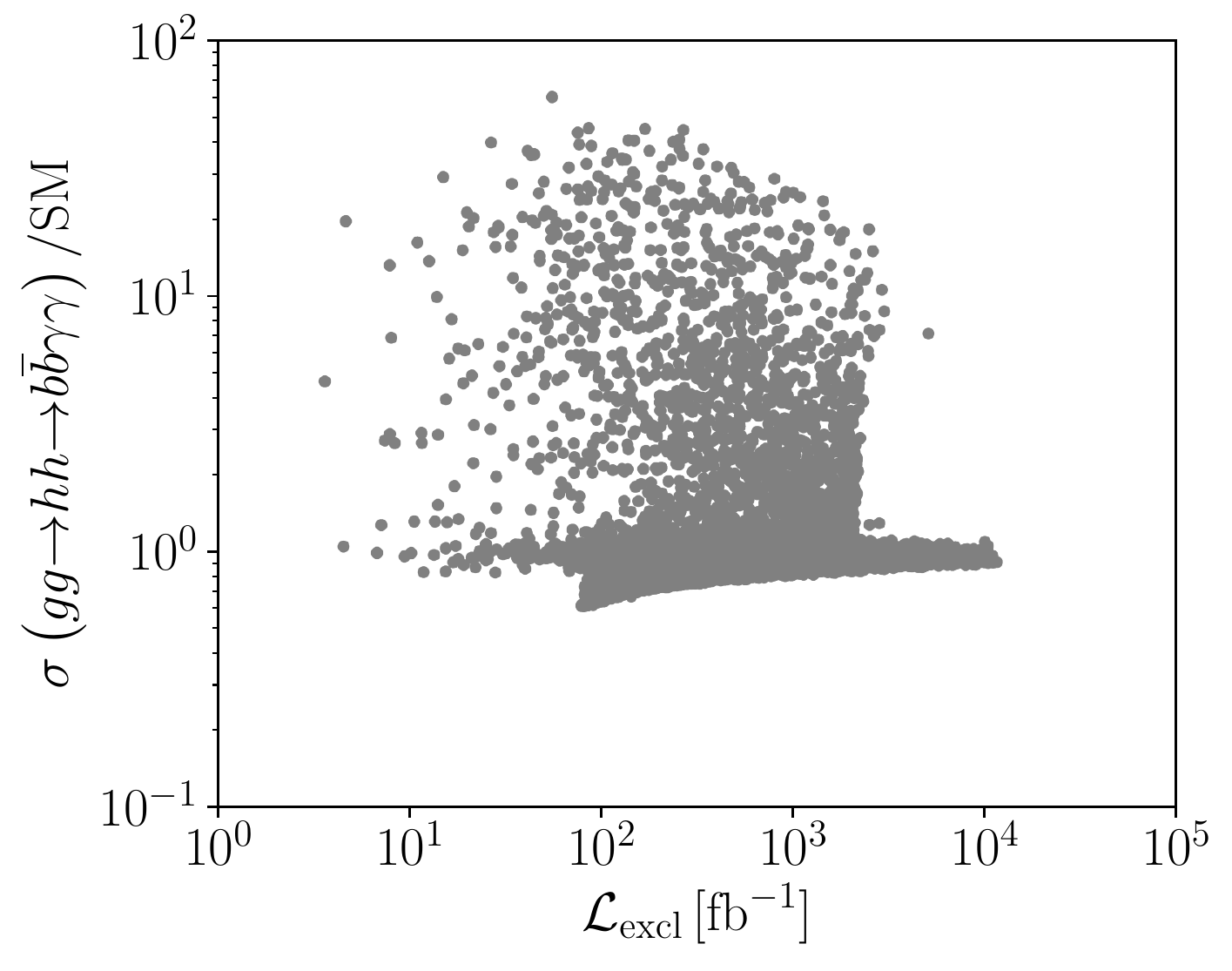}
  \includegraphics[width=0.47\linewidth]{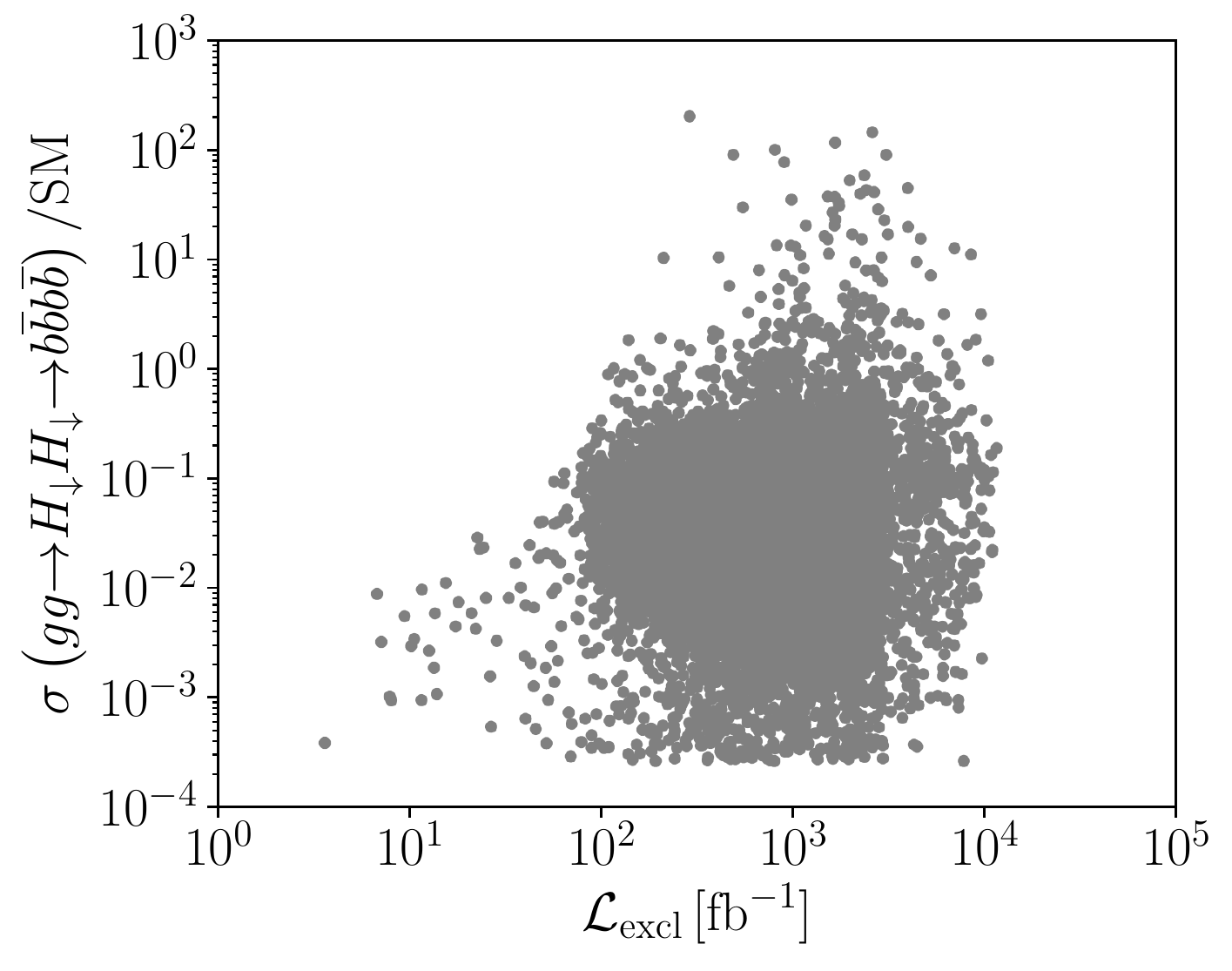}
  \caption{C2HDM T1: Scatter plots for scenarios passing our applied
    constraints: Higgs pair production cross sections normalized to
    the SM value for SM-like Higgs pairs decaying into
    $(b\bar{b})(\gamma\gamma)$ (left) and light-non-SM-like Higgs pairs
    decaying into $(b\bar{b})(b\bar{b})$ (right) as a function of the
    exclusion luminosity. \label{fig:finalversuslumi}}
\end{figure*}

The SM Higgs pair production cross section in gluon fusion amounts to
19.72~fb at LO and 38.19~fb at NLO\footnote{This value differs from
  the one given in~\cite{Borowka:2016ehy}. This is because we do not
  include top quark mass effects here and use a different pdf set.}
with a $K$-factor of $K=1.95$. Inspection of the Tab.~\ref{tab:maxcxn} shows, that
both in  the T1 and T2 scenarios, the maximum attained cross section
for $hh$ production can exceed the SM value, in T1 by a factor of
about 40 and in T2 by about a factor 1.7. This is also the case at
NLO, with NLO cross sections and $K$-factors for $hh$ production in
the T1 and T2 models given by 
\beq
\begin{array}{lll}
\mbox{T1:} & \quad \sigma(hh)^\text{NLO} = 1.64 \mbox{ pb}, & \quad K = 2.06,  \\
\mbox{T2:} & \quad \sigma(hh)^\text{NLO} = 64.96 \mbox{ fb}, & \quad K = 1.94. 
\end{array}
\eeq
The reason for the large enhancement in T1 is the resonant production
of the heavier Higgs bosons $H_\downarrow$ and $H_\uparrow$ with a
mass of 285~GeV and 287~GeV, respectively, that subsequently decay
into a pair of SM-like Higgs bosons. This is also the reason for the
enhancement in T2, where the masses of the non-SM-like Higgs bosons
amount to 794 and 798~GeV. The reason for the much smaller enhancement
in $hh$ production in T2 compared to T1 is the overall heavier Higgs
spectrum in T2. In particular, the intermediate heavy resonances in
the T2 scenarios that can produce $hh$ in their decay usually fall
into the heavy mass range where the ATLAS and CMS limits on the upper
cross section for $(4b)$ production drop rapidly, {\it
  cf.}~\cite{Aaboud:2018knk,CMS:2017xxp}. In T1, furthermore, the
maximum values of di-Higgs production processes involving
$H_\downarrow$ can compete with SM Higgs pair production or even
largely exceed it. Thus the production of a SM-like Higgs boson and
$H_\downarrow$ can be larger by a factor 2.5. This final state is
interesting as it is clearly a non-SM-like signature where the
experiments can use the SM-like Higgs boson to calibrate or ``tag''
this exotic configuration. This does not apply for T2, however, where
due to the experimental constraints, the non-SM-like Higgs bosons are
in general heavier than in T1, inducing small di-Higgs production
processes due to a much smaller phase space.   

\subsubsection*{Experimental accessibility and exclusion luminosity}
In order to assess the experimental accessibility of these cross
sections, however, we need to look at their decay products. We
therefore applied the narrow width approximation and multiplied the
produced Higgs bosons with their branching ratios in various SM final
states. The most promising final states for the investigation of Higgs
pair production at the LHC are the $(b\bar{b})
(\gamma\gamma)$~\cite{Baur:2003gp}, the $(b \bar{b})
(\tau\tau)$~\cite{Baur:2003gpa,Dolan:2012rv,Barr:2013tda} and
$(b\bar{b})(b\bar{b})$~\cite{Baur:2003gpa,deLima:2014dta,Wardrope:2014kya} final
states (for other final states see also~\cite{Baur:2002rb,Baur:2002qd,Papaefstathiou:2012qe}). In Fig.~\ref{fig:finalversuslumi} we  show for all
parameter points that pass our applied constraints for the C2HDM T1,
the cross section values of SM Higgs pair production in the
$(b\bar{b})(\gamma\gamma)$ final state (left) and for $H_\downarrow
H_\downarrow$ production in the $(b\bar{b})(b\bar{b})$ final state
(right) normalized to the corresponding SM values 
as a function of the exclusion luminosity. By the latter we define the
luminosity at which this process would be excluded experimentally,
based on the extrapolations described in
Subsection~\ref{subsec:extrapol}. First of all we notice that the
figures contain parameter points at lower luminosity that should have
been excluded by {\tt HiggsBounds}. The reason why they are there is
that {\tt HiggsBounds} relies on the published experimental results
and cannot check for certain signatures that become relevant in BSM
Higgs sectors. Thus there exist Higgs spectra with heavy Higgs bosons
that dominantly decay into top quark pairs. These would induce exotic
four-top final states in heavy Higgs pair production. Such  signatures
compete, however, with single heavy Higgs production and subsequent
decay into a top-quark pair. Applying our rough estimate on the
exclusion power of the experiments for this process, based on the $Z'$
data, such scenarios are excluded already, although they have been let
through by {\tt HiggsBounds} due to the lack of a dedicated experimental
analysis for this. This shows the importance of experimental analyses
investigating top pair final states from heavy Higgs production in
order to properly assess the exclusion limits for BSM Higgs sectors -
with dramatic effects on possible Higgs pair production
signatures. While our rough extrapolation excludes about 0.6\% of the T1
points for a luminosity of about
36~fb$^{-1}$, the effect is much larger for the T2 sample allowed
by {\tt HiggsBounds}\footnote{{\tt HiggsBounds} takes into account
data at 36~fb$^{-1}$.}. Here about 22\% of the points would be
excluded. This is because of the overall heavy non-SM-like Higgs bosons in T2
and their prominent decays into top-quark pairs. 

As can be inferred from the figures in the C2HDM T1, the production of
a SM-like Higgs pair with subsequent decay into
$(b\bar{b})(\gamma\gamma)$ can exceed the SM rates by up to a factor
60. This maximum enhancement factor is the same for all final states, as
the branching ratios of the SM-like Higgs boson $h$ are almost the
same as in the SM. In the following, we will use the quantity 
\begin{equation}
\Sigma_X=\sum\limits_{i\in {\text{SM}}\backslash\{h\}} \text{BR}(X\to i)\,,
\end{equation}
to classify whether a Higgs boson $X$ has a sizable non-SM branching
ratio and decay phenomenology. If $\Sigma_X\simeq 1$ then the exotic
states can be dominantly discovered in ``standard'' SM-Higgs-like
decay channels, {\it e.g.} $X\to b\bar b$ or $t\bar t$ if the mass of $X$
permits such a decay. 

In the $H_\downarrow H_\downarrow$ final state with both
$H_\downarrow$'s decaying into bottom quarks the enhancement can even
be up to a factor of about 200. The point with the maximum enhancement
corresponds to the one quoted in Tab.~\ref{tab:maxcxn} and the enhancement is due to the large di-Higgs
production process of 3.2~pb and a slightly enhanced branching ratio
into $b$-quarks as compared to the SM. The same factor is found for
the $(b\bar{b})(\tau\bar{\tau})$ final state. Due to a smaller
branching ratio into photons, however, the maximum allowed enhancement
in the $(b\bar{b})(\gamma\gamma)$ final state only amounts up to a
factor  of 40. The $H_\downarrow$ in this scenario has a mass of
$m_{H_\downarrow}=131$~GeV, and the mass of $H_\uparrow$ is
$m_{H_\uparrow}=313$~GeV. Its main branching ratios are BR($H_\uparrow
\to Z H_\downarrow$)= 0.53 and BR($H_\uparrow \to H_\downarrow
H_\downarrow$) = 0.46. The maximum branching ratios of the charged
Higgs boson with a mass of $m_{H^+} = 312$~GeV are BR($H^+ \to W^+
H_\downarrow$)=0.65 and BR($H^+ \to t \bar{b}$)=0.34. With its large
di-Higgs production cross section and the large non-SM-like branching
ratios, this parameter point is an interesting scenario for studying
new physics effects (also beyond the Higgs pair events that we
consider here).  

All remaining di-Higgs production processes are less promising. Thus
the enhancement factor for $h H_\downarrow$ production remains below 3
in the $4b$ and $2b2\tau$ final state and below 2 in the $2b2\gamma$
final state. All other final states range below the SM values.  

As can already be inferred from the maximum di-Higgs production values
in T2, given in Tab.~\ref{tab:maxcxn} the situation looks much less
promising in the C2HDM T2. There are  very few points in $hh$
production with subsequent decay into the $(2b)(2\tau)$ and $4b$ final
state that exceed the SM rate, and only by a factor of about 2.4. The maximum enhancement found in the
$(2b)(2\gamma)$ final state is about 2.4. All
other final states lead to smaller rates than in the SM. 

From these considerations we can conclude that there are promising
di-Higgs signatures with large rates in the C2HDM T1 both for SM-like
Higgs pair production but also for final states with non-SM-like Higgs
bosons. The exotic Higgs bosons appear in SM-like final states,
however, with different kinematic correlations due to different
masses. This highlights the need to conduct Higgs pair analyses in a
broad range of kinematic possibilities. 
Furthermore, the strict constraints on T2 scenarios, would exclude the model if
di-Higgs signatures much larger than in the SM are found. 

\subsubsection{NMSSM}
In Table~\ref{tab:nmssmmaxcxn} we summarize for the NMSSM the maximum
di-Higgs production cross section values found in the sample of valid
parameter points that are not excluded at a luminosity of
64~fb$^{-1}$. All cross sections are calculated at LO QCD and hence
still increase by approximately a factor two when QCD corrections are
included. By $A_\downarrow$ we denote the lighter of the two
pseudoscalar Higgs bosons. 
\begin{table}[h]
\vspace*{0.2cm}
\begin{tabular}{c|c} \toprule
$H_iH_j$ & NMSSM \\ \hline
$hh$ & 67 \\
$h H_\downarrow$ & 26 \\
$h A_\downarrow$ &  493 \\
$h H_\uparrow$ & 25  \\
$H_\downarrow H_\downarrow$ & 4114 \\
$H_\downarrow H_\uparrow$ & 1.20  \\
$H_\uparrow H_\uparrow$ & 0.09 \\
$A_\downarrow A_\downarrow$ & 15894 \\ \bottomrule
\end{tabular}
\caption{NMSSM: Maximum cross section values 
in fb for LO gluon fusion into Higgs pairs, $\sigma( gg \to H_i H_j)$
with an exclusion luminosity $\ge 64$~fb$^{-1}$ that satisfy
all theoretical and experimental constraints described above. \label{tab:nmssmmaxcxn}}
\end{table}
\begin{figure*}[ht!]
  \centering 
  \includegraphics[width=0.47\linewidth]{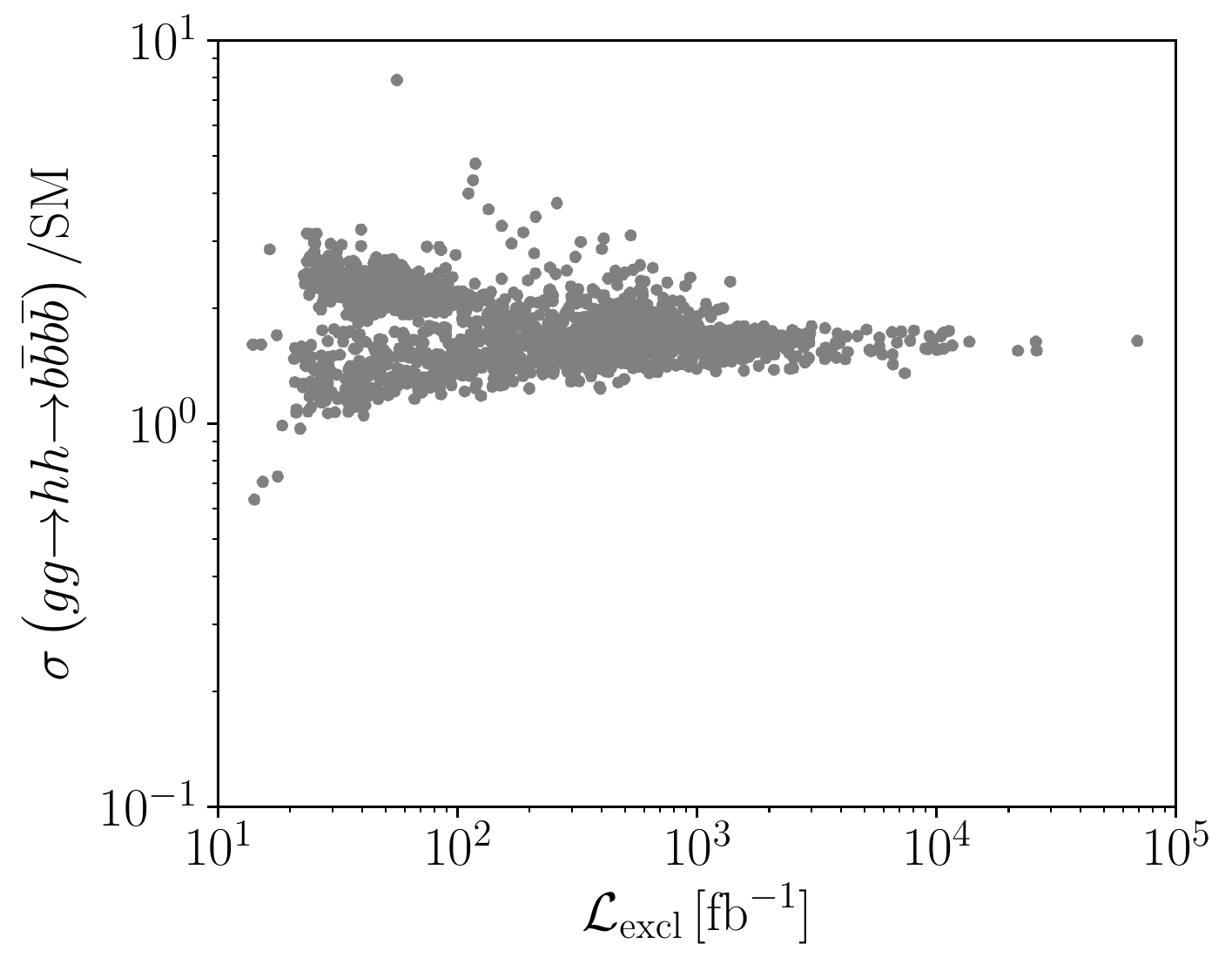}
  \includegraphics[width=0.47\linewidth]{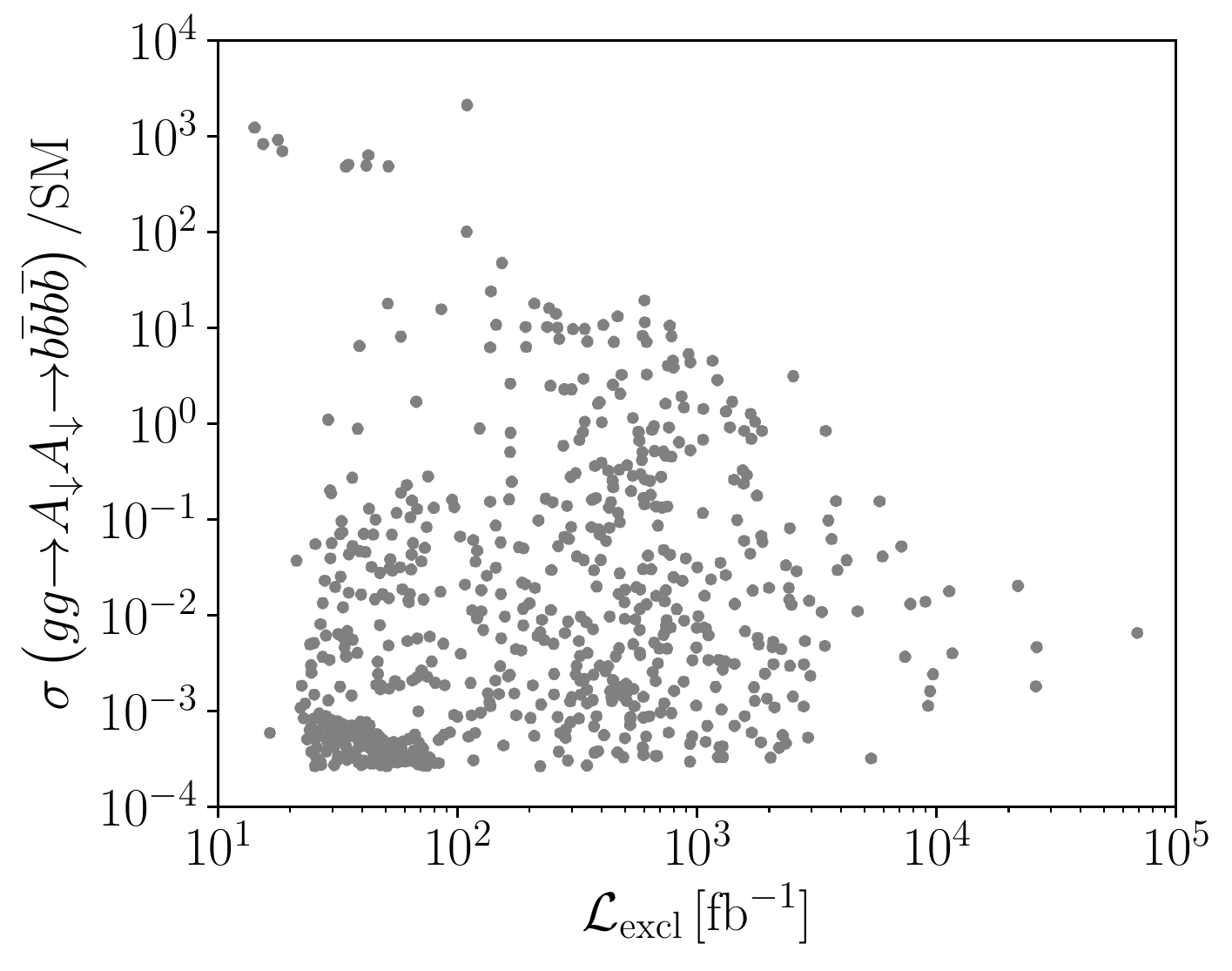}
  \caption{NMSSM: Scatter plots for scenarios passing our applied
    constraints: Higgs pair production cross sections normalized to
    the SM value for SM-like Higgs pairs decaying into
    $(b\bar{b})(b\bar{b})$ (left) and $A_\downarrow A_\downarrow$ Higgs pairs
    decaying into $(b\bar{b})(b\bar{b})$ (right) as a function of the
    exclusion luminosity. \label{fig:nmssmfinalversuslumi}} 
\end{figure*}

The reason for the large enhancement of $\sigma (gg\to hh)$ is the
intermediate resonant production of heavy Higgs bosons $H_\downarrow$
and $H_\uparrow$ with subsequent decay into a SM-like Higgs pair. 
The $H_\downarrow H_\downarrow$ production cross section
is so large because of the smallness of the $H_\downarrow$ mass,
$m_{H_\downarrow} = 39.52$~GeV. The enhancement in $h A_\downarrow$
production is due to the resonant $A_2 \equiv A_\uparrow$ production with subsequent
decay into $hA_\downarrow$. The huge enhancement in $A_\downarrow
A_\downarrow$ production is on the one hand due to the smallness of the
$A_\downarrow$ mass of $m_{A_\downarrow} = 37$~GeV, on the other hand
due to the resonant $H_\uparrow$ production with subsequent decay into
$A_\downarrow A_\downarrow$ (the resonant $H_\downarrow$ production
plays a minor role). Searches for relatively low mass states are performed in the $\gamma\gamma$~\cite{Sirunyan:2018aui} and $\tau\tau$~\cite{Sirunyan:2018zut} channels, however,
with rather limited sensitivity.

\subsubsection*{Experimental accessibility and exclusion luminosity}
\begin{figure}[t!]
  \centering
  \includegraphics[width=0.95\linewidth]{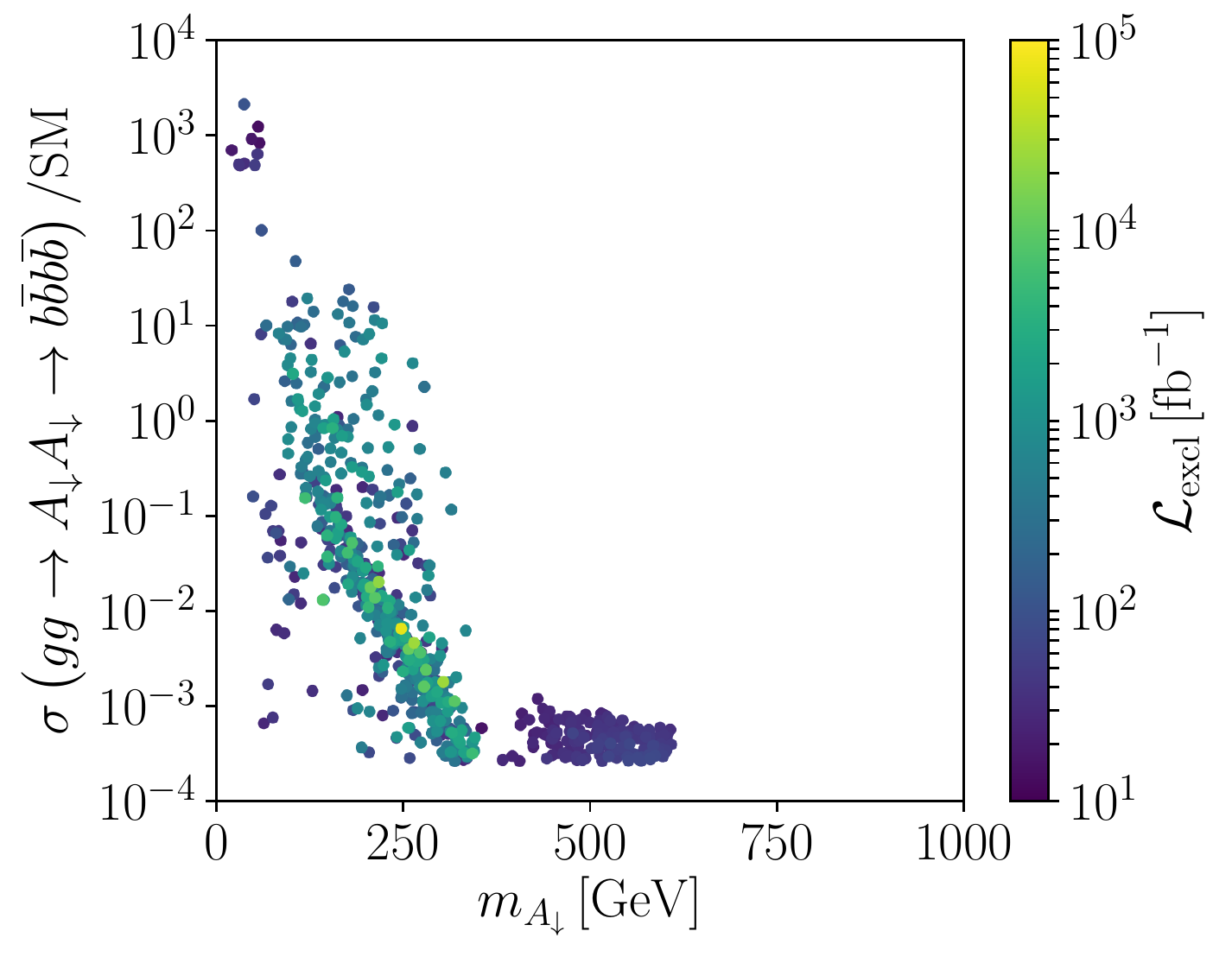}
  \caption{NMSSM: Scatter plots for $4b$ final state rates from
    $A_\downarrow A_\downarrow$ production normalized to the SM rate
    as a function of $m_{A_\downarrow}$. The colour code denotes the
    exclusion luminosity. \label{fig:diadownarrow}}
\end{figure}

In Fig.~\ref{fig:nmssmfinalversuslumi} we show for all parameter points that pass our applied constraints, the NMSSM cross section values of SM Higgs pair production in the $(b\bar{b})(b\bar{b})$ final state (left) and for $A_\downarrow A_\downarrow$ production in the $(b\bar{b})(b\bar{b})$ final state (right) normalized to the corresponding SM values 
as a function of the exclusion luminosity.

As can be inferred from Fig.~\ref{fig:nmssmfinalversuslumi} (left),
the $4b$ final state rates from SM-like Higgs pair production 
exceed the SM reference value by less than a factor 10 and only for
lower exclusion luminosities. 
Large enhancement factors are basically
limited by the LHC upper limits on heavy resonant scalar production
with subsequent decay into a SM-like Higgs pair.
The
situation looks even less promising in the production of an SM-like Higgs
boson together with the lighter of the CP-even non-SM-like Higgs
bosons, where only an enhancement factor slightly above 2.3 at most is
found. This is the case for high exclusion luminosities beyond
1~ab$^{-1}$ so that nevertheless this process might be accessible at
high luminosities. The situation is different in the
production of $h A_\downarrow$. Because the lighter pseudoscalar can be
relatively light and decays dominantly into $(b\bar{b})$\footnote{Note that typical trigger criteria
are too selective to directly observe $pp\to A_\downarrow \to b\bar{b}$.} we can have
enhancement factors above 10 up to about 45 in the $4b$
state. This makes it
particularly interesting, moreover in view of the exotic final state with two
different Higgs masses in di-Higgs production. The enhancement factors
can become huge in $A_\downarrow A_\downarrow$ production, which is mainly
due to the lightness of $A_\downarrow$. In $4b$ production it can be
up to 1000. For larger exclusion luminosities the enhancement factor
can still be a factor up to 10,  {\it
  cf.}~Fig.~\ref{fig:nmssmfinalversuslumi} (right). In the
$(b\bar{b})(\gamma\gamma)$ final state the enhancement can be larger
than 100 up to about 360.  

Figure~\ref{fig:diadownarrow} nicely summarizes
the situation of the enhanced di-Higgs cross sections involving very
light Higgs bosons. It shows the production of $A_\downarrow
A_\downarrow$ with subsequent decay in the $4b$ final state normalized
to the value of the corresponding process for the SM-like di-Higgs
production, as a function of the mass of the light pseudoscalar. The
color code denotes the exclusion luminosity. For very light masses
below 125~GeV (note the gap around 125~GeV is due to our scan
procedure) the rates are largely enhanced because of the large
di-Higgs production cross sections. With increasing mass the rates
decrease. The exclusion luminosities are high for exotic Higgs masses
above 125 GeV and below the top-pair threshold. Above the top-pair
threshold the exclusion luminosities are much lower due to the exclusion
limits in the top-pair final state. For masses below the SM-like Higgs
mass, however, there are parameter points where the exclusion
luminosities can exceed the 100~fb$^{-1}$ and even 1~ab$^{-1}$ while
still featuring large rates. The reason is that these points are not
excluded from single Higgs searches as light Higgs states with
dominant decays into $b\bar{b}$ final states are difficult to
probe. On the other hand this enhancement combined with the large
di-Higgs production cross section implies huge $4b$ final state rates, that
may be tested at the high luminosities, but with associated experimental
difficulties. This is a nice example of the
interplay between difficult single-Higgs searches and large exotic
di-Higgs rates, where new physics may be found.

\section{Benchmarks and Phenomenology}
\label{sec:bench}
\subsection{Type 1 Benchmarks}
We describe a representative set of benchmarks of the C2HDM T1 model
and their associated (exotic) multi-Higgs phenomenology. The input
parameters, the derived third neutral Higgs boson mass, the CP-odd
admixtures in terms of the squared mixing matrix elements $R_{i3}^2$
and the exclusion luminosity ${\cal L}_{\text{excl}}$ are given in
Tab.~\ref{tab:t1benchmarks1}. We also give the NLO QCD gluon fusion
$hh$ production cross section at $\sqrt{s}=14$~TeV together with its
$K$-factor, given by the ratio of the NLO cross section to the LO
one. In Table~\ref{tab:c2hdmrates1} we present the $4b$, $(2b)(2\tau)$
and $(2b)(2\gamma)$ rates from Higgs pair production normalized to the
rate expected in the SM from Higgs pairs relevant for the discussion
of the various benchmark points. 
\begin{table*}[p!]
 \begin{center}
 \begin{tabular}{l|cccccc}
     \toprule
    & {\tt T1BP1} & {\tt T1BP2} & {\tt T1BP3} & {\tt T1BP4} & {\tt T1BP5} & {\tt T1BP6} \\
   \hline
$m_{H_1}$ [GeV] &  125.09 &  125.09 & 125.09 & 119.73 & 125.09 & 62.67\\
$m_{H_2}$ [GeV] &  130.24 & 131.52 & 233.86 & 125.09 & 265.60 & 125.09\\
$m_{H^\pm}$ [GeV] & 169.99 & 282.75 & 164.87 & 185.41 & 307.47 & 164.35 \\
$\mbox{Re}(m_{12}^2)$ [GeV$^2$] & 679 & 12376 & 11473 & 7522 & 11435 &
   130 \\
$\alpha_1$ & 1.300 & 1.249 & 1.268 & 1.276 & 1.246 & -0.145 \\
$\alpha_2$ & -0.075 & -0.032 & 0.00262 & 1.494 & $7.125 \cdot 10^{-3}$ & -0.0536 \\
$\alpha_3$ & 1.306 & 1.570 & -0.809 & -1.460 & -1.478 & -0.0650\\
$\tan\beta$ & 4.05 & 3.23 & 3.32 & 5.30 & 5.54 & 8.26\\ \hline
$m_{H_3}$ [GeV] & 132.95 & 290.17 & 234.51 & 211.43 & 279.70 & 138.01\\
$R_{13}^2$ & $5.558\cdot 10^{-3}$ & $1.027 \cdot 10^{-3}$ & $6.863\cdot 10^{-6}$ &
   0.994 & $5.077 \cdot 10^{-5}$ & $2.870 \cdot 10^{-3}$\\
$R_{23}^2$ & $0.926$ & $0.999$ & 0.523 & $5.819 \cdot 10^{-3}$ & 0.991 & $4.212 \cdot 10^{-3}$\\
$R_{33}^2$ & 0.068 & 1.217$\cdot 10^{-6}$ & 0.477 & $7.140\cdot10^{-5}$ & $8.611
\cdot 10^{-3}$ & 0.993 \\
${\cal L}_{\text{excl}}$ [fb$^{-1}$] & 11500 & 1641 & 1365 & 2628 & 1082 & 2579 
\\ \hline
$\sigma_{hh}^{\text{NLO}}$ [fb] & 36.52 & 36.59 & 37.88
& 38.21 & 897.74 & 37.26 \\
$K$-factor & 1.95 &  1.95 & 1.95 & 1.95 & 2.06 & 1.95 \\ \bottomrule
  \end{tabular}
\caption{T1: Lines 1-8: The input parameters of the benchmark points {\tt T1BP1}-{\tt 6}. Lines
9 to 13: The derived 3rd neutral Higgs boson mass, the CP-odd
admixtures $R_{i3}^2$ and the exclusion luminosity ${\cal
  L}_{\text{excl}}$. Lines 14 and 15: The NLO QCD gluon fusion $hh$ production cross section at
$\sqrt{s}=14$~TeV and the corresponding $K$-factor.} \label{tab:t1benchmarks1}
   \end{center}
\vspace*{-0.2cm}
 \end{table*}
\begin{table*}[p!]
 \begin{center}
 \begin{tabular}{l|l|l|l|l|l|l}
     \toprule
  & {\tt T1BP1} & {\tt T1BP2} & {\tt T1BP3} & {\tt T1BP4} & {\tt T1BP5} & {\tt T1BP6} \\ \hline
   \midrule
$(b\bar{b})(b\bar{b})_{H_i H_j}$ [fb] & $[hh]$: 1.05 & $[hH_\downarrow]$: 1.69 &
$[h H_\downarrow]$: 0.026 &  $[h H_\downarrow]$: 0.66 & $[hh]$: 23.80
& $[H_\downarrow H_\downarrow]$: 145\\
$(b\bar{b})(\tau\bar{\tau})_{H_i H_j}$ [fb] & $[hh]$: 1.04
& $[hH_\downarrow]$: 1.70 & $[h H_\downarrow]$: 0.027 & $[h H_\downarrow]$: 0.66 & $[hh]$:
23.51  
& $[H_\downarrow H_\downarrow]$: 124
\\
$(b\bar{b})(\gamma\gamma)_{H_i H_j}$ [fb] & $[hh]$: 0.91 & $[hH_\downarrow]$: 0.97 &  
$[h H_\downarrow]$: 0.019 & $[h H_\downarrow]$: 0.41 & $[hh]$: 24.32
& $[H_\downarrow H_\downarrow]$: 0.29
\\ \bottomrule
  \end{tabular}
\caption{C2HDM T1: {\tt T1BP1}-{\tt 6}: The rates $(xx)(yy)_{H_i H_j}$ into the
  di-Higgs states $[H_iH_j]$ normalized to the SM
  from NLO QCD di-Higgs production through gluon fusion at
  $\sqrt{s}=14$~TeV in the final state 
  $(xx)(yy)$: $[pp \to H_i H_j \to (xx) (yy)]/[pp\to H^{\text{SM}}
  H^{\text{SM}} \to (xx) (yy)]$.}  \label{tab:c2hdmrates1} 
   \end{center}
\vspace*{-0.2cm}
 \end{table*}

\begin{table*}[p!]
 \begin{center}
 \begin{tabular}{l|cccccc}
     \toprule
    & {\tt T1BP1\_H} & {\tt T1BP2\_H} & {\tt T1BP5\_H} \\
   \hline
$m_{H_1}$ [GeV] &  125.09 & 125.09 & 125.09 \\
$m_{H_2}$ [GeV] &  407.30 & 364.98 & 397.67  \\
$m_{H^\pm}$ [GeV] & 410.24 & 436.72 & 428.96 \\
$\mbox{Re}(m_{12}^2)$ [GeV$^2$] & 17152 & 39827 & 17992 \\
$\alpha_1$ & 1.406 & 1.291 & 1.379 \\
$\alpha_2$ & -5.946$\cdot 10^{-3}$ & 3.132$\cdot 10^{-3}$ &
$ 5.435 \cdot 10^{-3}$ 
\\
$\alpha_3$ & 0.244 & -1.550 & -0.685 \\
$\tan\beta$ & 9.20 & 3.40 & 8.28 \\ \hline
$m_{H_3}$ [GeV] & 425.13 & 401.33 & 403.92 \\
$R_{13}^2$ & $3.535\cdot 10^{-5}$ & $9.809 \cdot 10^{-6}$ 
& $2.950 \cdot 10^{-5}$ \\
$R_{23}^2$ & $0.059$ & 0.999 & 0.400 \\
$R_{33}^2$ & 0.941 & 4.330$\cdot 10^{-4}$ & 0.600 \\ 
${\cal L}_{\text{excl}}$ [fb$^{-1}$] & 2461 & 1792 & 1590\\
\hline
$\sigma_{hh}^{\text{NLO}}$ [fb] & 206.53 & 43.81 & 400.60 \\
$K$-factor & 1.98 & 1.98 & 1.99 
\\ \bottomrule
  \end{tabular}
\caption{T1 heavy spectrum: Lines 1-8: The input parameters of the
  benchmark points {\tt T1BP1\_H}, {\tt T1BP2\_H} and {\tt T1BP5\_H}. Lines
9 to 13: The derived 3rd neutral Higgs boson mass, the CP-odd
admixtures $R_{i3}^2$ and the exclusion luminosity ${\cal
  L}_{\text{excl}}$. Lines 14 and 15: The NLO QCD gluon fusion $hh$
production cross section at $\sqrt{s}=14$~TeV and the corresponding $K$-factor.} \label{tab:t1benchmarks2}
   \end{center}
\vspace*{-0.2cm}
 \end{table*}

\begin{itemize}
\item {\tt T1BP1 - Highest\_exclusion\_lumi}: The point with the highest
  exclusion luminosity in the complete sample. 

  The exclusion
  luminosity for this point is found to be 11.5 ab$^{-1}$, {\it i.e.} well above the
  LHC design luminosity even after the high luminosity phase. All di-Higgs
  cross sections involving non-SM-like Higgs bosons have values below
  the SM reference point. Altogether this benchmark point behaves very SM-like as
  expected for such a high exclusion luminosity.
  The neutral Higgs mass spectrum 
  is relatively degenerate with all masses in the vicinity of the SM-like Higgs boson 
  at approximately 125 GeV. The SM-like Higgs pair production cross section for
  this point is SM-like; $b\bar b \gamma \gamma$ is about 10\% below
  the SM expectation 
  while $b\bar b \tau\tau$ and $b\bar b b\bar b $ are slightly enhanced
  by 5\%, {\it cf.}~Tab.~\ref{tab:c2hdmrates1}. The cases when a SM 
  Higgs is accompanied by an additional exotic Higgs are around 30\% of
  the SM expectation in $b\bar b \tau \tau$. With the latest
  improvements in hadronic tau
  tagging~\cite{Cadamuro:2015lbd,CMS-DP-2015-009,Mastrolorenzo:2016dyo}
  such a signature might be phenomenologically accessible at the LHC
  with large luminosity. 
\item {\tt T1BP2 - {HighLMaxHsmHl}}: Among the points with exclusion
  luminosities $\ge 1~\text{ab}^{-1}$ the point with the maximum cross
  section $gg \to h H_\downarrow$.
  
We have $\sigma (gg \to h H_\downarrow)=28.47$~fb at LO. As summarized
in Tab.~\ref{tab:c2hdmrates1}, for the final
states involving $b$'s, $\tau$'s and $\gamma$'s we find for this
di-Higgs final state relative to the production of a SM Higgs pair with
subsequent decay into the same final state: 
  \begin{eqnarray*}
  \sigma (gg \to h
  H_\downarrow \to 4b)/\mbox{SM} &= 1.69 \,,\\
  \sigma (gg \to h
  H_\downarrow \to (2b) (2\tau))/\mbox{SM} &= 1.70 \,,\\ 
  \sigma (gg \to h
  H_\downarrow \to (2b)(2\gamma))/\mbox{SM} &= 0.97\,.
  \end{eqnarray*}
The neutral Higgs mass spectrum in this case is slightly split
    while the pair production of the SM-like Higgs bosons follows
    largely the SM paradigm. The mass of the lighter neutral state is
    rather close to the SM boson, which allows us to compare the rates
    with the SM itself. $H_\downarrow$ has a decay phenomenology that
    is SM-like. At $3~\text{ab}^{-1}$ we can therefore expect around
    26k exotic $b\bar b \tau\tau$ events while the $b\bar b b\bar b$
    rate is enhanced by 70\% over the SM expectation. 

\item {\tt T1BP3 - NonSMHsmHL}: Among the
  points with exclusion 
  luminosities $\ge 1~\text{ab}^{-1}$ and $\Sigma_{H_\downarrow}  \le
  0.5$ ({\it i.e.}~dominantly non-SM-like decays for $H_\downarrow$)
  the point with the maximum cross section $gg \to h 
  H_\downarrow$.  

This point will have a highly non-SM decay phenomenology and
    cascade decays are an interesting avenue to look for such a
    scenario. All di-Higgs production cross sections involving
  non-SM-like Higgs 
bosons lie below 5~fb. While the point has interesting signatures for
non-SM-like single Higgs production, exotic di-Higgs
production is not a very promising avenue. The cross sections are far
below the SM value, as no 
resonance-enhancement is possible as the Higgs mass values are too
close. Additionally the branching ratios in SM-like Higgs states are
very small, as decays into non-SM-like final states  dominate. We
have $\Sigma_{H_\downarrow} = 0.180$, $\Sigma_{H_\uparrow} = 0.149$.
The SM-like Higgs pair production
  is consistent with the SM expectation within 10\% for this
  point. Exotic production of $H_\downarrow$ together 
  with the SM Higgs leads to around 220 exotic $(b\bar b) (\tau\tau)$
  events and around 1900 $(b\bar b) (b\bar b)$ at $3~\text{ab}^{-1}$. 
\item {\tt T1BP4 - MaxLEnhancedHsmHl}
: Among the points with $gg\to h
  H_\downarrow \ge 10$~fb the point with the maximum exclusion luminosity. 
  The SM-like Higgs pair production modes fall again within $\sim
    10\%$ of the SM expectation. The mass of the lightest neutral
    Higgs boson of about $120~\text{GeV}$ is again reasonably close to
    the SM-like Higgs to allow a direct comparison of expected rates,
    which are slightly smaller than the SM. Specifically, the exotic
    $b\bar b\tau\tau$ and $b\bar b b\bar b$ modes are 35\%
    smaller than what we would expect for the SM mode with $m_h\simeq
    125~\text{GeV}.$  
\item {\tt T1BP5 - EnhancedHsmHsm}: Among the points with exclusion
  luminosities $\ge 1$~ab$^{-1}$ the point with the maximum cross
  section $gg \to h h$. 
  
The exotic Higgs bosons lie well above the SM-like Higgs state which opens
  the possibility of enhancing the SM-like di-Higgs 
  production due to resonant enhancement of intermediate
  $H_{\downarrow/\uparrow}$ production. We find explicitly
  \begin{eqnarray*}
  \sigma (gg \to h
  h \to 4b)/\mbox{SM} = 23.80\,, \\ 
  \sigma (gg \to h
  h \to (2b) (2\tau))/\mbox{SM} = 23.51\,,\\
  \sigma (gg \to h
  h \to (2b)(2\gamma))/\mbox{SM} = 24.32\,.
  \end{eqnarray*}

\item {\tt T1BP6 - EnhancedHlHl}: Among the points with exclusion
  luminosities $\ge 1$~ab$^{-1}$ the point with the maximum cross
  section $gg \to H_\downarrow H_\downarrow$.

The exclusion luminosity is
2.58~$\text{ab}^{-1}$
   and we have a light $H_\downarrow$ with a
  mass just above half the SM-like mass so that the branching ratios
  of the latter remain in accordance with the LHC data. 
  The di-Higgs production of $H_\downarrow
  H_\downarrow$ amounts to 1.249~pb at LO QCD. Comparing to the SM
  expectation,
  we have   
  \begin{eqnarray*}
  \sigma (gg \to H_\downarrow
  H_\downarrow \to 4b)/\mbox{SM} &= 145 \,, \\
  \sigma (gg \to H_\downarrow
  H_\downarrow \to (2b) (2\tau))/\mbox{SM} &= 124 \,, \\
  \sigma (gg \to H_\downarrow
  H_\downarrow \to (2b)(2\gamma))/\mbox{SM} & = 0.29 \,, 
  \end{eqnarray*}
due to the suppressed decay $H_\downarrow \to \gamma\gamma$.
The light Higgs almost exclusively decays into $b$ pairs at a
branching fraction of $85\%$. This means that such a state is
difficult to observe in single Higgs production as trigger criteria
typically remove such events from the busy hadronic LHC
environment. There is a possibility to observe this state in its $\tau$
modes ($\simeq 8\%$). This point is a nice example how the $4b$ mode
can be an important BSM discovery tool when SM mass correlations are
relaxed. 
\end{itemize}

We note that none of these points features a strong first order electroweak phase transition. 
In general we observe that points with a strong first order phase
transition 
do not lead to enhanced rates in the four-particle final  
states; the cross section for SM-like di-Higgs production is close to
the SM value while other di-Higgs production cross sections are smaller
than the SM expectation. 
\medskip

In addition to these points which are all characterized by relatively
light exotic states we include benchmarks with heavy neutral
exotics. This is achieved by adding the additional requirement
$m_{H_\uparrow} \ge 400$~GeV to the defining criteria of the benchmark
points quoted above. We restrict ourselves to three benchmark points
highlighting special features. They are called   
{\tt T1BP1\_H}, {\tt T1BP2\_H} and {\tt T1BP5\_H} in analogy to the
their lighter mass spectrum counterparts {\tt T1BP1},{\tt 2},{\tt 5}.
Detailed information on the
benchmark points is summarized in Tab.~\ref{tab:t1benchmarks2}.

\begin{itemize}

\item {\tt T1BP1\_H - Highest\_exclusion\_lumi}: The exclusion
  luminosity for this point is found to be 2.46 ab$^{-1}$. All di-Higgs
  cross sections involving non-SM-like Higgs bosons have values below
  the SM reference as the Higgs pair production cross section steeply
  falls for heavy Higgs production. The additional resonant structures that are
  sourced in the SM-like Higgs pair production amount to an increase above the
  SM expectation by a factor of $\sim 5.8$ across the standard search channels
  $4b,(2b)(2\tau),(2b)(2\gamma)$.

\item {\tt T1BP2\_H - HighLMaxHsmHl}
: We
  have $\sigma (gg \to h 
  H_\downarrow)=2.34$~fb at LO, which is rather large given the mass
  of the exotic Higgs. The decay phenomenology of the additional
    Higgs is completely dominated by decays into top-final states. In
    this sense the single Higgs production and exotic $hH_\downarrow$
    production are fully correlated. The exclusion luminosity is
    $1.8~\text{ab}^{-1}$ and results from the extrapolation of the
    $t\bar t$ resonance search. This point, although not relevant for
    di-Higgs analyses shows how single Higgs measurements in the
    $t\bar t$ channel influence multi-Higgs final states. Such a
    benchmark could be adopted to further clarify the role of single
    Higgs measurements for exotic multi-Higgs final states.

\begin{table*}[ht!]
 \begin{center}
 \begin{tabular}{l|cccccc}
     \toprule
    & {\tt T2BP1} & {\tt T2BP2} \\
   \hline
$m_{H_1}$ [GeV] &  125.09 & 125.09\\
$m_{H_2}$ [GeV] &  858.09 &  814.56 \\
$m_{H^\pm}$ [GeV] & 835.85 & 894.84 \\
$\mbox{Re}(m_{12}^2)$ [GeV$^2$] & 252703 & 227697 \\
$\alpha_1$ & 1.141 & 1.042\\
$\alpha_2$ & $-5.268 \cdot 10^{-4}$ & $6.184 \cdot 10^{-4}$ \\
$\alpha_3$ & 1.198 & -1.157\\
$\tan\beta$ & 2.16 & 1.71\\ \hline
$m_{H_3}$ [GeV] & 858.65 & 814.94 \\
$R_{13}^2$ & $2.775\cdot 10^{-7}$ & $3.824 \cdot 10^{-7}$\\
$R_{23}^2$ & $0.867$ & 0.832\\
$R_{33}^2$ & 0.133 & 0.162 \\ 
${\cal L}_{\text{excl}}$ & 2664 & 2016 \\
\hline
$\sigma_{hh}^{\text{NLO}}$ [fb] & 37.82 & 38.02
   \\
$K$-factor & 1.95 & 1.95
\\ \bottomrule
  \end{tabular}
\caption{C2HDM T2: Lines 1-8: The input parameters of the benchmark points
  {\tt T2BP1} and {\tt T2BP2}. Lines
9 to 13: The derived 3rd neutral Higgs boson mass, the CP-odd
admixtures $R_{i3}^2$ and the exclusion luminosity ${\cal
  L}_{\text{excl}}$. Lines 14 and 15: The NLO QCD 
gluon fusion $hh$ production cross section at
$\sqrt{s}=14$~TeV and the corresponding $K$-factor.} \label{tab:t2benchmarks1}
   \end{center}
\vspace*{-0.2cm}
 \end{table*}

\item {\tt T1BP5\_H - EnhancedHsmHsm}: We have enhanced SM-like di-Higgs
  production due to resonant enhancement of intermediate
  $H_{\downarrow}$ and $H_{\uparrow}$ 
  production. We find 
   \begin{eqnarray*}
  \sigma (gg \to h
  h \to 4b)/\mbox{SM} &=& 11.22\,,\\
  \sigma (gg \to h
  h \to (2b) (2\tau))/\mbox{SM} &=& 11.08\,,\\
  \sigma (gg \to h
  h \to (2b)(2\gamma))/\mbox{SM} &=& 11.15\,.
  \end{eqnarray*}
  This point hence gives
  access to SM-like Higgs pair production even for a heavy Higgs
  spectrum, and is an immediate sign of BSM physics as the di-Higgs
  cross section is enhanced.
\end{itemize}

\subsection{Type 2 Benchmarks}
As already visible from Tab.~\ref{tab:maxcxn}, the C2HDM T2 model gives rise
to less spectacular signatures than the C2HDM T1. We give two 
representative scenarios below. In general, the spectrum is much
heavier than for T1. There is no scenario with the SM-like Higgs
  boson being $H_2$; the SM Higgs is always the lightest state $H_1$.
The input parameters for these points as well as further relevant
information are summarized in Tab.~\ref{tab:t2benchmarks1}. 

\begin{itemize}
\item {\tt T2BP1 - Highest\_exclusion\_lumi}: The exclusion
  luminosity for this point is found to be 
  2.66 ab$^{-1}$. All di-Higgs
  cross sections involving non-SM-like Higgs bosons have values below
  the SM reference value. Altogether this benchmark point behaves very SM-like as
  expected for such a high exclusion luminosity where $t\bar t$ resonance 
  searches become sensitive to this scenario.

\item {\tt T2BP2 - EnhancedHSMHSM}: 
  The di-Higgs production into a SM-like Higgs pair is enhanced 
  so that we get 
  \begin{eqnarray*}
  \sigma (gg \to hh \to
  4b)/\mbox{SM} &=& 1.101\,, \\
  \sigma (gg \to hh \to (2b)
  (2\tau))/\mbox{SM} &=& 1.088\,,\\ 
  \sigma (gg \to hh \to
  (2b)(2\gamma))/\mbox{SM} &=& 1.037\,.
  \end{eqnarray*}
  Despite a lower exclusion
  luminosity this point behaves also very SM like and barely exceeds
  the rates into SM-like final states of {\tt T2BP1} which has a higher
  exclusion luminosity. 
\end{itemize}

Overall, it will be difficult to probe the
  C2HDM T2, which features a heavy Higgs spectrum, in di-Higgs
  production. This is partly also due to the fact that enhanced
  SM-like Higgs pair production cross sections are already excluded by
  the LHC limits on resonant heavy scalar production with decay in a
  SM Higgs boson pair.

\subsection{NMSSM Benchmarks}
\begin{table*}[t!]
 \begin{center}
 \begin{tabular}{l|cccccc}
     \toprule
    & {\tt NMBP1} & {\tt NMBP2} & {\tt NMBP3} & {\tt NMBP4 }\\
   \hline
$M_1$ [GeV] & 638 &  457 & 608 & 313 & \\
$M_2$ [GeV] & 1254 &  386 & 546 & 569 & \\
$M_3$ [GeV] & 4169 & 6345 & 6778 & 3485 & \\
$A_t$ [GeV]  & 2456 & 5134 & 1092 & 532 & \\
$A_b$ [GeV]  & -2213 & -2908 & -4015 & 2009 & \\
$A_\tau$ [GeV] & 1443 &  -667 & 2370 & 354 & \\
$M_{\tilde{Q}_3}$ [GeV] & 1293 & 3175 & 2574 & 3581 & \\
$M_{\tilde{L}_3}$ [GeV] & 1147 & 1276 & 790 & 1188 & \\
$\tan\beta$ & 1.96 & 1.87 & 1.68 & 1.49 & \\
$\lambda$ & 0.55 &  0.50 & 0.60 & 0.54 & \\
$\kappa$ & 0.43 &  0.47 & 0.33 & 0.27 & \\
$A_\lambda$ [GeV] & -55 & 33 & 425 & -416 &  \\
$A_\kappa$ [GeV] & 373 & 358 & -672 & 667 & \\
$\mu_{\text{eff}}$ [GeV] & -293 & -299 & 321 & -327 & \\ \hline
$m_{H_1}$ [GeV] & 124.34 & 124.11 & 101.13 & 39.52 & \\
$m_{H_2}$ [GeV] &  335.27 & 409.70 & 125.88 & 125.64 & \\
$m_{H_3}$ [GeV] &  530.39 & 465.57 & 627.95 & 634.32 & \\
$m_{A_1}$ [GeV] &  487.34 & 406.66 & 608.57 & 580.42 & \\
$m_{A_2}$ [GeV] &  540.58 & 553.09 & 624.77 & 631.97 & \\
$m_{H^\pm}$ [GeV] & 520.47 & 426.32 & 621.81 & 628.97 & \\
${\cal L}_{\text{excl}}$ [fb$^{-1}$]  & 1370 & 118 & 1192 & 101 & \\ \hline
$\sigma_{hh}^{\text{NLO}}$ [fb] & 69.29 & 131.83 & 43.62 & 42.31 \\
$K$-factor & 1.97 & 1.97 & 1.96 & 1.96
\\ \bottomrule
  \end{tabular}
\caption{NMSSM: Lines 1-14: The input parameters of the benchmark
  points {\tt NMBP1}-{\tt 4}. Lines 15 to 20: The derived Higgs boson masses. Line
  21: The exclusion luminosity. Lines 22-23: The NLO QCD gluon fusion
  $hh$ production cross section at $\sqrt{s}=14$~TeV and its
  corresponding $K$-factor.} \label{tab:nmssmbenchmarks1} 
   \end{center}
\vspace*{-0.2cm}
 \end{table*}

\begin{table*}[t!]
 \begin{center}
 \begin{tabular}{l|l|l|l|l|l}
     \toprule
    & {\tt NMBP1} & {\tt NMBP2} & {\tt NMBP3} & {\tt NMBP4} \\ \hline
   \midrule
$(b\bar{b})(b\bar{b})_{H_i H_j}$ [fb] & $[hh]$: 2.35 & $[hh]$: 4.77 &
$[h H_\downarrow]$: 2.33 &  $[H_\downarrow H_\downarrow]$: 541.60 & \\
$(b\bar{b})(\tau\bar{\tau})_{H_i H_j}$ [fb] & $[hh]$: 2.31 & $[hh]$: 4.70 & 
$[h H_\downarrow]$: 2.27 & $[H_\downarrow H_\downarrow]$: 432.18 & 
\\
$(b\bar{b})(\gamma\gamma)_{H_i H_j}$ [fb] & $[hh]$: 2.10 & $[hh]$: 3.79 & 
$[h H_\downarrow]$: 1.16 & $[H_\downarrow H_\downarrow]$: 7.11 
\\ \bottomrule
  \end{tabular}
\caption{NMSSM {\tt NMBP1}-{\tt 4}: The rates $(xx)(yy)_{H_i H_j}$ into the
  di-Higgs states $[H_iH_j]$ normalized to the SM
  from NLO QCD di-Higgs production through gluon fusion at
  $\sqrt{s}=14$~TeV in the final state 
  $(xx)(yy)$: $[pp \to H_i H_j \to (xx) (yy)]/[pp\to H^{\text{SM}}
  H^{\text{SM}} \to (xx) (yy)]$.}  \label{tab:nmssmrates1} 
   \end{center}
\vspace*{-0.2cm}
 \end{table*}

Let us finally turn to the NMSSM. The criteria for selecting the
benchmark points are as follows. 
\begin{itemize}
\item {\tt NMBP1}: The point with the largest $4b$ rate from SM-like Higgs boson pair
  production with an exclusion luminosity above 1~ab$^{-1}$.
\item {\tt NMBP2}: The same as {\tt BP1} but with an exclusion
  luminosity beyond 100~fb$^{-1}$.
\item {\tt NMBP3}: The point with the largest $4b$ rate from the
  production of an SM-like
  Higgs boson and the lighter of the  
  CP-even non-SM-like Higgs bosons, $H_\downarrow$, with an exclusion
  luminosity above 1~ab$^{-1}$. 
\item {\tt NMBP4}: The point with the largest $4b$ rate from
  $H_\downarrow H_\downarrow$ production with an exclusion
  luminosity above 100~fb$^{-1}$. 
\end{itemize}

We also provide benchmark points for di-Higgs final states involving a
light pseudoscalar $A_\downarrow$ in the final state:
\begin{itemize}
\item {\tt NMBP5}: The point with the largest $4b$ rate from $h A_\downarrow$
  production with an exclusion luminosity above 1~ab$^{-1}$.
\item {\tt NMBP6}: The same as {\tt BP5} but with an exclusion
  luminosity beyond 100~fb$^{-1}$.
\item {\tt NMBP7}: The point with the largest $4b$ rate from
  $A_\downarrow A_\downarrow$ production and an exclusion luminosity
  above 1~ab$^{-1}$.
It turns out that {\tt NMBP7} is identical to {\tt NMBP5}. 
\item {\tt NMBP8}: The same definition as for {\tt NMBP7} but with an exclusion
  luminosity beyond 100~fb$^{-1}$.
\end{itemize}

The input values, the derived Higgs boson masses, the exclusion
luminosity and the NLO QCD cross section for $hh$ production with its
$K$-factor for the various benchmark points are listed in
Tabs.~\ref{tab:nmssmbenchmarks1} and \ref{tab:nmssmbenchmarks2}.
Note, that we include benchmark points with an
exclusion luminosity around 100~fb$^{-1}$ in case the rates are much
enhanced compared to the SM as in this case a luminosity of
100~fb$^{-1}$ might be enough to test this parameter point. As the
rates for di-Higgs production involving a heavy scalar or pseudoscalar
are low we do not present benchmarks for these cases. 

\begin{table*}[t!]
 \begin{center}
 \begin{tabular}{l|cccccc}
     \toprule
    & {\tt NMBP5} & {\tt NMBP6} & {\tt NMBP8} \\
   \hline
$M_1$ [GeV] & 455 &  842 & 321 & \\
$M_2$ [GeV] & 1741 &  1510 & 749 & \\
$M_3$ [GeV] & 4179 & 1924 & 2060 & \\
$A_t$ [GeV]  & -5923 & -4901 & 5449 & \\
$A_b$ [GeV]  & -3994 & -3817 & 1232 & \\
$A_\tau$ [GeV] & -1773 & -226 & 2253 & \\
$M_{\tilde{Q}_3}$ [GeV] &  2391 & 3539 & 2344 & \\
$M_{\tilde{L}_3}$ [GeV] & 2117 & 1623 & 1163 & \\
$\tan\beta$ & 2.17 & 1.46 & 1.98 & \\
$\lambda$ & 0.53 &  0.55 & 0.49 & \\
$\kappa$ & 0.44 &  0.42 & 0.46 & \\
$A_\lambda$ [GeV] & -177 & -4.78 & 60.12 &  \\
$A_\kappa$ [GeV] & 47 & 6.32 & -0.81 & \\
$\mu_{\text{eff}}$ [GeV] & -327 & -307 & 418 & \\ \hline
$m_{H_1}$ [GeV] & 125.62 & 125.59 & 125.07 & \\
$m_{H_2}$ [GeV] &  504.14 & 428.17 & 618.30 & \\
$m_{H_3}$ [GeV] &  612.42 & 465.10 & 767.13 & \\
$m_{A_1}$ [GeV] &  220.91 & 106.06 & 37.38 & \\
$m_{A_2}$ [GeV] &  602.07 & 444.01 & 629.92 & \\
$m_{H^\pm}$ [GeV] & 597.20 & 432.53 & 620.36 & \\
${\cal L}_{\text{excl}}$ [fb$^{-1}$]  & 1153 & 152 & 109 & \\ \hline
$\sigma_{hh}^{\text{NLO}}$ [fb] & 53.42 & 97.53 
& 54.54\\
$K$-factor & 1.96 & 1.97 & 1.96
\\ \bottomrule
  \end{tabular}
\caption{NMSSM: Lines 1-14: The input parameters of the benchmark
  points {\tt NMBP5},{\tt 6},{\tt 8} ({\tt NMBP7}={\tt NMBP5}). Lines
15-20: The derived Higgs boson masses. Line 21: The exclusion
luminosity. Lines 22-23: The NLO QCD gluon fusion $hh$ production cross section at
$\sqrt{s}=14$~TeV and its corresponding $K$-factor.} \label{tab:nmssmbenchmarks2}
   \end{center}
\vspace*{-0.2cm}
 \end{table*}

%
\begin{table*}[t!]
 \begin{center}
 \begin{tabular}{l|l|l|l|l|l}
     \toprule
    & \multicolumn{2}{l|}{ {\tt NMBP5}} & {\tt NMBP6} & {\tt NMBP8} \\ \hline
   \midrule
$(b\bar{b})(b\bar{b})_{H_i H_j}$ [fb] & $[hA_\downarrow]$: 2.56 &
$[A_\downarrow A_\downarrow]$: 4.57 & $[hA_\downarrow]$: 44.37 &
$[A_\downarrow A_\downarrow]$: 2127 & \\
$(b\bar{b})(\tau\bar{\tau})_{H_i H_j}$ [fb] & $[hA_\downarrow]$: 2.68 & 
$[A_\downarrow A_\downarrow]$: 5.07
& $[hA_\downarrow]$: 43.22 & $[A_\downarrow A_\downarrow]$: 1710 & 
\\
$(b\bar{b})(\gamma\gamma)_{H_i H_j}$ [fb] & $[hA_\downarrow]$: 2.06 
&  $[A_\downarrow A_\downarrow]$: 3.20 & $[hA_\downarrow]$: 22.65 
& $[A_\downarrow A_\downarrow]$: 11.30
\\ \bottomrule
  \end{tabular}
\caption{NMSSM {\tt NMBP5},{\tt 6},{\tt 8} ({\tt NMBP7}={\tt NMBP5}): The rates
  $(xx)(yy)_{H_i H_j}$ into the 
  di-Higgs states $[H_iH_j]$ normalized to the SM
  from NLO QCD di-Higgs production through gluon fusion at
  $\sqrt{s}=14$~TeV in the final state 
  $(xx)(yy)$: $[pp \to H_i H_j \to (xx) (yy)]/[pp\to H^{\text{SM}}
  H^{\text{SM}} \to (xx) (yy)]$.}  \label{tab:nmssmrates2} 
   \end{center}
\vspace*{-0.2cm}
 \end{table*}

From Tab.~\ref{tab:nmssmrates1}  we can read off that the $4b$ rates
from $hh$ production for an exclusion luminosity above 100~fb$^{-1}$
({\tt NMBP2})
can be up almost a factor of about 4.8, so that this process will be
difficult to be accessed at the lower luminosity. Assuming an
exclusion luminosity above 1~ab$^{-1}$ ({\tt NMBP1}) the 
enhancement compared to the SM rate is around 2.4. Again, larger
enhancements in the final state with an SM-like Higgs boson pair are
excluded by the limits provided from ATLAS and CMS
\cite{CMS:2017xxp,Sirunyan:2017isc} .
In the
$hH_\downarrow$ final state the enhancement factor is only 
2.3 as the di-Higgs production cross section 
 \begin{align}
  {\tt NMBP3}: \quad &
  \sigma (h H_\downarrow) = 49.13~\mbox{fb} \; (K=1.92) \;,
  \end{align}  
is not much larger than in the SM. For $H_\downarrow H_\downarrow$
production we have, however,
  \begin{align}
  {\tt NMBP4}: \quad &
  \sigma (H_\downarrow H_\downarrow) = 8.96~\mbox{pb} \;
                    (K=2.30) \,,
  \end{align}  
leading to an enhancement factor of up to 540. The $H_\downarrow$ mass is much
lower here than in the former case inducing dominant branching ratios
into $b\bar{b}$. The large cross section is mainly due to the small
$H_\downarrow$ mass. The resonant enhancement plays a minor role here.
  Both benchmark points are special in the sense that the
  SM-like Higgs boson is not the lightest but the second lightest
  CP-even Higgs boson in the spectrum. We hence have a light CP-even
  Higgs boson in these scenarios. The $H_\downarrow$ is very
  singlet-like in both cases and decays with a branching ratio of
  about 0.9 into $b\bar{b}$.

From Tab.~\ref{tab:nmssmrates2} we can read off that the $4b$,
$(2b)(2\tau)$ and $(2b)(2\gamma)$ final
states from both $h A_\downarrow$ and $A_\downarrow A_\downarrow$
production can be enhanced above the SM rate even for an exclusion
luminosity of 1~ab$^{-1}$. For the lower exclusion luminosity the
enhancement can be huge, in particular in the 4 fermion final state
from $A_\downarrow A_\downarrow$ production ({\tt NMBP8}). These enhancements are
due to large di-Higgs production cross sections which at NLO QCD
amount to  
\begin{align}
{\tt NMBP5}: \quad & \sigma (h A_\downarrow) = 56.04~\mbox{fb} \; 
                    (K=1.93) \,, \\
& \sigma (A_\downarrow A_\downarrow) = 74.34~\mbox{pb} \; 
                    (K=1.94) \,, \\
{\tt NMBP6}: \quad & \sigma (h A_\downarrow) = 988~\mbox{fb} \; 
                    (K=1.99) \,, \\
{\tt NMBP8}: \quad & \sigma (A_\downarrow A_\downarrow) = 34.59~\mbox{pb} \; 
                    (K= 2.30) \,. 
\end{align} 

The enhanced di-Higgs cross section values are on the one hand due to
the light pseudoscalar masses and on the other hand due to resonant
scalar production for $A_\downarrow A_\downarrow$ or pseudoscalar
production for $hA_\downarrow$ production. As already noted in the
discussion of Fig.~\ref{fig:diadownarrow} this is an example where new
physics may lead to huge measurable effects in Higgs pair production
while the single Higgs process, here $A_\downarrow$ production in
gluon fusion, is difficult to access. This is a prime example that
demonstrates that despite the very SM-like nature of the 125~GeV Higgs
boson Higgs pair production can be far from being SM-like. 

We finally remark that in all NMSSM scenarios the stop masses are
quite large, of the order of 1~TeV and larger. 

\section{Conclusions}
\label{sec:conc}
Multi-Higgs final states are statistically limited at the LHC, but
are key processes to gain a precise understanding of the mechanism of 
electroweak symmetry breaking. Phenomenologically, they
are highly correlated with measurements in single Higgs
final states. The question of how much additional information can be
gained from the investigation of multi-Higgs final states is therefore
best addressed using concrete BSM extensions. 

Another particularly relevant
question when considering di-Higgs final states is whether they could play a key
discovery tool for BSM interactions. This could happen at the LHC in
situations when single Higgs analyses are simply not competitive due
to, {\it e.g.}, trigger thresholds that might be mitigated in more complex
multi-Higgs final states. 
Large branching ratios of additional scalars
into SM-like Higgs bosons serve as an additional avenue to
observe resonantly enhanced SM-like Higgs production. In these
scenarios, the kinematic correlations are often significantly modified
compared to the SM. 

\medskip

In this work we have performed a comprehensive scan over the complex
2HDM and the NMSSM, with a particular emphasis on
the expected di-Higgs phenomenology in these models, taking into
account a variety of current constraints and future projections. We
find that in particular in the C2HDM Type 1 models,  the di-Higgs
phenomenology can significantly differ from the SM expectation. 
The differences range from new signatures in SM-like search channels
for light Higgs bosons all the way to new resonant structures in
di-Higgs final states for larger exotic Higgs masses.
In particular, in
final states involving light Higgs bosons the multi-fermion final
states can be significantly enhanced compared to the SM case. This is also
the case in the NMSSM where we can have light scalars or pseudoscalars
in the spectrum.
In the C2HDM Type 2 models 
with its heavy Higgs spectrum,
the multi-Higgs final states play a more
subdominant role as new physics discoveries typically occur in $t\bar t $
resonance searches before (exotic) di-Higgs production becomes
relevant. 

We have distilled our scans into a representative number of benchmark
points that not only reflect the phenomenological possibilities that
present themselves in non-standard Higgs sectors, but also point to
a particular range of phenomenological situations:

\medskip

Firstly, the expected sensitivity of the  $t\bar t$ resonance search
  is a crucial factor in deciding on the relevance of di-Higgs
  searches. The multi-Higgs signal is typically driven by
  top-mediated gluon fusion. Therefore, the decay to top final states is directly
  correlated with a large di-Higgs cross section for resonant production as well as
  enhancements in the decay to photons for non-resonant
  production.

\medskip

Secondly, when exotic Higgs masses fall below the $t\bar t$
  threshold, di-Higgs final states typically follow the SM decay rates
  with compressed neutral Higgs masses. This highlights the necessity
  to achieve a high mass resolution in the standard search channels
  $(2b)(2 b),(2 b)(2\tau),(2b)(2\gamma)$, even when SM-mass correlations are
  abandoned. While di-Higgs production would be enhanced in this instance, providing
  clear evidence of the presence of BSM interactions, their precise nature would remain elusive
  to some extent.
  
  Alternatively, additional Higgs exotics can create multiple
  resonant features leading to a large enhancement of the total SM Higgs pair
  production rate. The extrapolated signal-strength constraints
  locate viable candidates for enhanced di-Higgs production in $b\bar b$ final states, which 
  are difficult to access experimentally in single Higgs production. Here di-Higgs production
  can play a significant role as a discovery tool for BSM interactions due to smaller backgrounds and better
  kinematical handles.

\medskip

Thirdly, relatively light Higgs bosons with significant branching ratios
can lead to a strong enhancement in multi-fermion final states. Such signatures
are already studied by the experimental collaborations. Our
results indicate the importance of these analyses in the future.

\medskip

While we have specifically focused on di-Higgs production, it is
clear that these scenarios can have interesting non-SM signatures that
can be exploited to observe or constrain a certain parameter point in
a more targeted, yet parameter point-dependent way. We will leave this
for future work.

\acknowledgements
We want to thank Jonas Wittbrodt for providing us with
the C2HDM data sample. 

P.B. acknowledges the support by the DFG-funded Doctoral School
Karlsruhe School of Elementary and Astroparticle Physics: Science and
Technology (KSETA).
S.D. is  supported by the U.S. Department of Energy under Grant Contract de-sc0012704. 
C.E. is supported by the IPPP Associateship scheme and by the UK
Science and Technology Facilities Council (STFC) under grant
ST/P000746/1. 


\begin{thebibliography}{160}
\expandafter\ifx\csname natexlab\endcsname\relax\def\natexlab#1{#1}\fi
\expandafter\ifx\csname bibnamefont\endcsname\relax
  \def\bibnamefont#1{#1}\fi
\expandafter\ifx\csname bibfnamefont\endcsname\relax
  \def\bibfnamefont#1{#1}\fi
\expandafter\ifx\csname citenamefont\endcsname\relax
  \def\citenamefont#1{#1}\fi
\expandafter\ifx\csname url\endcsname\relax
  \def\url#1{\texttt{#1}}\fi
\expandafter\ifx\csname urlprefix\endcsname\relax\def\urlprefix{URL }\fi
\providecommand{\bibinfo}[2]{#2}
\providecommand{\eprint}[2][]{\url{#2}}

\bibitem[{\citenamefont{Aad et~al.}(2012)}]{Aad:2012tfa}
\bibinfo{author}{\bibfnamefont{G.}~\bibnamefont{Aad}} \bibnamefont{et~al.}
  (\bibinfo{collaboration}{ATLAS}), \bibinfo{journal}{Phys. Lett.}
  \textbf{\bibinfo{volume}{B716}}, \bibinfo{pages}{1} (\bibinfo{year}{2012}),
  \eprint{1207.7214}.

\bibitem[{\citenamefont{Chatrchyan et~al.}(2012)}]{Chatrchyan:2012xdj}
\bibinfo{author}{\bibfnamefont{S.}~\bibnamefont{Chatrchyan}}
  \bibnamefont{et~al.} (\bibinfo{collaboration}{CMS}), \bibinfo{journal}{Phys.
  Lett.} \textbf{\bibinfo{volume}{B716}}, \bibinfo{pages}{30}
  (\bibinfo{year}{2012}), \eprint{1207.7235}.

\bibitem[{\citenamefont{Cornwall et~al.}(1973)\citenamefont{Cornwall, Levin,
  and Tiktopoulos}}]{Cornwall:1973tb}
\bibinfo{author}{\bibfnamefont{J.~M.} \bibnamefont{Cornwall}},
  \bibinfo{author}{\bibfnamefont{D.~N.} \bibnamefont{Levin}}, \bibnamefont{and}
  \bibinfo{author}{\bibfnamefont{G.}~\bibnamefont{Tiktopoulos}},
  \bibinfo{journal}{Phys. Rev. Lett.} \textbf{\bibinfo{volume}{30}},
  \bibinfo{pages}{1268} (\bibinfo{year}{1973}), \bibinfo{note}{[Erratum: Phys.
  Rev. Lett.31,572(1973)]}.

\bibitem[{\citenamefont{Cornwall et~al.}(1974)\citenamefont{Cornwall, Levin,
  and Tiktopoulos}}]{Cornwall:1974km}
\bibinfo{author}{\bibfnamefont{J.~M.} \bibnamefont{Cornwall}},
  \bibinfo{author}{\bibfnamefont{D.~N.} \bibnamefont{Levin}}, \bibnamefont{and}
  \bibinfo{author}{\bibfnamefont{G.}~\bibnamefont{Tiktopoulos}},
  \bibinfo{journal}{Phys. Rev.} \textbf{\bibinfo{volume}{D10}},
  \bibinfo{pages}{1145} (\bibinfo{year}{1974}), \bibinfo{note}{[Erratum: Phys.
  Rev.D11,972(1975)]}.

\bibitem[{\citenamefont{Lee et~al.}(1977{\natexlab{a}})\citenamefont{Lee,
  Quigg, and Thacker}}]{Lee:1977yc}
\bibinfo{author}{\bibfnamefont{B.~W.} \bibnamefont{Lee}},
  \bibinfo{author}{\bibfnamefont{C.}~\bibnamefont{Quigg}}, \bibnamefont{and}
  \bibinfo{author}{\bibfnamefont{H.~B.} \bibnamefont{Thacker}},
  \bibinfo{journal}{Phys. Rev. Lett.} \textbf{\bibinfo{volume}{38}},
  \bibinfo{pages}{883} (\bibinfo{year}{1977}{\natexlab{a}}).

\bibitem[{\citenamefont{Lee et~al.}(1977{\natexlab{b}})\citenamefont{Lee,
  Quigg, and Thacker}}]{Lee:1977eg}
\bibinfo{author}{\bibfnamefont{B.~W.} \bibnamefont{Lee}},
  \bibinfo{author}{\bibfnamefont{C.}~\bibnamefont{Quigg}}, \bibnamefont{and}
  \bibinfo{author}{\bibfnamefont{H.~B.} \bibnamefont{Thacker}},
  \bibinfo{journal}{Phys. Rev.} \textbf{\bibinfo{volume}{D16}},
  \bibinfo{pages}{1519} (\bibinfo{year}{1977}{\natexlab{b}}).

\bibitem[{\citenamefont{Veltman}(1981)}]{Veltman:1980mj}
\bibinfo{author}{\bibfnamefont{M.~J.~G.} \bibnamefont{Veltman}},
  \bibinfo{journal}{Acta Phys. Polon.} \textbf{\bibinfo{volume}{B12}},
  \bibinfo{pages}{437} (\bibinfo{year}{1981}).

\bibitem[{\citenamefont{Sakharov}(1967)}]{Sakharov:1967dj}
\bibinfo{author}{\bibfnamefont{A.~D.} \bibnamefont{Sakharov}},
  \bibinfo{journal}{Pisma Zh. Eksp. Teor. Fiz.} \textbf{\bibinfo{volume}{5}},
  \bibinfo{pages}{32} (\bibinfo{year}{1967}), \bibinfo{note}{[Usp. Fiz.
  Nauk161,no.5,61(1991)]}.

\bibitem[{\citenamefont{Grzadkowski et~al.}(2010)\citenamefont{Grzadkowski,
  Iskrzynski, Misiak, and Rosiek}}]{Grzadkowski:2010es}
\bibinfo{author}{\bibfnamefont{B.}~\bibnamefont{Grzadkowski}},
  \bibinfo{author}{\bibfnamefont{M.}~\bibnamefont{Iskrzynski}},
  \bibinfo{author}{\bibfnamefont{M.}~\bibnamefont{Misiak}}, \bibnamefont{and}
  \bibinfo{author}{\bibfnamefont{J.}~\bibnamefont{Rosiek}},
  \bibinfo{journal}{JHEP} \textbf{\bibinfo{volume}{10}}, \bibinfo{pages}{085}
  (\bibinfo{year}{2010}), \eprint{1008.4884}.

\bibitem[{\citenamefont{de~Florian et~al.}(2016)}]{deFlorian:2016spz}
\bibinfo{author}{\bibfnamefont{D.}~\bibnamefont{de~Florian}}
  \bibnamefont{et~al.} (\bibinfo{collaboration}{LHC Higgs Cross Section Working
  Group}) (\bibinfo{year}{2016}), \eprint{1610.07922}.

\bibitem[{\citenamefont{Dawson et~al.}(2018)\citenamefont{Dawson, Englert, and
  Plehn}}]{Dawson:2018dcd}
\bibinfo{author}{\bibfnamefont{S.}~\bibnamefont{Dawson}},
  \bibinfo{author}{\bibfnamefont{C.}~\bibnamefont{Englert}}, \bibnamefont{and}
  \bibinfo{author}{\bibfnamefont{T.}~\bibnamefont{Plehn}}
  (\bibinfo{year}{2018}), \eprint{1808.01324}.

\bibitem[{\citenamefont{Ellis et~al.}(1976)\citenamefont{Ellis, Gaillard, and
  Nanopoulos}}]{Ellis:1975ap}
\bibinfo{author}{\bibfnamefont{J.~R.} \bibnamefont{Ellis}},
  \bibinfo{author}{\bibfnamefont{M.~K.} \bibnamefont{Gaillard}},
  \bibnamefont{and} \bibinfo{author}{\bibfnamefont{D.~V.}
  \bibnamefont{Nanopoulos}}, \bibinfo{journal}{Nucl. Phys.}
  \textbf{\bibinfo{volume}{B106}}, \bibinfo{pages}{292} (\bibinfo{year}{1976}).

\bibitem[{\citenamefont{Shifman et~al.}(1979)\citenamefont{Shifman, Vainshtein,
  Voloshin, and Zakharov}}]{Shifman:1979eb}
\bibinfo{author}{\bibfnamefont{M.~A.} \bibnamefont{Shifman}},
  \bibinfo{author}{\bibfnamefont{A.~I.} \bibnamefont{Vainshtein}},
  \bibinfo{author}{\bibfnamefont{M.~B.} \bibnamefont{Voloshin}},
  \bibnamefont{and} \bibinfo{author}{\bibfnamefont{V.~I.}
  \bibnamefont{Zakharov}}, \bibinfo{journal}{Sov. J. Nucl. Phys.}
  \textbf{\bibinfo{volume}{30}}, \bibinfo{pages}{711} (\bibinfo{year}{1979}),
  \bibinfo{note}{[Yad. Fiz.30,1368(1979)]}.

\bibitem[{\citenamefont{Vainshtein et~al.}(1980)\citenamefont{Vainshtein,
  Zakharov, and Shifman}}]{Vainshtein:1980ea}
\bibinfo{author}{\bibfnamefont{A.~I.} \bibnamefont{Vainshtein}},
  \bibinfo{author}{\bibfnamefont{V.~I.} \bibnamefont{Zakharov}},
  \bibnamefont{and} \bibinfo{author}{\bibfnamefont{M.~A.}
  \bibnamefont{Shifman}}, \bibinfo{journal}{Sov. Phys. Usp.}
  \textbf{\bibinfo{volume}{23}}, \bibinfo{pages}{429} (\bibinfo{year}{1980}),
  \bibinfo{note}{[Usp. Fiz. Nauk131,537(1980)]}.

\bibitem[{\citenamefont{Voloshin}(1986)}]{Voloshin:1985tc}
\bibinfo{author}{\bibfnamefont{M.~B.} \bibnamefont{Voloshin}},
  \bibinfo{journal}{Sov. J. Nucl. Phys.} \textbf{\bibinfo{volume}{44}},
  \bibinfo{pages}{478} (\bibinfo{year}{1986}), \bibinfo{note}{[Yad.
  Fiz.44,738(1986)]}.

\bibitem[{\citenamefont{Kniehl and Spira}(1995)}]{Kniehl:1995tn}
\bibinfo{author}{\bibfnamefont{B.~A.} \bibnamefont{Kniehl}} \bibnamefont{and}
  \bibinfo{author}{\bibfnamefont{M.}~\bibnamefont{Spira}}, \bibinfo{journal}{Z.
  Phys.} \textbf{\bibinfo{volume}{C69}}, \bibinfo{pages}{77}
  (\bibinfo{year}{1995}), \eprint{hep-ph/9505225}.

\bibitem[{\citenamefont{Baglio et~al.}(2018)\citenamefont{Baglio, Campanario,
  Glaus, Muhlleitner, Spira, and Streicher}}]{Baglio:2018lrj}
\bibinfo{author}{\bibfnamefont{J.}~\bibnamefont{Baglio}},
  \bibinfo{author}{\bibfnamefont{F.}~\bibnamefont{Campanario}},
  \bibinfo{author}{\bibfnamefont{S.}~\bibnamefont{Glaus}},
  \bibinfo{author}{\bibfnamefont{M.}~\bibnamefont{Muhlleitner}},
  \bibinfo{author}{\bibfnamefont{M.}~\bibnamefont{Spira}}, \bibnamefont{and}
  \bibinfo{author}{\bibfnamefont{J.}~\bibnamefont{Streicher}}
  (\bibinfo{year}{2018}), \eprint{1811.05692}.

\bibitem[{\citenamefont{Borowka
  et~al.}(2016{\natexlab{a}})\citenamefont{Borowka, Greiner, Heinrich, Jones,
  Kerner, Schlenk, Schubert, and Zirke}}]{Borowka:2016ehy}
\bibinfo{author}{\bibfnamefont{S.}~\bibnamefont{Borowka}},
  \bibinfo{author}{\bibfnamefont{N.}~\bibnamefont{Greiner}},
  \bibinfo{author}{\bibfnamefont{G.}~\bibnamefont{Heinrich}},
  \bibinfo{author}{\bibfnamefont{S.}~\bibnamefont{Jones}},
  \bibinfo{author}{\bibfnamefont{M.}~\bibnamefont{Kerner}},
  \bibinfo{author}{\bibfnamefont{J.}~\bibnamefont{Schlenk}},
  \bibinfo{author}{\bibfnamefont{U.}~\bibnamefont{Schubert}}, \bibnamefont{and}
  \bibinfo{author}{\bibfnamefont{T.}~\bibnamefont{Zirke}},
  \bibinfo{journal}{Phys. Rev. Lett.} \textbf{\bibinfo{volume}{117}},
  \bibinfo{pages}{012001} (\bibinfo{year}{2016}{\natexlab{a}}),
  \bibinfo{note}{[Erratum: Phys. Rev. Lett.117,no.7,079901(2016)]},
  \eprint{1604.06447}.

\bibitem[{\citenamefont{Borowka
  et~al.}(2016{\natexlab{b}})\citenamefont{Borowka, Greiner, Heinrich, Jones,
  Kerner, Schlenk, and Zirke}}]{Borowka:2016ypz}
\bibinfo{author}{\bibfnamefont{S.}~\bibnamefont{Borowka}},
  \bibinfo{author}{\bibfnamefont{N.}~\bibnamefont{Greiner}},
  \bibinfo{author}{\bibfnamefont{G.}~\bibnamefont{Heinrich}},
  \bibinfo{author}{\bibfnamefont{S.~P.} \bibnamefont{Jones}},
  \bibinfo{author}{\bibfnamefont{M.}~\bibnamefont{Kerner}},
  \bibinfo{author}{\bibfnamefont{J.}~\bibnamefont{Schlenk}}, \bibnamefont{and}
  \bibinfo{author}{\bibfnamefont{T.}~\bibnamefont{Zirke}},
  \bibinfo{journal}{JHEP} \textbf{\bibinfo{volume}{10}}, \bibinfo{pages}{107}
  (\bibinfo{year}{2016}{\natexlab{b}}), \eprint{1608.04798}.

\bibitem[{\citenamefont{Heinrich et~al.}(2017)\citenamefont{Heinrich, Jones,
  Kerner, Luisoni, and Vryonidou}}]{Heinrich:2017kxx}
\bibinfo{author}{\bibfnamefont{G.}~\bibnamefont{Heinrich}},
  \bibinfo{author}{\bibfnamefont{S.~P.} \bibnamefont{Jones}},
  \bibinfo{author}{\bibfnamefont{M.}~\bibnamefont{Kerner}},
  \bibinfo{author}{\bibfnamefont{G.}~\bibnamefont{Luisoni}}, \bibnamefont{and}
  \bibinfo{author}{\bibfnamefont{E.}~\bibnamefont{Vryonidou}},
  \bibinfo{journal}{JHEP} \textbf{\bibinfo{volume}{08}}, \bibinfo{pages}{088}
  (\bibinfo{year}{2017}), \eprint{1703.09252}.

\bibitem[{\citenamefont{Davies et~al.}(2018)\citenamefont{Davies, Mishima,
  Steinhauser, and Wellmann}}]{Davies:2018qvx}
\bibinfo{author}{\bibfnamefont{J.}~\bibnamefont{Davies}},
  \bibinfo{author}{\bibfnamefont{G.}~\bibnamefont{Mishima}},
  \bibinfo{author}{\bibfnamefont{M.}~\bibnamefont{Steinhauser}},
  \bibnamefont{and} \bibinfo{author}{\bibfnamefont{D.}~\bibnamefont{Wellmann}}
  (\bibinfo{year}{2018}), \eprint{1811.05489}.

\bibitem[{\citenamefont{Bonciani et~al.}(2018)\citenamefont{Bonciani, Degrassi,
  Giardino, and Gröber}}]{Bonciani:2018omm}
\bibinfo{author}{\bibfnamefont{R.}~\bibnamefont{Bonciani}},
  \bibinfo{author}{\bibfnamefont{G.}~\bibnamefont{Degrassi}},
  \bibinfo{author}{\bibfnamefont{P.~P.} \bibnamefont{Giardino}},
  \bibnamefont{and} \bibinfo{author}{\bibfnamefont{R.}~\bibnamefont{Gröber}},
  \bibinfo{journal}{Phys. Rev. Lett.} \textbf{\bibinfo{volume}{121}},
  \bibinfo{pages}{162003} (\bibinfo{year}{2018}), \eprint{1806.11564}.

\bibitem[{\citenamefont{Chen et~al.}(2015)\citenamefont{Chen, Dawson, and
  Lewis}}]{Chen:2014ask}
\bibinfo{author}{\bibfnamefont{C.-Y.} \bibnamefont{Chen}},
  \bibinfo{author}{\bibfnamefont{S.}~\bibnamefont{Dawson}}, \bibnamefont{and}
  \bibinfo{author}{\bibfnamefont{I.~M.} \bibnamefont{Lewis}},
  \bibinfo{journal}{Phys. Rev.} \textbf{\bibinfo{volume}{D91}},
  \bibinfo{pages}{035015} (\bibinfo{year}{2015}), \eprint{1410.5488}.

\bibitem[{\citenamefont{Barger et~al.}(2003)\citenamefont{Barger, Han,
  Langacker, McElrath, and Zerwas}}]{Barger:2003rs}
\bibinfo{author}{\bibfnamefont{V.}~\bibnamefont{Barger}},
  \bibinfo{author}{\bibfnamefont{T.}~\bibnamefont{Han}},
  \bibinfo{author}{\bibfnamefont{P.}~\bibnamefont{Langacker}},
  \bibinfo{author}{\bibfnamefont{B.}~\bibnamefont{McElrath}}, \bibnamefont{and}
  \bibinfo{author}{\bibfnamefont{P.}~\bibnamefont{Zerwas}},
  \bibinfo{journal}{Phys. Rev.} \textbf{\bibinfo{volume}{D67}},
  \bibinfo{pages}{115001} (\bibinfo{year}{2003}), \eprint{hep-ph/0301097}.

\bibitem[{\citenamefont{Baglio et~al.}(2013)\citenamefont{Baglio, Djouadi,
  Grober, Muhlleitner, Quevillon, and Spira}}]{Baglio:2012np}
\bibinfo{author}{\bibfnamefont{J.}~\bibnamefont{Baglio}},
  \bibinfo{author}{\bibfnamefont{A.}~\bibnamefont{Djouadi}},
  \bibinfo{author}{\bibfnamefont{R.}~\bibnamefont{Grober}},
  \bibinfo{author}{\bibfnamefont{M.~M.} \bibnamefont{Muhlleitner}},
  \bibinfo{author}{\bibfnamefont{J.}~\bibnamefont{Quevillon}},
  \bibnamefont{and} \bibinfo{author}{\bibfnamefont{M.}~\bibnamefont{Spira}},
  \bibinfo{journal}{JHEP} \textbf{\bibinfo{volume}{04}}, \bibinfo{pages}{151}
  (\bibinfo{year}{2013}), \eprint{1212.5581}.

\bibitem[{\citenamefont{Gupta et~al.}(2013)\citenamefont{Gupta, Rzehak, and
  Wells}}]{Gupta:2013zza}
\bibinfo{author}{\bibfnamefont{R.~S.} \bibnamefont{Gupta}},
  \bibinfo{author}{\bibfnamefont{H.}~\bibnamefont{Rzehak}}, \bibnamefont{and}
  \bibinfo{author}{\bibfnamefont{J.~D.} \bibnamefont{Wells}},
  \bibinfo{journal}{Phys. Rev.} \textbf{\bibinfo{volume}{D88}},
  \bibinfo{pages}{055024} (\bibinfo{year}{2013}), \eprint{1305.6397}.

\bibitem[{\citenamefont{Di~Vita et~al.}(2017)\citenamefont{Di~Vita, Grojean,
  Panico, Riembau, and Vantalon}}]{DiVita:2017eyz}
\bibinfo{author}{\bibfnamefont{S.}~\bibnamefont{Di~Vita}},
  \bibinfo{author}{\bibfnamefont{C.}~\bibnamefont{Grojean}},
  \bibinfo{author}{\bibfnamefont{G.}~\bibnamefont{Panico}},
  \bibinfo{author}{\bibfnamefont{M.}~\bibnamefont{Riembau}}, \bibnamefont{and}
  \bibinfo{author}{\bibfnamefont{T.}~\bibnamefont{Vantalon}},
  \bibinfo{journal}{JHEP} \textbf{\bibinfo{volume}{09}}, \bibinfo{pages}{069}
  (\bibinfo{year}{2017}), \eprint{1704.01953}.

\bibitem[{\citenamefont{Di~Luzio et~al.}(2017)\citenamefont{Di~Luzio, Grober,
  and Spannowsky}}]{DiLuzio:2017tfn}
\bibinfo{author}{\bibfnamefont{L.}~\bibnamefont{Di~Luzio}},
  \bibinfo{author}{\bibfnamefont{R.}~\bibnamefont{Grober}}, \bibnamefont{and}
  \bibinfo{author}{\bibfnamefont{M.}~\bibnamefont{Spannowsky}},
  \bibinfo{journal}{Eur. Phys. J.} \textbf{\bibinfo{volume}{C77}},
  \bibinfo{pages}{788} (\bibinfo{year}{2017}), \eprint{1704.02311}.

\bibitem[{\citenamefont{Khachatryan et~al.}(2017)}]{CMS:2017cwx}
\bibinfo{author}{\bibfnamefont{V.}~\bibnamefont{Khachatryan}}
  \bibnamefont{et~al.} (\bibinfo{collaboration}{CMS}) (\bibinfo{year}{2017}).

\bibitem[{\citenamefont{Adhikary et~al.}(2018)\citenamefont{Adhikary, Banerjee,
  Barman, Bhattacherjee, and Niyogi}}]{Adhikary:2017jtu}
\bibinfo{author}{\bibfnamefont{A.}~\bibnamefont{Adhikary}},
  \bibinfo{author}{\bibfnamefont{S.}~\bibnamefont{Banerjee}},
  \bibinfo{author}{\bibfnamefont{R.~K.} \bibnamefont{Barman}},
  \bibinfo{author}{\bibfnamefont{B.}~\bibnamefont{Bhattacherjee}},
  \bibnamefont{and} \bibinfo{author}{\bibfnamefont{S.}~\bibnamefont{Niyogi}},
  \bibinfo{journal}{JHEP} \textbf{\bibinfo{volume}{07}}, \bibinfo{pages}{116}
  (\bibinfo{year}{2018}), \eprint{1712.05346}.

\bibitem[{\citenamefont{Goncalves et~al.}(2018)\citenamefont{Goncalves, Han,
  Kling, Plehn, and Takeuchi}}]{Goncalves:2018yva}
\bibinfo{author}{\bibfnamefont{D.}~\bibnamefont{Goncalves}},
  \bibinfo{author}{\bibfnamefont{T.}~\bibnamefont{Han}},
  \bibinfo{author}{\bibfnamefont{F.}~\bibnamefont{Kling}},
  \bibinfo{author}{\bibfnamefont{T.}~\bibnamefont{Plehn}}, \bibnamefont{and}
  \bibinfo{author}{\bibfnamefont{M.}~\bibnamefont{Takeuchi}},
  \bibinfo{journal}{Phys. Rev.} \textbf{\bibinfo{volume}{D97}},
  \bibinfo{pages}{113004} (\bibinfo{year}{2018}), \eprint{1802.04319}.

\bibitem[{\citenamefont{Homiller and Meade}(2018)}]{Homiller:2018dgu}
\bibinfo{author}{\bibfnamefont{S.}~\bibnamefont{Homiller}} \bibnamefont{and}
  \bibinfo{author}{\bibfnamefont{P.}~\bibnamefont{Meade}}
  (\bibinfo{year}{2018}), \eprint{1811.02572}.

\bibitem[{\citenamefont{Azatov et~al.}(2015)\citenamefont{Azatov, Contino,
  Panico, and Son}}]{Azatov:2015oxa}
\bibinfo{author}{\bibfnamefont{A.}~\bibnamefont{Azatov}},
  \bibinfo{author}{\bibfnamefont{R.}~\bibnamefont{Contino}},
  \bibinfo{author}{\bibfnamefont{G.}~\bibnamefont{Panico}}, \bibnamefont{and}
  \bibinfo{author}{\bibfnamefont{M.}~\bibnamefont{Son}},
  \bibinfo{journal}{Phys. Rev.} \textbf{\bibinfo{volume}{D92}},
  \bibinfo{pages}{035001} (\bibinfo{year}{2015}), \eprint{1502.00539}.

\bibitem[{\citenamefont{Yao}(2013)}]{Yao:2013ika}
\bibinfo{author}{\bibfnamefont{W.}~\bibnamefont{Yao}}, in
  \emph{\bibinfo{booktitle}{{Snowmass on the Mississippi (CSS2013)}}}
  (\bibinfo{year}{2013}), \eprint{1308.6302},
  \urlprefix\url{http://www.slac.stanford.edu/econf/C1307292/docs/submittedArxivFiles/1308.6302.pdf}.

\bibitem[{\citenamefont{Barr et~al.}(2015)\citenamefont{Barr, Dolan, Englert,
  Ferreira~de Lima, and Spannowsky}}]{Barr:2014sga}
\bibinfo{author}{\bibfnamefont{A.~J.} \bibnamefont{Barr}},
  \bibinfo{author}{\bibfnamefont{M.~J.} \bibnamefont{Dolan}},
  \bibinfo{author}{\bibfnamefont{C.}~\bibnamefont{Englert}},
  \bibinfo{author}{\bibfnamefont{D.~E.} \bibnamefont{Ferreira~de Lima}},
  \bibnamefont{and}
  \bibinfo{author}{\bibfnamefont{M.}~\bibnamefont{Spannowsky}},
  \bibinfo{journal}{JHEP} \textbf{\bibinfo{volume}{02}}, \bibinfo{pages}{016}
  (\bibinfo{year}{2015}), \eprint{1412.7154}.

\bibitem[{\citenamefont{Papaefstathiou}(2015)}]{Papaefstathiou:2015iba}
\bibinfo{author}{\bibfnamefont{A.}~\bibnamefont{Papaefstathiou}},
  \bibinfo{journal}{Phys. Rev.} \textbf{\bibinfo{volume}{D91}},
  \bibinfo{pages}{113016} (\bibinfo{year}{2015}), \eprint{1504.04621}.

\bibitem[{\citenamefont{Zhao et~al.}(2017)\citenamefont{Zhao, Li, Li, and
  Yan}}]{Zhao:2016tai}
\bibinfo{author}{\bibfnamefont{X.}~\bibnamefont{Zhao}},
  \bibinfo{author}{\bibfnamefont{Q.}~\bibnamefont{Li}},
  \bibinfo{author}{\bibfnamefont{Z.}~\bibnamefont{Li}}, \bibnamefont{and}
  \bibinfo{author}{\bibfnamefont{Q.-S.} \bibnamefont{Yan}},
  \bibinfo{journal}{Chin. Phys.} \textbf{\bibinfo{volume}{C41}},
  \bibinfo{pages}{023105} (\bibinfo{year}{2017}), \eprint{1604.04329}.

\bibitem[{\citenamefont{Contino et~al.}(2017)}]{Contino:2016spe}
\bibinfo{author}{\bibfnamefont{R.}~\bibnamefont{Contino}} \bibnamefont{et~al.},
  \bibinfo{journal}{CERN Yellow Report} pp. \bibinfo{pages}{255--440}
  (\bibinfo{year}{2017}), \eprint{1606.09408}.

\bibitem[{\citenamefont{Banerjee et~al.}(2018)\citenamefont{Banerjee, Englert,
  Mangano, Selvaggi, and Spannowsky}}]{Banerjee:2018yxy}
\bibinfo{author}{\bibfnamefont{S.}~\bibnamefont{Banerjee}},
  \bibinfo{author}{\bibfnamefont{C.}~\bibnamefont{Englert}},
  \bibinfo{author}{\bibfnamefont{M.~L.} \bibnamefont{Mangano}},
  \bibinfo{author}{\bibfnamefont{M.}~\bibnamefont{Selvaggi}}, \bibnamefont{and}
  \bibinfo{author}{\bibfnamefont{M.}~\bibnamefont{Spannowsky}},
  \bibinfo{journal}{Eur. Phys. J.} \textbf{\bibinfo{volume}{C78}},
  \bibinfo{pages}{322} (\bibinfo{year}{2018}), \eprint{1802.01607}.

\bibitem[{\citenamefont{Grober and Muhlleitner}(2011)}]{Grober:2010yv}
\bibinfo{author}{\bibfnamefont{R.}~\bibnamefont{Grober}} \bibnamefont{and}
  \bibinfo{author}{\bibfnamefont{M.}~\bibnamefont{Muhlleitner}},
  \bibinfo{journal}{JHEP} \textbf{\bibinfo{volume}{06}}, \bibinfo{pages}{020}
  (\bibinfo{year}{2011}), \eprint{1012.1562}.

\bibitem[{\citenamefont{Dolan et~al.}(2013)\citenamefont{Dolan, Englert, and
  Spannowsky}}]{Dolan:2012ac}
\bibinfo{author}{\bibfnamefont{M.~J.} \bibnamefont{Dolan}},
  \bibinfo{author}{\bibfnamefont{C.}~\bibnamefont{Englert}}, \bibnamefont{and}
  \bibinfo{author}{\bibfnamefont{M.}~\bibnamefont{Spannowsky}},
  \bibinfo{journal}{Phys. Rev.} \textbf{\bibinfo{volume}{D87}},
  \bibinfo{pages}{055002} (\bibinfo{year}{2013}), \eprint{1210.8166}.

\bibitem[{\citenamefont{Chen et~al.}(2014)\citenamefont{Chen, Dawson, and
  Lewis}}]{Chen:2014xwa}
\bibinfo{author}{\bibfnamefont{C.-Y.} \bibnamefont{Chen}},
  \bibinfo{author}{\bibfnamefont{S.}~\bibnamefont{Dawson}}, \bibnamefont{and}
  \bibinfo{author}{\bibfnamefont{I.~M.} \bibnamefont{Lewis}},
  \bibinfo{journal}{Phys. Rev.} \textbf{\bibinfo{volume}{D90}},
  \bibinfo{pages}{035016} (\bibinfo{year}{2014}), \eprint{1406.3349}.

\bibitem[{\citenamefont{Grober et~al.}(2016)\citenamefont{Grober, Muhlleitner,
  and Spira}}]{Grober:2016wmf}
\bibinfo{author}{\bibfnamefont{R.}~\bibnamefont{Grober}},
  \bibinfo{author}{\bibfnamefont{M.}~\bibnamefont{Muhlleitner}},
  \bibnamefont{and} \bibinfo{author}{\bibfnamefont{M.}~\bibnamefont{Spira}},
  \bibinfo{journal}{JHEP} \textbf{\bibinfo{volume}{06}}, \bibinfo{pages}{080}
  (\bibinfo{year}{2016}), \eprint{1602.05851}.

\bibitem[{\citenamefont{Vryonidou and Zhang}(2018)}]{Vryonidou:2018eyv}
\bibinfo{author}{\bibfnamefont{E.}~\bibnamefont{Vryonidou}} \bibnamefont{and}
  \bibinfo{author}{\bibfnamefont{C.}~\bibnamefont{Zhang}},
  \bibinfo{journal}{JHEP} \textbf{\bibinfo{volume}{08}}, \bibinfo{pages}{036}
  (\bibinfo{year}{2018}), \eprint{1804.09766}.

\bibitem[{\citenamefont{Lee}(1973)}]{Lee:1973iz}
\bibinfo{author}{\bibfnamefont{T.~D.} \bibnamefont{Lee}},
  \bibinfo{journal}{Phys. Rev.} \textbf{\bibinfo{volume}{D8}},
  \bibinfo{pages}{1226} (\bibinfo{year}{1973}), \bibinfo{note}{[,516(1973)]}.

\bibitem[{\citenamefont{Gunion et~al.}(2000)\citenamefont{Gunion, Haber, Kane,
  and Dawson}}]{Gunion:1989we}
\bibinfo{author}{\bibfnamefont{J.~F.} \bibnamefont{Gunion}},
  \bibinfo{author}{\bibfnamefont{H.~E.} \bibnamefont{Haber}},
  \bibinfo{author}{\bibfnamefont{G.~L.} \bibnamefont{Kane}}, \bibnamefont{and}
  \bibinfo{author}{\bibfnamefont{S.}~\bibnamefont{Dawson}},
  \bibinfo{journal}{Front. Phys.} \textbf{\bibinfo{volume}{80}},
  \bibinfo{pages}{1} (\bibinfo{year}{2000}).

\bibitem[{\citenamefont{Branco et~al.}(2012)\citenamefont{Branco, Ferreira,
  Lavoura, Rebelo, Sher, and Silva}}]{Branco:2011iw}
\bibinfo{author}{\bibfnamefont{G.~C.} \bibnamefont{Branco}},
  \bibinfo{author}{\bibfnamefont{P.~M.} \bibnamefont{Ferreira}},
  \bibinfo{author}{\bibfnamefont{L.}~\bibnamefont{Lavoura}},
  \bibinfo{author}{\bibfnamefont{M.~N.} \bibnamefont{Rebelo}},
  \bibinfo{author}{\bibfnamefont{M.}~\bibnamefont{Sher}}, \bibnamefont{and}
  \bibinfo{author}{\bibfnamefont{J.~P.} \bibnamefont{Silva}},
  \bibinfo{journal}{Phys. Rept.} \textbf{\bibinfo{volume}{516}},
  \bibinfo{pages}{1} (\bibinfo{year}{2012}), \eprint{1106.0034}.

\bibitem[{\citenamefont{Ginzburg et~al.}(2002)\citenamefont{Ginzburg, Krawczyk,
  and Osland}}]{Ginzburg:2002wt}
\bibinfo{author}{\bibfnamefont{I.~F.} \bibnamefont{Ginzburg}},
  \bibinfo{author}{\bibfnamefont{M.}~\bibnamefont{Krawczyk}}, \bibnamefont{and}
  \bibinfo{author}{\bibfnamefont{P.}~\bibnamefont{Osland}}, in
  \emph{\bibinfo{booktitle}{{Linear colliders. Proceedings, LCWS 2002}}}
  (\bibinfo{year}{2002}), pp. \bibinfo{pages}{703--706},
  \bibinfo{note}{[,703(2002)]}, \eprint{hep-ph/0211371},
  \urlprefix\url{http://weblib.cern.ch/abstract?CERN-TH-2002-330}.

\bibitem[{\citenamefont{Fontes et~al.}(2014)\citenamefont{Fontes, Romao, and
  Silva}}]{Fontes:2014xva}
\bibinfo{author}{\bibfnamefont{D.}~\bibnamefont{Fontes}},
  \bibinfo{author}{\bibfnamefont{J.~C.} \bibnamefont{Romao}}, \bibnamefont{and}
  \bibinfo{author}{\bibfnamefont{J.~P.} \bibnamefont{Silva}},
  \bibinfo{journal}{JHEP} \textbf{\bibinfo{volume}{12}}, \bibinfo{pages}{043}
  (\bibinfo{year}{2014}), \eprint{1408.2534}.

\bibitem[{\citenamefont{Lavoura and Silva}(1994)}]{Lavoura:1994fv}
\bibinfo{author}{\bibfnamefont{L.}~\bibnamefont{Lavoura}} \bibnamefont{and}
  \bibinfo{author}{\bibfnamefont{J.~P.} \bibnamefont{Silva}},
  \bibinfo{journal}{Phys. Rev.} \textbf{\bibinfo{volume}{D50}},
  \bibinfo{pages}{4619} (\bibinfo{year}{1994}), \eprint{hep-ph/9404276}.

\bibitem[{\citenamefont{Botella and Silva}(1995)}]{Botella:1994cs}
\bibinfo{author}{\bibfnamefont{F.~J.} \bibnamefont{Botella}} \bibnamefont{and}
  \bibinfo{author}{\bibfnamefont{J.~P.} \bibnamefont{Silva}},
  \bibinfo{journal}{Phys. Rev.} \textbf{\bibinfo{volume}{D51}},
  \bibinfo{pages}{3870} (\bibinfo{year}{1995}), \eprint{hep-ph/9411288}.

\bibitem[{\citenamefont{El~Kaffas et~al.}(2007)\citenamefont{El~Kaffas, Osland,
  and Ogreid}}]{ElKaffas:2007rq}
\bibinfo{author}{\bibfnamefont{A.~W.} \bibnamefont{El~Kaffas}},
  \bibinfo{author}{\bibfnamefont{P.}~\bibnamefont{Osland}}, \bibnamefont{and}
  \bibinfo{author}{\bibfnamefont{O.~M.} \bibnamefont{Ogreid}},
  \bibinfo{journal}{Nonlin. Phenom. Complex Syst.}
  \textbf{\bibinfo{volume}{10}}, \bibinfo{pages}{347} (\bibinfo{year}{2007}),
  \eprint{hep-ph/0702097}.

\bibitem[{\citenamefont{Fontes et~al.}(2018)\citenamefont{Fontes, Muhlleitner,
  Romao, Santos, Silva, and Wittbrodt}}]{Fontes:2017zfn}
\bibinfo{author}{\bibfnamefont{D.}~\bibnamefont{Fontes}},
  \bibinfo{author}{\bibfnamefont{M.}~\bibnamefont{Muhlleitner}},
  \bibinfo{author}{\bibfnamefont{J.~C.} \bibnamefont{Romao}},
  \bibinfo{author}{\bibfnamefont{R.}~\bibnamefont{Santos}},
  \bibinfo{author}{\bibfnamefont{J.~P.} \bibnamefont{Silva}}, \bibnamefont{and}
  \bibinfo{author}{\bibfnamefont{J.}~\bibnamefont{Wittbrodt}},
  \bibinfo{journal}{JHEP} \textbf{\bibinfo{volume}{02}}, \bibinfo{pages}{073}
  (\bibinfo{year}{2018}), \eprint{1711.09419}.

\bibitem[{\citenamefont{Ellwanger et~al.}(2010)\citenamefont{Ellwanger,
  Hugonie, and Teixeira}}]{Ellwanger:2009dp}
\bibinfo{author}{\bibfnamefont{U.}~\bibnamefont{Ellwanger}},
  \bibinfo{author}{\bibfnamefont{C.}~\bibnamefont{Hugonie}}, \bibnamefont{and}
  \bibinfo{author}{\bibfnamefont{A.~M.} \bibnamefont{Teixeira}},
  \bibinfo{journal}{Phys. Rept.} \textbf{\bibinfo{volume}{496}},
  \bibinfo{pages}{1} (\bibinfo{year}{2010}), \eprint{0910.1785}.

\bibitem[{\citenamefont{Maniatis}(2010)}]{Maniatis:2009re}
\bibinfo{author}{\bibfnamefont{M.}~\bibnamefont{Maniatis}},
  \bibinfo{journal}{Int. J. Mod. Phys.} \textbf{\bibinfo{volume}{A25}},
  \bibinfo{pages}{3505} (\bibinfo{year}{2010}), \eprint{0906.0777}.

\bibitem[{\citenamefont{Haber and Stål}(2015)}]{Haber:2015pua}
\bibinfo{author}{\bibfnamefont{H.~E.} \bibnamefont{Haber}} \bibnamefont{and}
  \bibinfo{author}{\bibfnamefont{O.}~\bibnamefont{Stål}},
  \bibinfo{journal}{Eur. Phys. J.} \textbf{\bibinfo{volume}{C75}},
  \bibinfo{pages}{491} (\bibinfo{year}{2015}), \bibinfo{note}{[Erratum: Eur.
  Phys. J.C76,no.6,312(2016)]}, \eprint{1507.04281}.

\bibitem[{\citenamefont{Baglio et~al.}(2014{\natexlab{a}})\citenamefont{Baglio,
  Eberhardt, Nierste, and Wiebusch}}]{Baglio:2014nea}
\bibinfo{author}{\bibfnamefont{J.}~\bibnamefont{Baglio}},
  \bibinfo{author}{\bibfnamefont{O.}~\bibnamefont{Eberhardt}},
  \bibinfo{author}{\bibfnamefont{U.}~\bibnamefont{Nierste}}, \bibnamefont{and}
  \bibinfo{author}{\bibfnamefont{M.}~\bibnamefont{Wiebusch}},
  \bibinfo{journal}{Phys. Rev.} \textbf{\bibinfo{volume}{D90}},
  \bibinfo{pages}{015008} (\bibinfo{year}{2014}{\natexlab{a}}),
  \eprint{1403.1264}.

\bibitem[{\citenamefont{Aad et~al.}(2015{\natexlab{a}})}]{Aad:2015zhl}
\bibinfo{author}{\bibfnamefont{G.}~\bibnamefont{Aad}} \bibnamefont{et~al.}
  (\bibinfo{collaboration}{ATLAS, CMS}), \bibinfo{journal}{Phys. Rev. Lett.}
  \textbf{\bibinfo{volume}{114}}, \bibinfo{pages}{191803}
  (\bibinfo{year}{2015}{\natexlab{a}}), \eprint{1503.07589}.

\bibitem[{\citenamefont{Haber and Logan}(2000)}]{Haber:1999zh}
\bibinfo{author}{\bibfnamefont{H.~E.} \bibnamefont{Haber}} \bibnamefont{and}
  \bibinfo{author}{\bibfnamefont{H.~E.} \bibnamefont{Logan}},
  \bibinfo{journal}{Phys. Rev.} \textbf{\bibinfo{volume}{D62}},
  \bibinfo{pages}{015011} (\bibinfo{year}{2000}), \eprint{hep-ph/9909335}.

\bibitem[{\citenamefont{Deschamps et~al.}(2010)\citenamefont{Deschamps,
  Descotes-Genon, Monteil, Niess, T'Jampens, and Tisserand}}]{Deschamps:2009rh}
\bibinfo{author}{\bibfnamefont{O.}~\bibnamefont{Deschamps}},
  \bibinfo{author}{\bibfnamefont{S.}~\bibnamefont{Descotes-Genon}},
  \bibinfo{author}{\bibfnamefont{S.}~\bibnamefont{Monteil}},
  \bibinfo{author}{\bibfnamefont{V.}~\bibnamefont{Niess}},
  \bibinfo{author}{\bibfnamefont{S.}~\bibnamefont{T'Jampens}},
  \bibnamefont{and}
  \bibinfo{author}{\bibfnamefont{V.}~\bibnamefont{Tisserand}},
  \bibinfo{journal}{Phys. Rev.} \textbf{\bibinfo{volume}{D82}},
  \bibinfo{pages}{073012} (\bibinfo{year}{2010}), \eprint{0907.5135}.

\bibitem[{\citenamefont{Mahmoudi and Stal}(2010)}]{Mahmoudi:2009zx}
\bibinfo{author}{\bibfnamefont{F.}~\bibnamefont{Mahmoudi}} \bibnamefont{and}
  \bibinfo{author}{\bibfnamefont{O.}~\bibnamefont{Stal}},
  \bibinfo{journal}{Phys. Rev.} \textbf{\bibinfo{volume}{D81}},
  \bibinfo{pages}{035016} (\bibinfo{year}{2010}), \eprint{0907.1791}.

\bibitem[{\citenamefont{Hermann et~al.}(2012)\citenamefont{Hermann, Misiak, and
  Steinhauser}}]{Hermann:2012fc}
\bibinfo{author}{\bibfnamefont{T.}~\bibnamefont{Hermann}},
  \bibinfo{author}{\bibfnamefont{M.}~\bibnamefont{Misiak}}, \bibnamefont{and}
  \bibinfo{author}{\bibfnamefont{M.}~\bibnamefont{Steinhauser}},
  \bibinfo{journal}{JHEP} \textbf{\bibinfo{volume}{11}}, \bibinfo{pages}{036}
  (\bibinfo{year}{2012}), \eprint{1208.2788}.

\bibitem[{\citenamefont{Misiak et~al.}(2015)}]{Misiak:2015xwa}
\bibinfo{author}{\bibfnamefont{M.}~\bibnamefont{Misiak}} \bibnamefont{et~al.},
  \bibinfo{journal}{Phys. Rev. Lett.} \textbf{\bibinfo{volume}{114}},
  \bibinfo{pages}{221801} (\bibinfo{year}{2015}), \eprint{1503.01789}.

\bibitem[{\citenamefont{Misiak and Steinhauser}(2017)}]{Misiak:2017bgg}
\bibinfo{author}{\bibfnamefont{M.}~\bibnamefont{Misiak}} \bibnamefont{and}
  \bibinfo{author}{\bibfnamefont{M.}~\bibnamefont{Steinhauser}},
  \bibinfo{journal}{Eur. Phys. J.} \textbf{\bibinfo{volume}{C77}},
  \bibinfo{pages}{201} (\bibinfo{year}{2017}), \eprint{1702.04571}.

\bibitem[{\citenamefont{Dawson and Sullivan}(2018)}]{Dawson:2017jja}
\bibinfo{author}{\bibfnamefont{S.}~\bibnamefont{Dawson}} \bibnamefont{and}
  \bibinfo{author}{\bibfnamefont{M.}~\bibnamefont{Sullivan}},
  \bibinfo{journal}{Phys. Rev.} \textbf{\bibinfo{volume}{D97}},
  \bibinfo{pages}{015022} (\bibinfo{year}{2018}), \eprint{1711.06683}.

\bibitem[{\citenamefont{Baak et~al.}(2014)\citenamefont{Baak, CÃºth, Haller,
  Hoecker, Kogler, MÃ¶nig, Schott, and Stelzer}}]{Baak:2014ora}
\bibinfo{author}{\bibfnamefont{M.}~\bibnamefont{Baak}},
  \bibinfo{author}{\bibfnamefont{J.}~\bibnamefont{CÃºth}},
  \bibinfo{author}{\bibfnamefont{J.}~\bibnamefont{Haller}},
  \bibinfo{author}{\bibfnamefont{A.}~\bibnamefont{Hoecker}},
  \bibinfo{author}{\bibfnamefont{R.}~\bibnamefont{Kogler}},
  \bibinfo{author}{\bibfnamefont{K.}~\bibnamefont{MÃ¶nig}},
  \bibinfo{author}{\bibfnamefont{M.}~\bibnamefont{Schott}}, \bibnamefont{and}
  \bibinfo{author}{\bibfnamefont{J.}~\bibnamefont{Stelzer}}
  (\bibinfo{collaboration}{Gfitter Group}), \bibinfo{journal}{Eur. Phys. J.}
  \textbf{\bibinfo{volume}{C74}}, \bibinfo{pages}{3046} (\bibinfo{year}{2014}),
  \eprint{1407.3792}.

\bibitem[{\citenamefont{Tanabashi et~al.}(2018)}]{PhysRevD.98.030001}
\bibinfo{author}{\bibfnamefont{M.}~\bibnamefont{Tanabashi}}
  \bibnamefont{et~al.} (\bibinfo{collaboration}{Particle Data Group}),
  \bibinfo{journal}{Phys. Rev. D} \textbf{\bibinfo{volume}{98}},
  \bibinfo{pages}{030001} (\bibinfo{year}{2018}),
  \urlprefix\url{https://link.aps.org/doi/10.1103/PhysRevD.98.030001}.

\bibitem[{\citenamefont{Denner}(2015)}]{Dennerlhcnote}
\bibinfo{author}{\bibfnamefont{A.~e.~a.} \bibnamefont{Denner}}
  (\bibinfo{year}{2015}), \eprint{LHCHXSWG-INT-2015-006}.

\bibitem[{\citenamefont{Coimbra et~al.}(2013)\citenamefont{Coimbra, Sampaio,
  and Santos}}]{Coimbra:2013qq}
\bibinfo{author}{\bibfnamefont{R.}~\bibnamefont{Coimbra}},
  \bibinfo{author}{\bibfnamefont{M.~O.~P.} \bibnamefont{Sampaio}},
  \bibnamefont{and} \bibinfo{author}{\bibfnamefont{R.}~\bibnamefont{Santos}},
  \bibinfo{journal}{Eur. Phys. J.} \textbf{\bibinfo{volume}{C73}},
  \bibinfo{pages}{2428} (\bibinfo{year}{2013}), \eprint{1301.2599}.

\bibitem[{\citenamefont{Ferreira et~al.}(2014)\citenamefont{Ferreira, Guedes,
  Sampaio, and Santos}}]{Ferreira:2014dya}
\bibinfo{author}{\bibfnamefont{P.~M.} \bibnamefont{Ferreira}},
  \bibinfo{author}{\bibfnamefont{R.}~\bibnamefont{Guedes}},
  \bibinfo{author}{\bibfnamefont{M.~O.~P.} \bibnamefont{Sampaio}},
  \bibnamefont{and} \bibinfo{author}{\bibfnamefont{R.}~\bibnamefont{Santos}},
  \bibinfo{journal}{JHEP} \textbf{\bibinfo{volume}{12}}, \bibinfo{pages}{067}
  (\bibinfo{year}{2014}), \eprint{1409.6723}.

\bibitem[{\citenamefont{Ivanov and Silva}(2015)}]{Ivanov:2015nea}
\bibinfo{author}{\bibfnamefont{I.~P.} \bibnamefont{Ivanov}} \bibnamefont{and}
  \bibinfo{author}{\bibfnamefont{J.~P.} \bibnamefont{Silva}},
  \bibinfo{journal}{Phys. Rev.} \textbf{\bibinfo{volume}{D92}},
  \bibinfo{pages}{055017} (\bibinfo{year}{2015}), \eprint{1507.05100}.

\bibitem[{\citenamefont{Bechtle et~al.}(2010)\citenamefont{Bechtle, Brein,
  Heinemeyer, Weiglein, and Williams}}]{Bechtle:2008jh}
\bibinfo{author}{\bibfnamefont{P.}~\bibnamefont{Bechtle}},
  \bibinfo{author}{\bibfnamefont{O.}~\bibnamefont{Brein}},
  \bibinfo{author}{\bibfnamefont{S.}~\bibnamefont{Heinemeyer}},
  \bibinfo{author}{\bibfnamefont{G.}~\bibnamefont{Weiglein}}, \bibnamefont{and}
  \bibinfo{author}{\bibfnamefont{K.~E.} \bibnamefont{Williams}},
  \bibinfo{journal}{Comput. Phys. Commun.} \textbf{\bibinfo{volume}{181}},
  \bibinfo{pages}{138} (\bibinfo{year}{2010}), \eprint{0811.4169}.

\bibitem[{\citenamefont{Bechtle et~al.}(2011)\citenamefont{Bechtle, Brein,
  Heinemeyer, Weiglein, and Williams}}]{Bechtle:2011sb}
\bibinfo{author}{\bibfnamefont{P.}~\bibnamefont{Bechtle}},
  \bibinfo{author}{\bibfnamefont{O.}~\bibnamefont{Brein}},
  \bibinfo{author}{\bibfnamefont{S.}~\bibnamefont{Heinemeyer}},
  \bibinfo{author}{\bibfnamefont{G.}~\bibnamefont{Weiglein}}, \bibnamefont{and}
  \bibinfo{author}{\bibfnamefont{K.~E.} \bibnamefont{Williams}},
  \bibinfo{journal}{Comput. Phys. Commun.} \textbf{\bibinfo{volume}{182}},
  \bibinfo{pages}{2605} (\bibinfo{year}{2011}), \eprint{1102.1898}.

\bibitem[{\citenamefont{Bechtle
  et~al.}(2014{\natexlab{a}})\citenamefont{Bechtle, Brein, Heinemeyer, Stï¿½l,
  Stefaniak, Weiglein, and Williams}}]{Bechtle:2013wla}
\bibinfo{author}{\bibfnamefont{P.}~\bibnamefont{Bechtle}},
  \bibinfo{author}{\bibfnamefont{O.}~\bibnamefont{Brein}},
  \bibinfo{author}{\bibfnamefont{S.}~\bibnamefont{Heinemeyer}},
  \bibinfo{author}{\bibfnamefont{O.}~\bibnamefont{Stï¿½l}},
  \bibinfo{author}{\bibfnamefont{T.}~\bibnamefont{Stefaniak}},
  \bibinfo{author}{\bibfnamefont{G.}~\bibnamefont{Weiglein}}, \bibnamefont{and}
  \bibinfo{author}{\bibfnamefont{K.~E.} \bibnamefont{Williams}},
  \bibinfo{journal}{Eur. Phys. J.} \textbf{\bibinfo{volume}{C74}},
  \bibinfo{pages}{2693} (\bibinfo{year}{2014}{\natexlab{a}}),
  \eprint{1311.0055}.

\bibitem[{\citenamefont{Bechtle
  et~al.}(2014{\natexlab{b}})\citenamefont{Bechtle, Heinemeyer, StÃ¥l,
  Stefaniak, and Weiglein}}]{Bechtle:2013xfa}
\bibinfo{author}{\bibfnamefont{P.}~\bibnamefont{Bechtle}},
  \bibinfo{author}{\bibfnamefont{S.}~\bibnamefont{Heinemeyer}},
  \bibinfo{author}{\bibfnamefont{O.}~\bibnamefont{StÃ¥l}},
  \bibinfo{author}{\bibfnamefont{T.}~\bibnamefont{Stefaniak}},
  \bibnamefont{and} \bibinfo{author}{\bibfnamefont{G.}~\bibnamefont{Weiglein}},
  \bibinfo{journal}{Eur. Phys. J.} \textbf{\bibinfo{volume}{C74}},
  \bibinfo{pages}{2711} (\bibinfo{year}{2014}{\natexlab{b}}),
  \eprint{1305.1933}.

\bibitem[{\citenamefont{Djouadi et~al.}(1998)\citenamefont{Djouadi, Kalinowski,
  and Spira}}]{Djouadi:1997yw}
\bibinfo{author}{\bibfnamefont{A.}~\bibnamefont{Djouadi}},
  \bibinfo{author}{\bibfnamefont{J.}~\bibnamefont{Kalinowski}},
  \bibnamefont{and} \bibinfo{author}{\bibfnamefont{M.}~\bibnamefont{Spira}},
  \bibinfo{journal}{Comput. Phys. Commun.} \textbf{\bibinfo{volume}{108}},
  \bibinfo{pages}{56} (\bibinfo{year}{1998}), \eprint{hep-ph/9704448}.

\bibitem[{\citenamefont{Djouadi et~al.}(2018)\citenamefont{Djouadi, Kalinowski,
  Muehlleitner, and Spira}}]{Djouadi:2018xqq}
\bibinfo{author}{\bibfnamefont{A.}~\bibnamefont{Djouadi}},
  \bibinfo{author}{\bibfnamefont{J.}~\bibnamefont{Kalinowski}},
  \bibinfo{author}{\bibfnamefont{M.}~\bibnamefont{Muehlleitner}},
  \bibnamefont{and} \bibinfo{author}{\bibfnamefont{M.}~\bibnamefont{Spira}}
  (\bibinfo{year}{2018}), \eprint{1801.09506}.

\bibitem[{\citenamefont{Muhlleitner et~al.}(2017)\citenamefont{Muhlleitner,
  Sampaio, Santos, and Wittbrodt}}]{Muhlleitner:2017dkd}
\bibinfo{author}{\bibfnamefont{M.}~\bibnamefont{Muhlleitner}},
  \bibinfo{author}{\bibfnamefont{M.~O.~P.} \bibnamefont{Sampaio}},
  \bibinfo{author}{\bibfnamefont{R.}~\bibnamefont{Santos}}, \bibnamefont{and}
  \bibinfo{author}{\bibfnamefont{J.}~\bibnamefont{Wittbrodt}},
  \bibinfo{journal}{JHEP} \textbf{\bibinfo{volume}{08}}, \bibinfo{pages}{132}
  (\bibinfo{year}{2017}), \eprint{1703.07750}.

\bibitem[{\citenamefont{Inoue et~al.}(2014)\citenamefont{Inoue, Ramsey-Musolf,
  and Zhang}}]{Inoue:2014nva}
\bibinfo{author}{\bibfnamefont{S.}~\bibnamefont{Inoue}},
  \bibinfo{author}{\bibfnamefont{M.~J.} \bibnamefont{Ramsey-Musolf}},
  \bibnamefont{and} \bibinfo{author}{\bibfnamefont{Y.}~\bibnamefont{Zhang}},
  \bibinfo{journal}{Phys. Rev.} \textbf{\bibinfo{volume}{D89}},
  \bibinfo{pages}{115023} (\bibinfo{year}{2014}), \eprint{1403.4257}.

\bibitem[{\citenamefont{Andreev et~al.}(2018)}]{Andreev:2018ayy}
\bibinfo{author}{\bibfnamefont{V.}~\bibnamefont{Andreev}} \bibnamefont{et~al.}
  (\bibinfo{collaboration}{ACME}), \bibinfo{journal}{Nature}
  \textbf{\bibinfo{volume}{562}}, \bibinfo{pages}{355} (\bibinfo{year}{2018}).

\bibitem[{\citenamefont{Quiros}(1994)}]{Quiros:1994dr}
\bibinfo{author}{\bibfnamefont{M.}~\bibnamefont{Quiros}},
  \bibinfo{journal}{Helv. Phys. Acta} \textbf{\bibinfo{volume}{67}},
  \bibinfo{pages}{451} (\bibinfo{year}{1994}).

\bibitem[{\citenamefont{Moore}(1999)}]{Moore:1998swa}
\bibinfo{author}{\bibfnamefont{G.~D.} \bibnamefont{Moore}},
  \bibinfo{journal}{Phys. Rev.} \textbf{\bibinfo{volume}{D59}},
  \bibinfo{pages}{014503} (\bibinfo{year}{1999}), \eprint{hep-ph/9805264}.

\bibitem[{\citenamefont{Basler and Muhlleitner}(2018)}]{Basler:2018cwe}
\bibinfo{author}{\bibfnamefont{P.}~\bibnamefont{Basler}} \bibnamefont{and}
  \bibinfo{author}{\bibfnamefont{M.}~\bibnamefont{Muhlleitner}}
  (\bibinfo{year}{2018}), \eprint{1803.02846}.

\bibitem[{\citenamefont{King et~al.}(2014)\citenamefont{King, M{\"u}hlleitner,
  Nevzorov, and Walz}}]{King:2014xwa}
\bibinfo{author}{\bibfnamefont{S.~F.} \bibnamefont{King}},
  \bibinfo{author}{\bibfnamefont{M.}~\bibnamefont{M{\"u}hlleitner}},
  \bibinfo{author}{\bibfnamefont{R.}~\bibnamefont{Nevzorov}}, \bibnamefont{and}
  \bibinfo{author}{\bibfnamefont{K.}~\bibnamefont{Walz}},
  \bibinfo{journal}{Phys. Rev.} \textbf{\bibinfo{volume}{D90}},
  \bibinfo{pages}{095014} (\bibinfo{year}{2014}), \eprint{1408.1120}.

\bibitem[{\citenamefont{Costa et~al.}(2016)\citenamefont{Costa, Muhlleitner,
  Sampaio, and Santos}}]{Costa:2015llh}
\bibinfo{author}{\bibfnamefont{R.}~\bibnamefont{Costa}},
  \bibinfo{author}{\bibfnamefont{M.}~\bibnamefont{Muhlleitner}},
  \bibinfo{author}{\bibfnamefont{M.~O.~P.} \bibnamefont{Sampaio}},
  \bibnamefont{and} \bibinfo{author}{\bibfnamefont{R.}~\bibnamefont{Santos}},
  \bibinfo{journal}{JHEP} \textbf{\bibinfo{volume}{06}}, \bibinfo{pages}{034}
  (\bibinfo{year}{2016}), \eprint{1512.05355}.

\bibitem[{\citenamefont{Azevedo et~al.}(2018)\citenamefont{Azevedo, Ferreira,
  Margarete~MÃ¼hlleitner, Santos, and Wittbrodt}}]{Azevedo:2018llq}
\bibinfo{author}{\bibfnamefont{D.}~\bibnamefont{Azevedo}},
  \bibinfo{author}{\bibfnamefont{P.}~\bibnamefont{Ferreira}},
  \bibinfo{author}{\bibfnamefont{M.}~\bibnamefont{Margarete~MÃ¼hlleitner}},
  \bibinfo{author}{\bibfnamefont{R.}~\bibnamefont{Santos}}, \bibnamefont{and}
  \bibinfo{author}{\bibfnamefont{J.}~\bibnamefont{Wittbrodt}}
  (\bibinfo{year}{2018}), \eprint{1808.00755}.

\bibitem[{\citenamefont{Skands et~al.}(2004)}]{Skands:2003cj}
\bibinfo{author}{\bibfnamefont{P.~Z.} \bibnamefont{Skands}}
  \bibnamefont{et~al.}, \bibinfo{journal}{JHEP} \textbf{\bibinfo{volume}{07}},
  \bibinfo{pages}{036} (\bibinfo{year}{2004}), \eprint{hep-ph/0311123}.

\bibitem[{\citenamefont{Allanach et~al.}(2009)}]{Allanach:2008qq}
\bibinfo{author}{\bibfnamefont{B.~C.} \bibnamefont{Allanach}}
  \bibnamefont{et~al.}, \bibinfo{journal}{Comput. Phys. Commun.}
  \textbf{\bibinfo{volume}{180}}, \bibinfo{pages}{8} (\bibinfo{year}{2009}),
  \eprint{0801.0045}.

\bibitem[{\citenamefont{Ellwanger et~al.}(2005)\citenamefont{Ellwanger, Gunion,
  and Hugonie}}]{Ellwanger:2004xm}
\bibinfo{author}{\bibfnamefont{U.}~\bibnamefont{Ellwanger}},
  \bibinfo{author}{\bibfnamefont{J.~F.} \bibnamefont{Gunion}},
  \bibnamefont{and} \bibinfo{author}{\bibfnamefont{C.}~\bibnamefont{Hugonie}},
  \bibinfo{journal}{JHEP} \textbf{\bibinfo{volume}{02}}, \bibinfo{pages}{066}
  (\bibinfo{year}{2005}), \eprint{hep-ph/0406215}.

\bibitem[{\citenamefont{Ellwanger and Hugonie}(2006)}]{Ellwanger:2005dv}
\bibinfo{author}{\bibfnamefont{U.}~\bibnamefont{Ellwanger}} \bibnamefont{and}
  \bibinfo{author}{\bibfnamefont{C.}~\bibnamefont{Hugonie}},
  \bibinfo{journal}{Comput. Phys. Commun.} \textbf{\bibinfo{volume}{175}},
  \bibinfo{pages}{290} (\bibinfo{year}{2006}), \eprint{hep-ph/0508022}.

\bibitem[{\citenamefont{Ellwanger and Hugonie}(2007)}]{Ellwanger:2006rn}
\bibinfo{author}{\bibfnamefont{U.}~\bibnamefont{Ellwanger}} \bibnamefont{and}
  \bibinfo{author}{\bibfnamefont{C.}~\bibnamefont{Hugonie}},
  \bibinfo{journal}{Comput. Phys. Commun.} \textbf{\bibinfo{volume}{177}},
  \bibinfo{pages}{399} (\bibinfo{year}{2007}), \eprint{hep-ph/0612134}.

\bibitem[{\citenamefont{Das et~al.}(2012)\citenamefont{Das, Ellwanger, and
  Teixeira}}]{Das:2011dg}
\bibinfo{author}{\bibfnamefont{D.}~\bibnamefont{Das}},
  \bibinfo{author}{\bibfnamefont{U.}~\bibnamefont{Ellwanger}},
  \bibnamefont{and} \bibinfo{author}{\bibfnamefont{A.~M.}
  \bibnamefont{Teixeira}}, \bibinfo{journal}{Comput. Phys. Commun.}
  \textbf{\bibinfo{volume}{183}}, \bibinfo{pages}{774} (\bibinfo{year}{2012}),
  \eprint{1106.5633}.

\bibitem[{\citenamefont{M{\"u}hlleitner
  et~al.}(2005)\citenamefont{M{\"u}hlleitner, Djouadi, and
  Mambrini}}]{Muhlleitner:2003vg}
\bibinfo{author}{\bibfnamefont{M.}~\bibnamefont{M{\"u}hlleitner}},
  \bibinfo{author}{\bibfnamefont{A.}~\bibnamefont{Djouadi}}, \bibnamefont{and}
  \bibinfo{author}{\bibfnamefont{Y.}~\bibnamefont{Mambrini}},
  \bibinfo{journal}{Comput. Phys. Commun.} \textbf{\bibinfo{volume}{168}},
  \bibinfo{pages}{46} (\bibinfo{year}{2005}), \eprint{hep-ph/0311167}.

\bibitem[{\citenamefont{Belanger et~al.}(2005)\citenamefont{Belanger, Boudjema,
  Hugonie, Pukhov, and Semenov}}]{Belanger:2005kh}
\bibinfo{author}{\bibfnamefont{G.}~\bibnamefont{Belanger}},
  \bibinfo{author}{\bibfnamefont{F.}~\bibnamefont{Boudjema}},
  \bibinfo{author}{\bibfnamefont{C.}~\bibnamefont{Hugonie}},
  \bibinfo{author}{\bibfnamefont{A.}~\bibnamefont{Pukhov}}, \bibnamefont{and}
  \bibinfo{author}{\bibfnamefont{A.}~\bibnamefont{Semenov}},
  \bibinfo{journal}{JCAP} \textbf{\bibinfo{volume}{0509}}, \bibinfo{pages}{001}
  (\bibinfo{year}{2005}), \eprint{hep-ph/0505142}.

\bibitem[{\citenamefont{Aad et~al.}(2016)}]{Khachatryan:2016vau}
\bibinfo{author}{\bibfnamefont{G.}~\bibnamefont{Aad}} \bibnamefont{et~al.}
  (\bibinfo{collaboration}{ATLAS, CMS}), \bibinfo{journal}{JHEP}
  \textbf{\bibinfo{volume}{08}}, \bibinfo{pages}{045} (\bibinfo{year}{2016}),
  \eprint{1606.02266}.

\bibitem[{\citenamefont{Harlander et~al.}(2013)\citenamefont{Harlander,
  Liebler, and Mantler}}]{Harlander:2012pb}
\bibinfo{author}{\bibfnamefont{R.~V.} \bibnamefont{Harlander}},
  \bibinfo{author}{\bibfnamefont{S.}~\bibnamefont{Liebler}}, \bibnamefont{and}
  \bibinfo{author}{\bibfnamefont{H.}~\bibnamefont{Mantler}},
  \bibinfo{journal}{Comput. Phys. Commun.} \textbf{\bibinfo{volume}{184}},
  \bibinfo{pages}{1605} (\bibinfo{year}{2013}), \eprint{1212.3249}.

\bibitem[{\citenamefont{Harlander et~al.}(2017)\citenamefont{Harlander,
  Liebler, and Mantler}}]{Harlander:2016hcx}
\bibinfo{author}{\bibfnamefont{R.~V.} \bibnamefont{Harlander}},
  \bibinfo{author}{\bibfnamefont{S.}~\bibnamefont{Liebler}}, \bibnamefont{and}
  \bibinfo{author}{\bibfnamefont{H.}~\bibnamefont{Mantler}},
  \bibinfo{journal}{Comput. Phys. Commun.} \textbf{\bibinfo{volume}{212}},
  \bibinfo{pages}{239} (\bibinfo{year}{2017}), \eprint{1605.03190}.

\bibitem[{\citenamefont{Spira et~al.}(1995)\citenamefont{Spira, Djouadi,
  Graudenz, and Zerwas}}]{Spira:1995rr}
\bibinfo{author}{\bibfnamefont{M.}~\bibnamefont{Spira}},
  \bibinfo{author}{\bibfnamefont{A.}~\bibnamefont{Djouadi}},
  \bibinfo{author}{\bibfnamefont{D.}~\bibnamefont{Graudenz}}, \bibnamefont{and}
  \bibinfo{author}{\bibfnamefont{P.~M.} \bibnamefont{Zerwas}},
  \bibinfo{journal}{Nucl. Phys.} \textbf{\bibinfo{volume}{B453}},
  \bibinfo{pages}{17} (\bibinfo{year}{1995}), \eprint{hep-ph/9504378}.

\bibitem[{\citenamefont{Harlander and
  Kilgore}(2002{\natexlab{a}})}]{Harlander:2002wh}
\bibinfo{author}{\bibfnamefont{R.~V.} \bibnamefont{Harlander}}
  \bibnamefont{and} \bibinfo{author}{\bibfnamefont{W.~B.}
  \bibnamefont{Kilgore}}, \bibinfo{journal}{Phys. Rev. Lett.}
  \textbf{\bibinfo{volume}{88}}, \bibinfo{pages}{201801}
  (\bibinfo{year}{2002}{\natexlab{a}}), \eprint{hep-ph/0201206}.

\bibitem[{\citenamefont{Anastasiou and Melnikov}(2002)}]{Anastasiou:2002yz}
\bibinfo{author}{\bibfnamefont{C.}~\bibnamefont{Anastasiou}} \bibnamefont{and}
  \bibinfo{author}{\bibfnamefont{K.}~\bibnamefont{Melnikov}},
  \bibinfo{journal}{Nucl. Phys.} \textbf{\bibinfo{volume}{B646}},
  \bibinfo{pages}{220} (\bibinfo{year}{2002}), \eprint{hep-ph/0207004}.

\bibitem[{\citenamefont{Harlander and
  Kilgore}(2002{\natexlab{b}})}]{Harlander:2002vv}
\bibinfo{author}{\bibfnamefont{R.~V.} \bibnamefont{Harlander}}
  \bibnamefont{and} \bibinfo{author}{\bibfnamefont{W.~B.}
  \bibnamefont{Kilgore}}, \bibinfo{journal}{JHEP}
  \textbf{\bibinfo{volume}{10}}, \bibinfo{pages}{017}
  (\bibinfo{year}{2002}{\natexlab{b}}), \eprint{hep-ph/0208096}.

\bibitem[{\citenamefont{Anastasiou and Melnikov}(2003)}]{Anastasiou:2002wq}
\bibinfo{author}{\bibfnamefont{C.}~\bibnamefont{Anastasiou}} \bibnamefont{and}
  \bibinfo{author}{\bibfnamefont{K.}~\bibnamefont{Melnikov}},
  \bibinfo{journal}{Phys. Rev.} \textbf{\bibinfo{volume}{D67}},
  \bibinfo{pages}{037501} (\bibinfo{year}{2003}), \eprint{hep-ph/0208115}.

\bibitem[{\citenamefont{Ravindran et~al.}(2003)\citenamefont{Ravindran, Smith,
  and van Neerven}}]{Ravindran:2003um}
\bibinfo{author}{\bibfnamefont{V.}~\bibnamefont{Ravindran}},
  \bibinfo{author}{\bibfnamefont{J.}~\bibnamefont{Smith}}, \bibnamefont{and}
  \bibinfo{author}{\bibfnamefont{W.~L.} \bibnamefont{van Neerven}},
  \bibinfo{journal}{Nucl. Phys.} \textbf{\bibinfo{volume}{B665}},
  \bibinfo{pages}{325} (\bibinfo{year}{2003}), \eprint{hep-ph/0302135}.

\bibitem[{\citenamefont{Anastasiou
  et~al.}(2015{\natexlab{a}})\citenamefont{Anastasiou, Duhr, Dulat, Furlan,
  Gehrmann, Herzog, and Mistlberger}}]{Anastasiou:2014lda}
\bibinfo{author}{\bibfnamefont{C.}~\bibnamefont{Anastasiou}},
  \bibinfo{author}{\bibfnamefont{C.}~\bibnamefont{Duhr}},
  \bibinfo{author}{\bibfnamefont{F.}~\bibnamefont{Dulat}},
  \bibinfo{author}{\bibfnamefont{E.}~\bibnamefont{Furlan}},
  \bibinfo{author}{\bibfnamefont{T.}~\bibnamefont{Gehrmann}},
  \bibinfo{author}{\bibfnamefont{F.}~\bibnamefont{Herzog}}, \bibnamefont{and}
  \bibinfo{author}{\bibfnamefont{B.}~\bibnamefont{Mistlberger}},
  \bibinfo{journal}{JHEP} \textbf{\bibinfo{volume}{03}}, \bibinfo{pages}{091}
  (\bibinfo{year}{2015}{\natexlab{a}}), \eprint{1411.3584}.

\bibitem[{\citenamefont{Anastasiou
  et~al.}(2015{\natexlab{b}})\citenamefont{Anastasiou, Duhr, Dulat, Furlan,
  Herzog, and Mistlberger}}]{Anastasiou:2015yha}
\bibinfo{author}{\bibfnamefont{C.}~\bibnamefont{Anastasiou}},
  \bibinfo{author}{\bibfnamefont{C.}~\bibnamefont{Duhr}},
  \bibinfo{author}{\bibfnamefont{F.}~\bibnamefont{Dulat}},
  \bibinfo{author}{\bibfnamefont{E.}~\bibnamefont{Furlan}},
  \bibinfo{author}{\bibfnamefont{F.}~\bibnamefont{Herzog}}, \bibnamefont{and}
  \bibinfo{author}{\bibfnamefont{B.}~\bibnamefont{Mistlberger}},
  \bibinfo{journal}{JHEP} \textbf{\bibinfo{volume}{08}}, \bibinfo{pages}{051}
  (\bibinfo{year}{2015}{\natexlab{b}}), \eprint{1505.04110}.

\bibitem[{\citenamefont{Anastasiou et~al.}(2016)\citenamefont{Anastasiou, Duhr,
  Dulat, Furlan, Gehrmann, Herzog, Lazopoulos, and
  Mistlberger}}]{Anastasiou:2016cez}
\bibinfo{author}{\bibfnamefont{C.}~\bibnamefont{Anastasiou}},
  \bibinfo{author}{\bibfnamefont{C.}~\bibnamefont{Duhr}},
  \bibinfo{author}{\bibfnamefont{F.}~\bibnamefont{Dulat}},
  \bibinfo{author}{\bibfnamefont{E.}~\bibnamefont{Furlan}},
  \bibinfo{author}{\bibfnamefont{T.}~\bibnamefont{Gehrmann}},
  \bibinfo{author}{\bibfnamefont{F.}~\bibnamefont{Herzog}},
  \bibinfo{author}{\bibfnamefont{A.}~\bibnamefont{Lazopoulos}},
  \bibnamefont{and}
  \bibinfo{author}{\bibfnamefont{B.}~\bibnamefont{Mistlberger}},
  \bibinfo{journal}{JHEP} \textbf{\bibinfo{volume}{05}}, \bibinfo{pages}{058}
  (\bibinfo{year}{2016}), \eprint{1602.00695}.

\bibitem[{\citenamefont{Mistlberger}(2018)}]{Mistlberger:2018etf}
\bibinfo{author}{\bibfnamefont{B.}~\bibnamefont{Mistlberger}},
  \bibinfo{journal}{JHEP} \textbf{\bibinfo{volume}{05}}, \bibinfo{pages}{028}
  (\bibinfo{year}{2018}), \eprint{1802.00833}.

\bibitem[{\citenamefont{Bonvini et~al.}(2015)\citenamefont{Bonvini,
  Papanastasiou, and Tackmann}}]{Bonvini:2015pxa}
\bibinfo{author}{\bibfnamefont{M.}~\bibnamefont{Bonvini}},
  \bibinfo{author}{\bibfnamefont{A.~S.} \bibnamefont{Papanastasiou}},
  \bibnamefont{and} \bibinfo{author}{\bibfnamefont{F.~J.}
  \bibnamefont{Tackmann}}, \bibinfo{journal}{JHEP}
  \textbf{\bibinfo{volume}{11}}, \bibinfo{pages}{196} (\bibinfo{year}{2015}),
  \eprint{1508.03288}.

\bibitem[{\citenamefont{Bonvini et~al.}(2016)\citenamefont{Bonvini,
  Papanastasiou, and Tackmann}}]{Bonvini:2016fgf}
\bibinfo{author}{\bibfnamefont{M.}~\bibnamefont{Bonvini}},
  \bibinfo{author}{\bibfnamefont{A.~S.} \bibnamefont{Papanastasiou}},
  \bibnamefont{and} \bibinfo{author}{\bibfnamefont{F.~J.}
  \bibnamefont{Tackmann}}, \bibinfo{journal}{JHEP}
  \textbf{\bibinfo{volume}{10}}, \bibinfo{pages}{053} (\bibinfo{year}{2016}),
  \eprint{1605.01733}.

\bibitem[{\citenamefont{Forte et~al.}(2015)\citenamefont{Forte, Napoletano, and
  Ubiali}}]{Forte:2015hba}
\bibinfo{author}{\bibfnamefont{S.}~\bibnamefont{Forte}},
  \bibinfo{author}{\bibfnamefont{D.}~\bibnamefont{Napoletano}},
  \bibnamefont{and} \bibinfo{author}{\bibfnamefont{M.}~\bibnamefont{Ubiali}},
  \bibinfo{journal}{Phys. Lett.} \textbf{\bibinfo{volume}{B751}},
  \bibinfo{pages}{331} (\bibinfo{year}{2015}), \eprint{1508.01529}.

\bibitem[{\citenamefont{Forte et~al.}(2016)\citenamefont{Forte, Napoletano, and
  Ubiali}}]{Forte:2016sja}
\bibinfo{author}{\bibfnamefont{S.}~\bibnamefont{Forte}},
  \bibinfo{author}{\bibfnamefont{D.}~\bibnamefont{Napoletano}},
  \bibnamefont{and} \bibinfo{author}{\bibfnamefont{M.}~\bibnamefont{Ubiali}},
  \bibinfo{journal}{Phys. Lett.} \textbf{\bibinfo{volume}{B763}},
  \bibinfo{pages}{190} (\bibinfo{year}{2016}), \eprint{1607.00389}.

\bibitem[{\citenamefont{Harlander et~al.}(2011)\citenamefont{Harlander, Kramer,
  and Schumacher}}]{Harlander:2011aa}
\bibinfo{author}{\bibfnamefont{R.}~\bibnamefont{Harlander}},
  \bibinfo{author}{\bibfnamefont{M.}~\bibnamefont{Kramer}}, \bibnamefont{and}
  \bibinfo{author}{\bibfnamefont{M.}~\bibnamefont{Schumacher}}
  (\bibinfo{year}{2011}), \eprint{1112.3478}.

\bibitem[{\citenamefont{Harlander and Kilgore}(2003)}]{Harlander:2003ai}
\bibinfo{author}{\bibfnamefont{R.~V.} \bibnamefont{Harlander}}
  \bibnamefont{and} \bibinfo{author}{\bibfnamefont{W.~B.}
  \bibnamefont{Kilgore}}, \bibinfo{journal}{Phys. Rev.}
  \textbf{\bibinfo{volume}{D68}}, \bibinfo{pages}{013001}
  (\bibinfo{year}{2003}), \eprint{hep-ph/0304035}.

\bibitem[{\citenamefont{Dittmaier et~al.}(2004)\citenamefont{Dittmaier,
  KrÃ¤mer, and Spira}}]{Dittmaier:2003ej}
\bibinfo{author}{\bibfnamefont{S.}~\bibnamefont{Dittmaier}},
  \bibinfo{author}{\bibfnamefont{M.}~\bibnamefont{KrÃ¤mer}}, \bibnamefont{and}
  \bibinfo{author}{\bibfnamefont{M.}~\bibnamefont{Spira}},
  \bibinfo{journal}{Phys. Rev.} \textbf{\bibinfo{volume}{D70}},
  \bibinfo{pages}{074010} (\bibinfo{year}{2004}), \eprint{hep-ph/0309204}.

\bibitem[{\citenamefont{Dawson et~al.}(2004)\citenamefont{Dawson, Jackson,
  Reina, and Wackeroth}}]{Dawson:2003kb}
\bibinfo{author}{\bibfnamefont{S.}~\bibnamefont{Dawson}},
  \bibinfo{author}{\bibfnamefont{C.~B.} \bibnamefont{Jackson}},
  \bibinfo{author}{\bibfnamefont{L.}~\bibnamefont{Reina}}, \bibnamefont{and}
  \bibinfo{author}{\bibfnamefont{D.}~\bibnamefont{Wackeroth}},
  \bibinfo{journal}{Phys. Rev.} \textbf{\bibinfo{volume}{D69}},
  \bibinfo{pages}{074027} (\bibinfo{year}{2004}), \eprint{hep-ph/0311067}.

\bibitem[{\citenamefont{Wiesemann et~al.}(2015)\citenamefont{Wiesemann,
  Frederix, Frixione, Hirschi, Maltoni, and Torrielli}}]{Wiesemann:2014ioa}
\bibinfo{author}{\bibfnamefont{M.}~\bibnamefont{Wiesemann}},
  \bibinfo{author}{\bibfnamefont{R.}~\bibnamefont{Frederix}},
  \bibinfo{author}{\bibfnamefont{S.}~\bibnamefont{Frixione}},
  \bibinfo{author}{\bibfnamefont{V.}~\bibnamefont{Hirschi}},
  \bibinfo{author}{\bibfnamefont{F.}~\bibnamefont{Maltoni}}, \bibnamefont{and}
  \bibinfo{author}{\bibfnamefont{P.}~\bibnamefont{Torrielli}},
  \bibinfo{journal}{JHEP} \textbf{\bibinfo{volume}{02}}, \bibinfo{pages}{132}
  (\bibinfo{year}{2015}), \eprint{1409.5301}.

\bibitem[{\citenamefont{Baglio et~al.}(2014{\natexlab{b}})\citenamefont{Baglio,
  Grober, Muhlleitner, Nhung, Rzehak, Spira, Streicher, and
  Walz}}]{Baglio:2013iia}
\bibinfo{author}{\bibfnamefont{J.}~\bibnamefont{Baglio}},
  \bibinfo{author}{\bibfnamefont{R.}~\bibnamefont{Grober}},
  \bibinfo{author}{\bibfnamefont{M.}~\bibnamefont{Muhlleitner}},
  \bibinfo{author}{\bibfnamefont{D.~T.} \bibnamefont{Nhung}},
  \bibinfo{author}{\bibfnamefont{H.}~\bibnamefont{Rzehak}},
  \bibinfo{author}{\bibfnamefont{M.}~\bibnamefont{Spira}},
  \bibinfo{author}{\bibfnamefont{J.}~\bibnamefont{Streicher}},
  \bibnamefont{and} \bibinfo{author}{\bibfnamefont{K.}~\bibnamefont{Walz}},
  \bibinfo{journal}{Comput. Phys. Commun.} \textbf{\bibinfo{volume}{185}},
  \bibinfo{pages}{3372} (\bibinfo{year}{2014}{\natexlab{b}}),
  \eprint{1312.4788}.

\bibitem[{\citenamefont{Aad et~al.}(2015{\natexlab{b}})}]{Aad:2015iea}
\bibinfo{author}{\bibfnamefont{G.}~\bibnamefont{Aad}} \bibnamefont{et~al.}
  (\bibinfo{collaboration}{ATLAS}), \bibinfo{journal}{JHEP}
  \textbf{\bibinfo{volume}{10}}, \bibinfo{pages}{054}
  (\bibinfo{year}{2015}{\natexlab{b}}), \eprint{1507.05525}.

\bibitem[{\citenamefont{Aaboud et~al.}(2016)}]{Aaboud:2016lwz}
\bibinfo{author}{\bibfnamefont{M.}~\bibnamefont{Aaboud}} \bibnamefont{et~al.}
  (\bibinfo{collaboration}{ATLAS}), \bibinfo{journal}{Phys. Rev.}
  \textbf{\bibinfo{volume}{D94}}, \bibinfo{pages}{052009}
  (\bibinfo{year}{2016}), \eprint{1606.03903}.

\bibitem[{\citenamefont{Aad et~al.}(2015{\natexlab{c}})}]{Aad:2015jqa}
\bibinfo{author}{\bibfnamefont{G.}~\bibnamefont{Aad}} \bibnamefont{et~al.}
  (\bibinfo{collaboration}{ATLAS}), \bibinfo{journal}{Eur. Phys. J.}
  \textbf{\bibinfo{volume}{C75}}, \bibinfo{pages}{208}
  (\bibinfo{year}{2015}{\natexlab{c}}), \eprint{1501.07110}.

\bibitem[{\citenamefont{Ade et~al.}(2014)}]{Ade:2013zuv}
\bibinfo{author}{\bibfnamefont{P.~A.~R.} \bibnamefont{Ade}}
  \bibnamefont{et~al.} (\bibinfo{collaboration}{Planck}),
  \bibinfo{journal}{Astron. Astrophys.} \textbf{\bibinfo{volume}{571}},
  \bibinfo{pages}{A16} (\bibinfo{year}{2014}), \eprint{1303.5076}.

\bibitem[{\citenamefont{Akerib et~al.}(2017)}]{Akerib:2016vxi}
\bibinfo{author}{\bibfnamefont{D.~S.} \bibnamefont{Akerib}}
  \bibnamefont{et~al.} (\bibinfo{collaboration}{LUX}), \bibinfo{journal}{Phys.
  Rev. Lett.} \textbf{\bibinfo{volume}{118}}, \bibinfo{pages}{021303}
  (\bibinfo{year}{2017}), \eprint{1608.07648}.

\bibitem[{\citenamefont{Aprile et~al.}(2018)}]{Aprile:2018dbl}
\bibinfo{author}{\bibfnamefont{E.}~\bibnamefont{Aprile}} \bibnamefont{et~al.}
  (\bibinfo{collaboration}{XENON}), \bibinfo{journal}{Phys. Rev. Lett.}
  \textbf{\bibinfo{volume}{121}}, \bibinfo{pages}{111302}
  (\bibinfo{year}{2018}), \eprint{1805.12562}.

\bibitem[{\citenamefont{Ackermann et~al.}(2015)}]{Ackermann:2015zua}
\bibinfo{author}{\bibfnamefont{M.}~\bibnamefont{Ackermann}}
  \bibnamefont{et~al.} (\bibinfo{collaboration}{Fermi-LAT}),
  \bibinfo{journal}{Phys. Rev. Lett.} \textbf{\bibinfo{volume}{115}},
  \bibinfo{pages}{231301} (\bibinfo{year}{2015}), \eprint{1503.02641}.

\bibitem[{\citenamefont{Aad et~al.}(2015{\natexlab{d}})}]{ATLAS:2014aga}
\bibinfo{author}{\bibfnamefont{G.}~\bibnamefont{Aad}} \bibnamefont{et~al.}
  (\bibinfo{collaboration}{ATLAS}), \bibinfo{journal}{Phys. Rev.}
  \textbf{\bibinfo{volume}{D92}}, \bibinfo{pages}{012006}
  (\bibinfo{year}{2015}{\natexlab{d}}), \eprint{1412.2641}.

\bibitem[{\citenamefont{Dittmaier et~al.}(2011)}]{Dittmaier:2011ti}
\bibinfo{author}{\bibfnamefont{S.}~\bibnamefont{Dittmaier}}
  \bibnamefont{et~al.} (\bibinfo{collaboration}{LHC Higgs Cross Section Working
  Group}) (\bibinfo{year}{2011}), \eprint{1101.0593}.

\bibitem[{\citenamefont{Aaboud et~al.}(2017{\natexlab{a}})}]{Aaboud:2017xsd}
\bibinfo{author}{\bibfnamefont{M.}~\bibnamefont{Aaboud}} \bibnamefont{et~al.}
  (\bibinfo{collaboration}{ATLAS}), \bibinfo{journal}{JHEP}
  \textbf{\bibinfo{volume}{12}}, \bibinfo{pages}{024}
  (\bibinfo{year}{2017}{\natexlab{a}}), \eprint{1708.03299}.

\bibitem[{\citenamefont{Sirunyan
  et~al.}(2018{\natexlab{a}})}]{Sirunyan:2017elk}
\bibinfo{author}{\bibfnamefont{A.~M.} \bibnamefont{Sirunyan}}
  \bibnamefont{et~al.} (\bibinfo{collaboration}{CMS}), \bibinfo{journal}{Phys.
  Lett.} \textbf{\bibinfo{volume}{B780}}, \bibinfo{pages}{501}
  (\bibinfo{year}{2018}{\natexlab{a}}), \eprint{1709.07497}.

\bibitem[{\citenamefont{Khachatryan
  et~al.}(2015{\natexlab{a}})}]{Khachatryan:2015bnx}
\bibinfo{author}{\bibfnamefont{V.}~\bibnamefont{Khachatryan}}
  \bibnamefont{et~al.} (\bibinfo{collaboration}{CMS}), \bibinfo{journal}{Phys.
  Rev.} \textbf{\bibinfo{volume}{D92}}, \bibinfo{pages}{032008}
  (\bibinfo{year}{2015}{\natexlab{a}}), \eprint{1506.01010}.

\bibitem[{\citenamefont{Sirunyan
  et~al.}(2018{\natexlab{b}})}]{Sirunyan:2017khh}
\bibinfo{author}{\bibfnamefont{A.~M.} \bibnamefont{Sirunyan}}
  \bibnamefont{et~al.} (\bibinfo{collaboration}{CMS}), \bibinfo{journal}{Phys.
  Lett.} \textbf{\bibinfo{volume}{B779}}, \bibinfo{pages}{283}
  (\bibinfo{year}{2018}{\natexlab{b}}), \eprint{1708.00373}.

\bibitem[{\citenamefont{Aaboud et~al.}(2017{\natexlab{b}})}]{Aaboud:2017yyg}
\bibinfo{author}{\bibfnamefont{M.}~\bibnamefont{Aaboud}} \bibnamefont{et~al.}
  (\bibinfo{collaboration}{ATLAS}), \bibinfo{journal}{Phys. Lett.}
  \textbf{\bibinfo{volume}{B775}}, \bibinfo{pages}{105}
  (\bibinfo{year}{2017}{\natexlab{b}}), \eprint{1707.04147}.

\bibitem[{\citenamefont{Aaboud et~al.}(2018{\natexlab{a}})}]{Aaboud:2017sjh}
\bibinfo{author}{\bibfnamefont{M.}~\bibnamefont{Aaboud}} \bibnamefont{et~al.}
  (\bibinfo{collaboration}{ATLAS}), \bibinfo{journal}{JHEP}
  \textbf{\bibinfo{volume}{01}}, \bibinfo{pages}{055}
  (\bibinfo{year}{2018}{\natexlab{a}}), \eprint{1709.07242}.

\bibitem[{\citenamefont{Aaboud et~al.}(2018{\natexlab{b}})}]{Aaboud:2017gsl}
\bibinfo{author}{\bibfnamefont{M.}~\bibnamefont{Aaboud}} \bibnamefont{et~al.}
  (\bibinfo{collaboration}{ATLAS}), \bibinfo{journal}{Eur. Phys. J.}
  \textbf{\bibinfo{volume}{C78}}, \bibinfo{pages}{24}
  (\bibinfo{year}{2018}{\natexlab{b}}), \eprint{1710.01123}.

\bibitem[{\citenamefont{Aaboud et~al.}(2018{\natexlab{c}})}]{Aaboud:2017rel}
\bibinfo{author}{\bibfnamefont{M.}~\bibnamefont{Aaboud}} \bibnamefont{et~al.}
  (\bibinfo{collaboration}{ATLAS}), \bibinfo{journal}{Eur. Phys. J.}
  \textbf{\bibinfo{volume}{C78}}, \bibinfo{pages}{293}
  (\bibinfo{year}{2018}{\natexlab{c}}), \eprint{1712.06386}.

\bibitem[{\citenamefont{Aaboud et~al.}(2018{\natexlab{d}})}]{Aaboud:2018mjh}
\bibinfo{author}{\bibfnamefont{M.}~\bibnamefont{Aaboud}} \bibnamefont{et~al.}
  (\bibinfo{collaboration}{ATLAS}), \bibinfo{journal}{Eur. Phys. J.}
  \textbf{\bibinfo{volume}{C78}}, \bibinfo{pages}{565}
  (\bibinfo{year}{2018}{\natexlab{d}}), \eprint{1804.10823}.

\bibitem[{\citenamefont{Aaboud et~al.}(2017{\natexlab{c}})}]{Aaboud:2017hnm}
\bibinfo{author}{\bibfnamefont{M.}~\bibnamefont{Aaboud}} \bibnamefont{et~al.}
  (\bibinfo{collaboration}{ATLAS}), \bibinfo{journal}{Phys. Rev. Lett.}
  \textbf{\bibinfo{volume}{119}}, \bibinfo{pages}{191803}
  (\bibinfo{year}{2017}{\natexlab{c}}), \eprint{1707.06025}.

\bibitem[{\citenamefont{Agostini et~al.}(2016)\citenamefont{Agostini, Degrassi,
  Grober, and Slavich}}]{Agostini:2016vze}
\bibinfo{author}{\bibfnamefont{A.}~\bibnamefont{Agostini}},
  \bibinfo{author}{\bibfnamefont{G.}~\bibnamefont{Degrassi}},
  \bibinfo{author}{\bibfnamefont{R.}~\bibnamefont{Grober}}, \bibnamefont{and}
  \bibinfo{author}{\bibfnamefont{P.}~\bibnamefont{Slavich}},
  \bibinfo{journal}{JHEP} \textbf{\bibinfo{volume}{04}}, \bibinfo{pages}{106}
  (\bibinfo{year}{2016}), \eprint{1601.03671}.

\bibitem[{\citenamefont{Grober et~al.}(2017)\citenamefont{Grober, Muhlleitner,
  and Spira}}]{Grober:2017gut}
\bibinfo{author}{\bibfnamefont{R.}~\bibnamefont{Grober}},
  \bibinfo{author}{\bibfnamefont{M.}~\bibnamefont{Muhlleitner}},
  \bibnamefont{and} \bibinfo{author}{\bibfnamefont{M.}~\bibnamefont{Spira}},
  \bibinfo{journal}{Nucl. Phys.} \textbf{\bibinfo{volume}{B925}},
  \bibinfo{pages}{1} (\bibinfo{year}{2017}), \eprint{1705.05314}.

\bibitem[{\citenamefont{Dulat et~al.}(2016)\citenamefont{Dulat, Hou, Gao,
  Guzzi, Huston, Nadolsky, Pumplin, Schmidt, Stump, and Yuan}}]{Dulat:2015mca}
\bibinfo{author}{\bibfnamefont{S.}~\bibnamefont{Dulat}},
  \bibinfo{author}{\bibfnamefont{T.-J.} \bibnamefont{Hou}},
  \bibinfo{author}{\bibfnamefont{J.}~\bibnamefont{Gao}},
  \bibinfo{author}{\bibfnamefont{M.}~\bibnamefont{Guzzi}},
  \bibinfo{author}{\bibfnamefont{J.}~\bibnamefont{Huston}},
  \bibinfo{author}{\bibfnamefont{P.}~\bibnamefont{Nadolsky}},
  \bibinfo{author}{\bibfnamefont{J.}~\bibnamefont{Pumplin}},
  \bibinfo{author}{\bibfnamefont{C.}~\bibnamefont{Schmidt}},
  \bibinfo{author}{\bibfnamefont{D.}~\bibnamefont{Stump}}, \bibnamefont{and}
  \bibinfo{author}{\bibfnamefont{C.~P.} \bibnamefont{Yuan}},
  \bibinfo{journal}{Phys. Rev.} \textbf{\bibinfo{volume}{D93}},
  \bibinfo{pages}{033006} (\bibinfo{year}{2016}), \eprint{1506.07443}.

\bibitem[{\citenamefont{Aaboud et~al.}(2018{\natexlab{e}})}]{Aaboud:2018knk}
\bibinfo{author}{\bibfnamefont{M.}~\bibnamefont{Aaboud}} \bibnamefont{et~al.}
  (\bibinfo{collaboration}{ATLAS}) (\bibinfo{year}{2018}{\natexlab{e}}),
  \eprint{1804.06174}.

\bibitem[{\citenamefont{Collaboration}(2017{\natexlab{a}})}]{CMS:2017xxp}
\bibinfo{author}{\bibfnamefont{C.}~\bibnamefont{Collaboration}}
  (\bibinfo{collaboration}{CMS}) (\bibinfo{year}{2017}{\natexlab{a}}).

\bibitem[{\citenamefont{Sirunyan
  et~al.}(2018{\natexlab{c}})}]{Sirunyan:2017isc}
\bibinfo{author}{\bibfnamefont{A.~M.} \bibnamefont{Sirunyan}}
  \bibnamefont{et~al.} (\bibinfo{collaboration}{CMS}), \bibinfo{journal}{Phys.
  Lett.} \textbf{\bibinfo{volume}{B781}}, \bibinfo{pages}{244}
  (\bibinfo{year}{2018}{\natexlab{c}}), \eprint{1710.04960}.

\bibitem[{\citenamefont{Collaboration}(2017{\natexlab{b}})}]{CMS:2017vdr}
\bibinfo{author}{\bibfnamefont{C.}~\bibnamefont{Collaboration}}
  (\bibinfo{collaboration}{CMS}) (\bibinfo{year}{2017}{\natexlab{b}}).

\bibitem[{\citenamefont{Sirunyan
  et~al.}(2018{\natexlab{d}})}]{Sirunyan:2017djm}
\bibinfo{author}{\bibfnamefont{A.~M.} \bibnamefont{Sirunyan}}
  \bibnamefont{et~al.} (\bibinfo{collaboration}{CMS}), \bibinfo{journal}{Phys.
  Lett.} \textbf{\bibinfo{volume}{B778}}, \bibinfo{pages}{101}
  (\bibinfo{year}{2018}{\natexlab{d}}), \eprint{1707.02909}.

\bibitem[{\citenamefont{Collaboration}(2017{\natexlab{c}})}]{CMS:2017ihs}
\bibinfo{author}{\bibfnamefont{C.}~\bibnamefont{Collaboration}}
  (\bibinfo{collaboration}{CMS}) (\bibinfo{year}{2017}{\natexlab{c}}).

\bibitem[{\citenamefont{Baur et~al.}(2004)\citenamefont{Baur, Plehn, and
  Rainwater}}]{Baur:2003gp}
\bibinfo{author}{\bibfnamefont{U.}~\bibnamefont{Baur}},
  \bibinfo{author}{\bibfnamefont{T.}~\bibnamefont{Plehn}}, \bibnamefont{and}
  \bibinfo{author}{\bibfnamefont{D.~L.} \bibnamefont{Rainwater}},
  \bibinfo{journal}{Phys. Rev.} \textbf{\bibinfo{volume}{D69}},
  \bibinfo{pages}{053004} (\bibinfo{year}{2004}), \eprint{hep-ph/0310056}.

\bibitem[{\citenamefont{Baur et~al.}(2003{\natexlab{a}})\citenamefont{Baur,
  Plehn, and Rainwater}}]{Baur:2003gpa}
\bibinfo{author}{\bibfnamefont{U.}~\bibnamefont{Baur}},
  \bibinfo{author}{\bibfnamefont{T.}~\bibnamefont{Plehn}}, \bibnamefont{and}
  \bibinfo{author}{\bibfnamefont{D.~L.} \bibnamefont{Rainwater}},
  \bibinfo{journal}{Phys. Rev.} \textbf{\bibinfo{volume}{D68}},
  \bibinfo{pages}{033001} (\bibinfo{year}{2003}{\natexlab{a}}),
  \eprint{hep-ph/0304015}.

\bibitem[{\citenamefont{Dolan et~al.}(2012)\citenamefont{Dolan, Englert, and
  Spannowsky}}]{Dolan:2012rv}
\bibinfo{author}{\bibfnamefont{M.~J.} \bibnamefont{Dolan}},
  \bibinfo{author}{\bibfnamefont{C.}~\bibnamefont{Englert}}, \bibnamefont{and}
  \bibinfo{author}{\bibfnamefont{M.}~\bibnamefont{Spannowsky}},
  \bibinfo{journal}{JHEP} \textbf{\bibinfo{volume}{10}}, \bibinfo{pages}{112}
  (\bibinfo{year}{2012}), \eprint{1206.5001}.

\bibitem[{\citenamefont{Barr et~al.}(2014)\citenamefont{Barr, Dolan, Englert,
  and Spannowsky}}]{Barr:2013tda}
\bibinfo{author}{\bibfnamefont{A.~J.} \bibnamefont{Barr}},
  \bibinfo{author}{\bibfnamefont{M.~J.} \bibnamefont{Dolan}},
  \bibinfo{author}{\bibfnamefont{C.}~\bibnamefont{Englert}}, \bibnamefont{and}
  \bibinfo{author}{\bibfnamefont{M.}~\bibnamefont{Spannowsky}},
  \bibinfo{journal}{Phys. Lett.} \textbf{\bibinfo{volume}{B728}},
  \bibinfo{pages}{308} (\bibinfo{year}{2014}), \eprint{1309.6318}.

\bibitem[{\citenamefont{Ferreira~de Lima et~al.}(2014)\citenamefont{Ferreira~de
  Lima, Papaefstathiou, and Spannowsky}}]{deLima:2014dta}
\bibinfo{author}{\bibfnamefont{D.~E.} \bibnamefont{Ferreira~de Lima}},
  \bibinfo{author}{\bibfnamefont{A.}~\bibnamefont{Papaefstathiou}},
  \bibnamefont{and}
  \bibinfo{author}{\bibfnamefont{M.}~\bibnamefont{Spannowsky}},
  \bibinfo{journal}{JHEP} \textbf{\bibinfo{volume}{08}}, \bibinfo{pages}{030}
  (\bibinfo{year}{2014}), \eprint{1404.7139}.

\bibitem[{\citenamefont{Wardrope et~al.}(2015)\citenamefont{Wardrope, Jansen,
  Konstantinidis, Cooper, Falla, and Norjoharuddeen}}]{Wardrope:2014kya}
\bibinfo{author}{\bibfnamefont{D.}~\bibnamefont{Wardrope}},
  \bibinfo{author}{\bibfnamefont{E.}~\bibnamefont{Jansen}},
  \bibinfo{author}{\bibfnamefont{N.}~\bibnamefont{Konstantinidis}},
  \bibinfo{author}{\bibfnamefont{B.}~\bibnamefont{Cooper}},
  \bibinfo{author}{\bibfnamefont{R.}~\bibnamefont{Falla}}, \bibnamefont{and}
  \bibinfo{author}{\bibfnamefont{N.}~\bibnamefont{Norjoharuddeen}},
  \bibinfo{journal}{Eur. Phys. J.} \textbf{\bibinfo{volume}{C75}},
  \bibinfo{pages}{219} (\bibinfo{year}{2015}), \eprint{1410.2794}.

\bibitem[{\citenamefont{Baur et~al.}(2002)\citenamefont{Baur, Plehn, and
  Rainwater}}]{Baur:2002rb}
\bibinfo{author}{\bibfnamefont{U.}~\bibnamefont{Baur}},
  \bibinfo{author}{\bibfnamefont{T.}~\bibnamefont{Plehn}}, \bibnamefont{and}
  \bibinfo{author}{\bibfnamefont{D.~L.} \bibnamefont{Rainwater}},
  \bibinfo{journal}{Phys. Rev. Lett.} \textbf{\bibinfo{volume}{89}},
  \bibinfo{pages}{151801} (\bibinfo{year}{2002}), \eprint{hep-ph/0206024}.

\bibitem[{\citenamefont{Baur et~al.}(2003{\natexlab{b}})\citenamefont{Baur,
  Plehn, and Rainwater}}]{Baur:2002qd}
\bibinfo{author}{\bibfnamefont{U.}~\bibnamefont{Baur}},
  \bibinfo{author}{\bibfnamefont{T.}~\bibnamefont{Plehn}}, \bibnamefont{and}
  \bibinfo{author}{\bibfnamefont{D.~L.} \bibnamefont{Rainwater}},
  \bibinfo{journal}{Phys. Rev.} \textbf{\bibinfo{volume}{D67}},
  \bibinfo{pages}{033003} (\bibinfo{year}{2003}{\natexlab{b}}),
  \eprint{hep-ph/0211224}.

\bibitem[{\citenamefont{Papaefstathiou
  et~al.}(2013)\citenamefont{Papaefstathiou, Yang, and
  Zurita}}]{Papaefstathiou:2012qe}
\bibinfo{author}{\bibfnamefont{A.}~\bibnamefont{Papaefstathiou}},
  \bibinfo{author}{\bibfnamefont{L.~L.} \bibnamefont{Yang}}, \bibnamefont{and}
  \bibinfo{author}{\bibfnamefont{J.}~\bibnamefont{Zurita}},
  \bibinfo{journal}{Phys. Rev.} \textbf{\bibinfo{volume}{D87}},
  \bibinfo{pages}{011301} (\bibinfo{year}{2013}), \eprint{1209.1489}.

\bibitem[{\citenamefont{Sirunyan
  et~al.}(2018{\natexlab{e}})}]{Sirunyan:2018aui}
\bibinfo{author}{\bibfnamefont{A.~M.} \bibnamefont{Sirunyan}}
  \bibnamefont{et~al.} (\bibinfo{collaboration}{CMS}),
  \bibinfo{journal}{Submitted to: Phys. Lett.}
  (\bibinfo{year}{2018}{\natexlab{e}}), \eprint{1811.08459}.

\bibitem[{\citenamefont{Sirunyan
  et~al.}(2018{\natexlab{f}})}]{Sirunyan:2018zut}
\bibinfo{author}{\bibfnamefont{A.~M.} \bibnamefont{Sirunyan}}
  \bibnamefont{et~al.} (\bibinfo{collaboration}{CMS}), \bibinfo{journal}{JHEP}
  \textbf{\bibinfo{volume}{09}}, \bibinfo{pages}{007}
  (\bibinfo{year}{2018}{\natexlab{f}}), \eprint{1803.06553}.

\bibitem[{\citenamefont{Cadamuro}(2015)}]{Cadamuro:2015lbd}
\bibinfo{author}{\bibfnamefont{L.}~\bibnamefont{Cadamuro}}
  (\bibinfo{collaboration}{CMS}), \bibinfo{journal}{PoS}
  \textbf{\bibinfo{volume}{EPS-HEP2015}}, \bibinfo{pages}{226}
  (\bibinfo{year}{2015}).

\bibitem[{\citenamefont{Khachatryan
  et~al.}(2015{\natexlab{b}})}]{CMS-DP-2015-009}
\bibinfo{author}{\bibfnamefont{V.}~\bibnamefont{Khachatryan}}
  \bibnamefont{et~al.} (\bibinfo{collaboration}{CMS})
  (\bibinfo{year}{2015}{\natexlab{b}}),
  \urlprefix\url{https://cds.cern.ch/record/2018400}.

\bibitem[{\citenamefont{Mastrolorenzo}(2016)}]{Mastrolorenzo:2016dyo}
\bibinfo{author}{\bibfnamefont{L.}~\bibnamefont{Mastrolorenzo}},
  \bibinfo{journal}{Nucl. Part. Phys. Proc.}
  \textbf{\bibinfo{volume}{273-275}}, \bibinfo{pages}{2518}
  (\bibinfo{year}{2016}).

\bibitem[{\citenamefont{Aaboud et~al.}(2018{\natexlab{f}})}]{Aaboud:2018sfw}
\bibinfo{author}{\bibfnamefont{M.}~\bibnamefont{Aaboud}} \bibnamefont{et~al.}
  (\bibinfo{collaboration}{ATLAS}), \bibinfo{journal}{Phys. Rev. Lett.}
  \textbf{\bibinfo{volume}{121}}, \bibinfo{pages}{191801}
  (\bibinfo{year}{2018}{\natexlab{f}}), \eprint{1808.00336}.

\end{thebibliography}


\end{document}